\documentclass[reprint,amsmath,amssymb,aps,prb,superscriptaddress,floatfix]{revtex4-2}

\pdfoutput=1
\usepackage{graphicx}
\usepackage{dcolumn}
\usepackage{bm}
\usepackage{braket}
\usepackage[mathlines]{lineno}
\usepackage{stackengine}
\usepackage{verbatim}
\usepackage{graphics}
\usepackage{hyperref}
\usepackage{xcolor}

\begin{document}

\preprint{APS/123-QED}

\title{Dynamics of spectral correlations in the entanglement Hamiltonian of the Aubry-Andr\'e-Harper model}

\author{Aamna Ahmed}
\affiliation{Department of Physics, Indian Institute of Science Education and Research, Bhopal, Madhya Pradesh 462066, India\\}
\author{Nilanjan Roy}
\affiliation{Department of Physics, Indian Institute of Science Education and Research, Bhopal, Madhya Pradesh 462066, India\\}
\affiliation{Department of Physics, Indian Institute of Science, Bangalore 560012, India\\}
\author{Auditya Sharma}%
\affiliation{Department of Physics, Indian Institute of Science Education and Research, Bhopal, Madhya Pradesh 462066, India\\}

\date{\today}

\begin{abstract}
We numerically study the evolution of spectral correlations in the
entanglement Hamiltonian (EH) of non-interacting fermions in the
Aubry-Andr\'e-Harper (AAH) model. We analyze the time evolution of the
EH spectrum in a nonequilibrium setting by studying several
quantities: spectral distribution, level statistics, entanglement
entropy, and spectral form factor (SFF) in the context of the
delocalization-localization transition in the AAH model. It is
observed that the SFF of the entanglement spectrum in the delocalized
phase and at the phase-transition point evolves in three time
intervals.  We make a systematic study of the emergence of these three
timescales for various initial states and find that the number of time
intervals remains three unless the Hamiltonian is tuned in the
localized phase or when the initial state is maximally entangled, when
there is a featureless time evolution. We find a broad direct
correlation between the entanglement entropy and the length of the
ramp of the SFF. We also find that in the delocalized phase the
spectral correlations are stronger in the center of the spectrum and
grow progressively weaker as more and more of the spectrum is
considered.
\end{abstract}

\maketitle


\section{INTRODUCTION}
The study of chaos in many-body quantum systems from a dynamical
viewpoint is of great current interest
\cite{chaos1,chaos2,chaos3,chaos4,chaos5,chaos6,chaos7,chaos8,chaos9,chaos10,chaos11}.
In the traditional approach which has mainly focused on short-range
correlations in the spectrum, level statistics, and universal spectral
properties through random matrix theory (RMT)~\cite{Haake, Mehta, RMT1} have been popular objects of
study. However other indicators of chaos such as the out-of-time order
correlators (OTOC)~\cite{OTOC1, OTOC2, OTOC3} and the spectral form
factor (SFF)~\cite{SFF1} are also of interest. For example, they have
been used to understand chaos in the Sachdev-Ye-Kitaev (SYK) model
which is important in the context of blackhole
physics~\cite{OTOC4,OTOC5,OTOC6,SFF2}.

The spectral form factor is a useful measure for describing spectral
statistics in quantum systems. It is defined as the Fourier transform
of the correlation function between two levels of the
spectrum~\cite{SFF3}:
\begin{equation}
g(\tau)=\left\langle\sum_{i,j}e^{-i\tau\left(\lambda_i-\lambda_j\right)}\right\rangle.
\end{equation}
For a chaotic quantum system, the SFF of the spectrum exhibits a
``ramp" signaling the presence of universal spectral correlations
whereas for an integrable system, the ramp is absent. Since it deals
with the long-range correlations in the system, it presents an
alternative picture from the known measures of level-statistics and
hence it is being studied in a variety of models such as Floquet
systems\cite{SFF5, SFF8, SFF9}, models featuring many-body localization etc~\cite{SFF4,
  SFF6, SFF7}. It has also been discussed in detail in the
context of various ensembles of RMT such as the Gaussian ensembles\cite{SFF10,SFF11} or
the Wishart ensembles~\cite{SFF3,SFF10} and hence can serve as
an indicator of quantum chaos\cite{Haake}.

In systems where delocalization-localization phase transitions can be
observed with a change in disorder strength, entanglement
entropy~\cite{model1, EE6} often provides valuable insights. In
general, the entanglement entropy is observed to be larger in the
delocalized phase due to the presence of more correlations than in the
localized phase. Thus one can study this quantity together with the
spectral form factor to understand the correlations of the spectrum
better. One such model which exhibits a delocalization-localization
transition even in one dimension is the Aubry-Andr\'e-Harper model
which is governed by a quasi-periodic disorder. All the
single-particle states of this system are either localized or
delocalized depending on the strength of the quasi-periodic
disorder~\cite{model2,model3,model4}.

It has been seen recently that the dynamics of the spectral form
factor serves as a useful probe in understanding the various stages of
approach to thermalization in chaotic (non-integrable) models
~\cite{model5} through the study of the correlations in the spectrum
of the entanglement Hamiltonian. It has been noticed that there is a
certain correspondence between the dynamics of the level repulsion,
entanglement entropy ~\cite{EED1,EED2,EED3}, and the development of
the spectral form factor. With the aid of a variety of
  quantities including gap ratio and entanglement entropy, Chang et
  al~\cite{model5} were able to identify three distinct timescales,
  which were observed in both the chaotic and MBL phases. In the
  backdrop of such interesting findings in an interacting model, it is
  of great interest to understand how these different timescales play
  out within a noninteracting model.

Here in this article, we consider the AAH model in which both
delocalized and localized phases are possible depending on the
strength of the disorder. It is a quadratic system with the
integrability weakly broken by the disorder. We study the development
of correlations in the entanglement Hamiltonian spectrum of the AAH
model through quench dynamics in its different phases. Tracking the
evolution of various quantities such as the gap ratio, Renyi
entropies, and spectral density together with the SFF, we find the
presence of three distinct timescales in both the delocalized phase
and at the critical point, while is predictably flat in the localized
phase. We also compare the SFF with the evolution of the higher-order
gap ratios and find correspondence between both the measures. From the
spectrum resolved study of the SFF, we find that the correlations in
the delocalized phase are dominant more at the center of the spectrum.
However, at the critical point the correlations are uniformly spread
out throughout the spectrum. A systematic study of the dynamics
starting from a variety of initial states allows us to conclude that
the length of the ramp (in the units of $\tau$) in the SFF is strongly
correlated with the magnitude of the entanglement entropy.
   
The paper is organized as follows. In Section~\ref{sec:level2}, the
model Hamiltonian, spectral form factor, and the entanglement
Hamiltonian are briefly introduced. The nonequilibrium dynamics of the
entanglement Hamiltonian is discussed in Section~\ref{sec:level3} for
various initial states: non-entangled and entangled. This section
comprises six subsections. In Subsections~\ref{subA}--\ref{subE} the
nonequilibrium dynamics for a non-entangled initial product state is
studied by computing various quantities such as the gap ratio,
entanglement entropy, and spectral form factor. Also, we look into the
(sub)system-size dependences of these quantities. In
Subsection~\ref{subF} we consider entangled initial states and study
the dynamics of the corresponding EH spectrum. Then we conclude in
Section~\ref{sec:level4}.

\section{\label{sec:level2}Model and Methods}
\subsection{Hamiltonian}
We consider the AAH model given by the Hamiltonian
\cite{model2,model3} :
\begin{equation}
 H=-t\sum_{i} (c_i^{\dagger}c_{i+1}+H.c.)+\sum_{i}2\lambda \cos(2\pi ib+\theta_p)c_i^{\dagger}c_{i}.
\end{equation}
Here $c^{\dagger}_i$ $(c_i)$ is the creation (annihilation) operator
on site $i$. The first term describes the nearest-neighbor hopping
along the chain where $t$ is the hopping parameter which we set to
unity. The second term describes the quasi-periodic on-site energy
where the strength of the quasi-periodic potential is $\lambda$, the
quasi-periodicity parameter `$b$' is taken to be an irrational number
and $\theta_p$ is an arbitrary global phase chosen randomly from a
uniform distribution in the range $[0,2\pi]$. The total number of
sites is $N$ and periodic boundary conditions are assumed. As is
well-known \cite{model7}, all the energy eigenstates are delocalized
when $\lambda< 1$, and all the energy eigenstates are localized when
$\lambda > 1$. $\lambda = 1$ is the critical point where all the
eigenstates are multifractal\cite{model8}. At $\lambda = 1$, the AAH model in
position space maps to itself in momentum space which makes the model
self-dual~\cite{model2,model6}. This is observed whenever the
quasiperiodicity parameter is chosen to be an irrational number. In
this article, following convention, we set $b$ to be the golden mean
$(\sqrt{5}-1)/2$.
 
\subsection{Spectral form factor}\label{SFF}

The spectral form factor is used to understand the correlations
present in a spectrum. Given a spectrum of $N$ eigenvalues, we consider the eigenvalue density:
\begin{equation}
\nu(\lambda)=\sum_{i=1}^{N} \delta(\lambda-\lambda_i).
\end{equation}
Defining the Fourier transform of the eigenvalue density as:
\begin{equation}
Z(\tau)=\sum_{i=1}^{N}e^{-i\tau\lambda_i},
\end{equation}
the spectral form factor is conveniently expressed in terms of $Z(\tau)$ as:
\begin{equation}
g(\tau)=\left\langle Z(\tau)Z^*(\tau)\right\rangle=\left\langle\sum_{i,j=1}^{N}e^{-i\tau\left(\lambda_i-\lambda_j\right)}\right\rangle,
\label{ab5}
\end{equation}
where the average is taken over an ensemble in order to remove
non-universal fluctuations in $g(\tau)$ (and
  $g_c(\tau)$), since it is not a self-averaging quantity
~\cite{avg1,avg2}. In our work, we average our data
  over a set of values of the random phase $\theta_p$, as described in
  the previous subsection. The spectral form factor exhibits a
typical structure corresponding to universal spectral correlations in
the system. Starting at $N^2$ at $\tau=0$ it starts to decrease until
it reaches a minimum, after which it shows a strictly linear growth
called the ``ramp" which is a signature of the presence of long-range
correlations in the system. The curve then becomes constant at long
times (at Heisenberg time $\tau_H$ defined as $2\pi$ times the inverse
of the mean level spacing), reaching its `plateau' value~\cite{avg3}:
$\lim_{\tau\rightarrow \infty} g(\tau)=N$.

Another quantity of interest is the connected spectral form factor (CSFF)
\begin{equation}
g_c(\tau)=\left\langle Z(\tau)Z^*(\tau)\right\rangle-\left\langle Z(\tau)\right\rangle\left\langle Z^*(\tau)\right\rangle,
\label{ab6}
\end{equation}
which is obtained by deducting the disconnected part from
$g(\tau)$. The disconnected part which is non-universal ~\cite{SFF3}
is active primarily in the smaller $\tau$ regime, and leads to a
reduction in the size of the ramp seen in $g(\tau)$. $g_c(\tau)$
explicitly removes the non-universal disconnected part, and exhibits a
longer ramp for the timescale that corresponds to energy differences
that lie in the universal regime. 

\subsection{Entanglement Hamiltonian and correlation matrix }
We consider non-interacting spinless fermions at half-filling in the
AAH model. To calculate the entanglement Hamiltonian
~\cite{EE1,EE2,EE3}, one can divide the system into two contiguous
parts, one with $N_A$ number of sites and the other with $N_B=N-N_A$ sites
where $N$ is the total number of sites. Given the density matrix $\rho$ the
reduced density matrix can be calculated as $\rho_A=Tr_B
(\rho)$. Using the reduced density matrix, the von Neumann entropy of
the subsystem can be calculated as $S_A=-Tr\left(\rho_A\log
\rho_A\right)$. But for a single Slater determinant many-body
state such as for free fermions, using Wick's theorem we can write
\begin{equation}
\rho_A= \frac{e^{-H_A}}{Z},
\label{ab4}
\end{equation}
 where $H_A =\sum_{ij}^{N_A}H_{ij}^A c_i^{\dagger}c_j$~\cite{EE1} is
 called the entanglement Hamiltonian, and $Z$ satisfies the condition
 Tr$(\rho_A) = 1$. The correlation matrix for the chosen subsystem $A$
 is given as: $C_{ij} = \left\langle c_i^{\dagger}c_j\right\rangle$
 where $i,j =1,2......N_A$. The information held in $\rho_A$ of size
 $2^{N_A} \times 2^{N_A}$ can be captured in terms of the correlation
 matrix $C$ of size $N_A \times N_A$~\cite{EE2}. This subsystem
 correlation matrix is related to the entanglement Hamiltonian as
 \begin{equation}
   H_A=\ln\frac{1-C}{C}.
   \label{eq:ab}
 \end{equation} 
 Thus, we can find the eigenvalues of the entanglement Hamiltonian
 from the eigenvalues of the correlation matrix. In
   the present work, we will study the spectral form factor of the
   entanglement Hamiltonian. While it is also meaningful to study the
   spectral correlations of the reduced density matrix of the
   subsystem ~\cite{model5,SFF3}, it is easier to consider the
   entanglement Hamiltonian directly due to its intimate relationship
   with the reduced density matrix (see Eq.~\ref{ab4}).
\begin{figure}[b]
\vspace{3mm}
\centering
\stackunder{\hspace{-3.5cm}(a)}{\includegraphics[height=3.3cm, width=4cm]{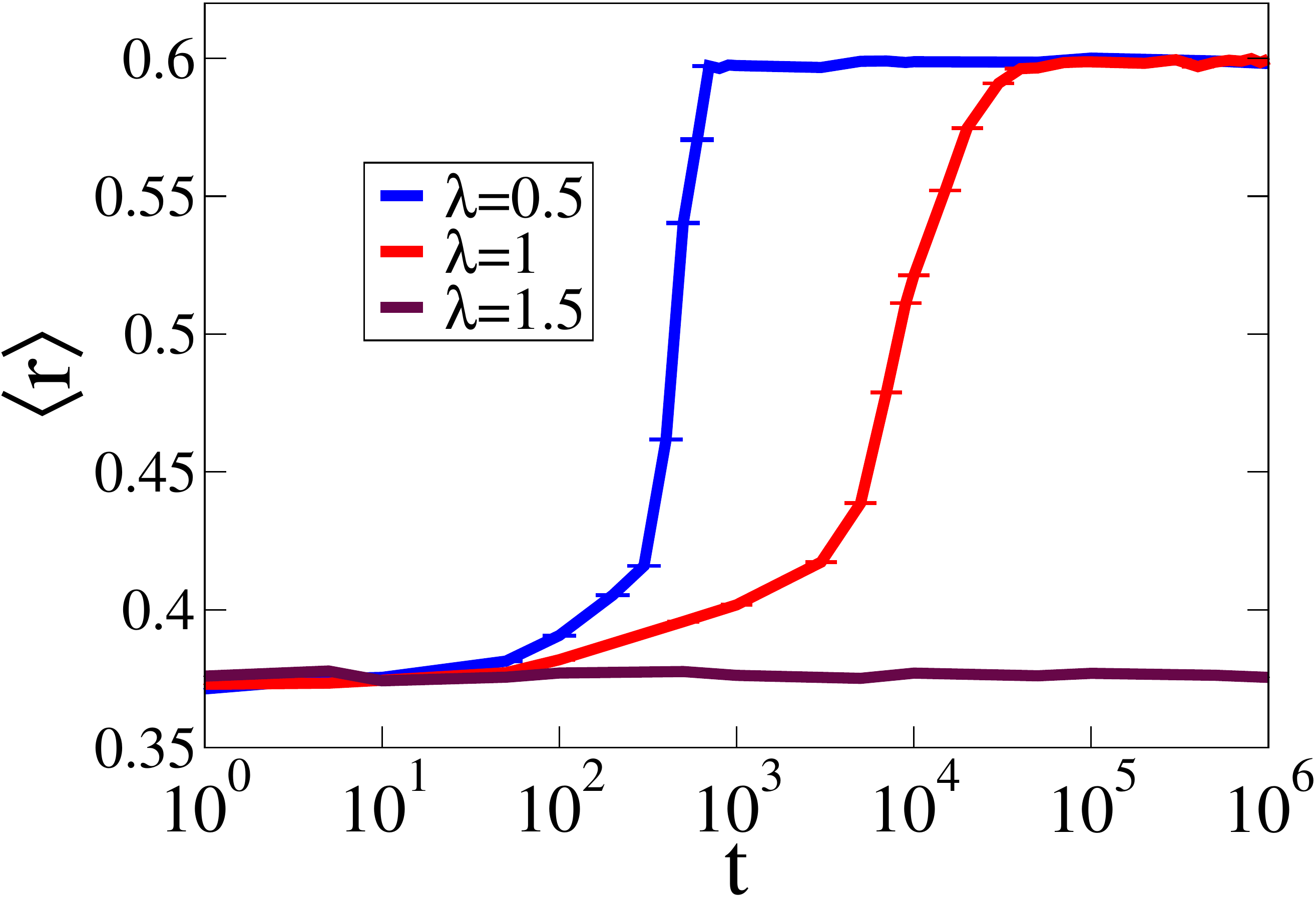}}\hspace{1mm}
\stackunder{\hspace{-3.5cm}(b)}{\includegraphics[height=3.3cm, width=4cm]{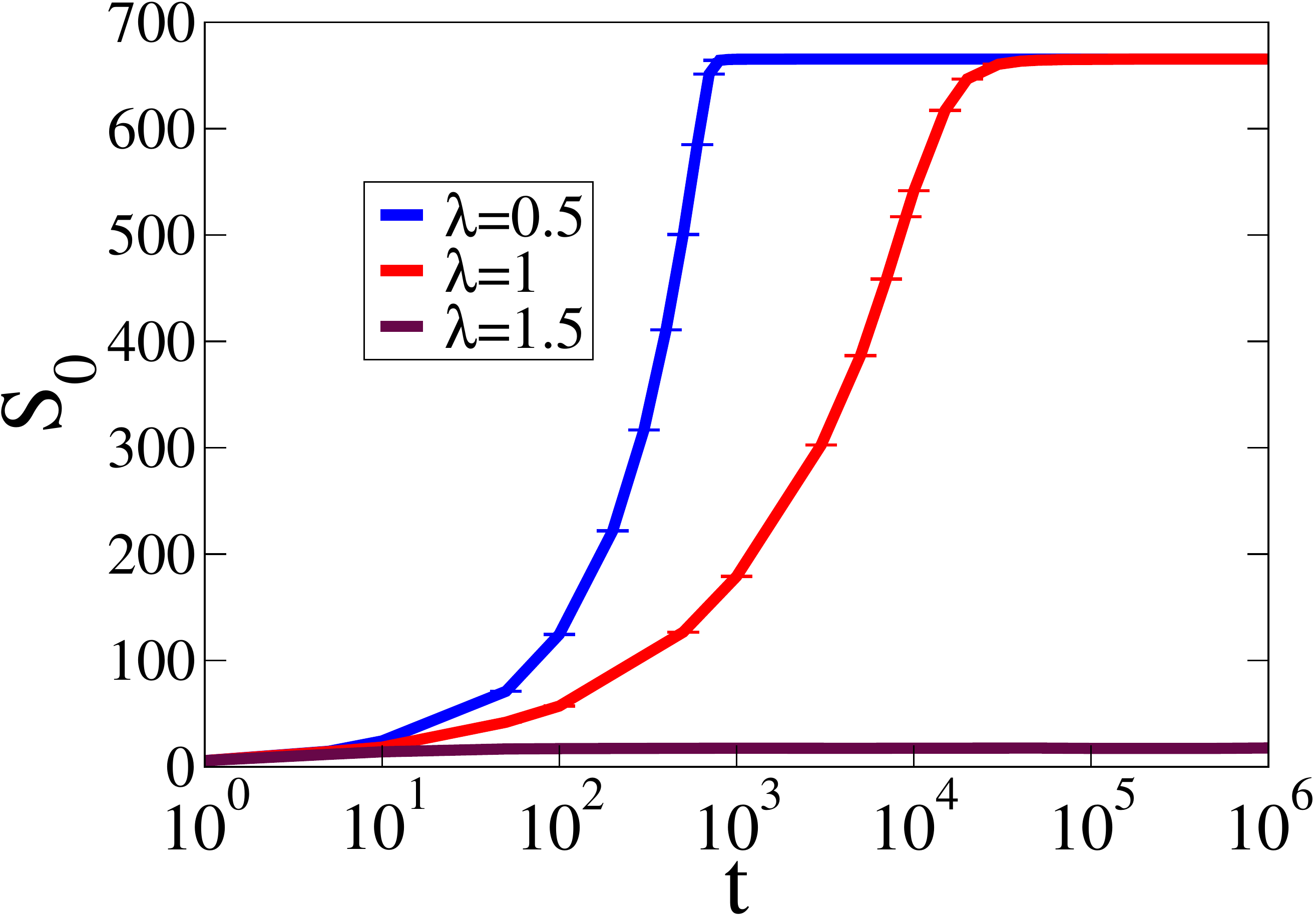}}
\stackunder{\hspace{-3.5cm}(c)}{\includegraphics[height=3.3cm, width=4cm]{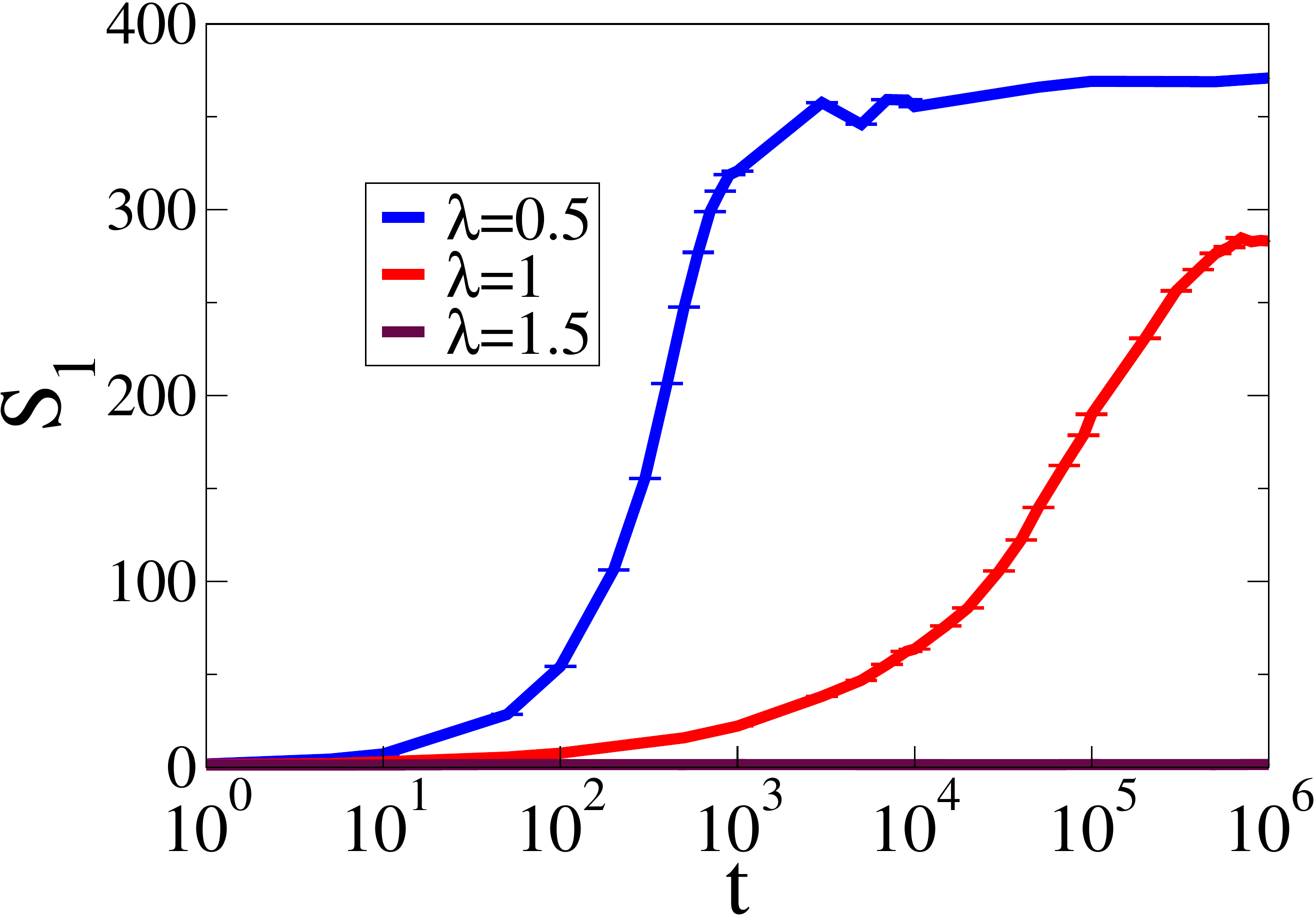}}\hspace{1mm}
\stackunder{\hspace{-3.5cm}(d)}{\includegraphics[height=3.3cm, width=4cm]{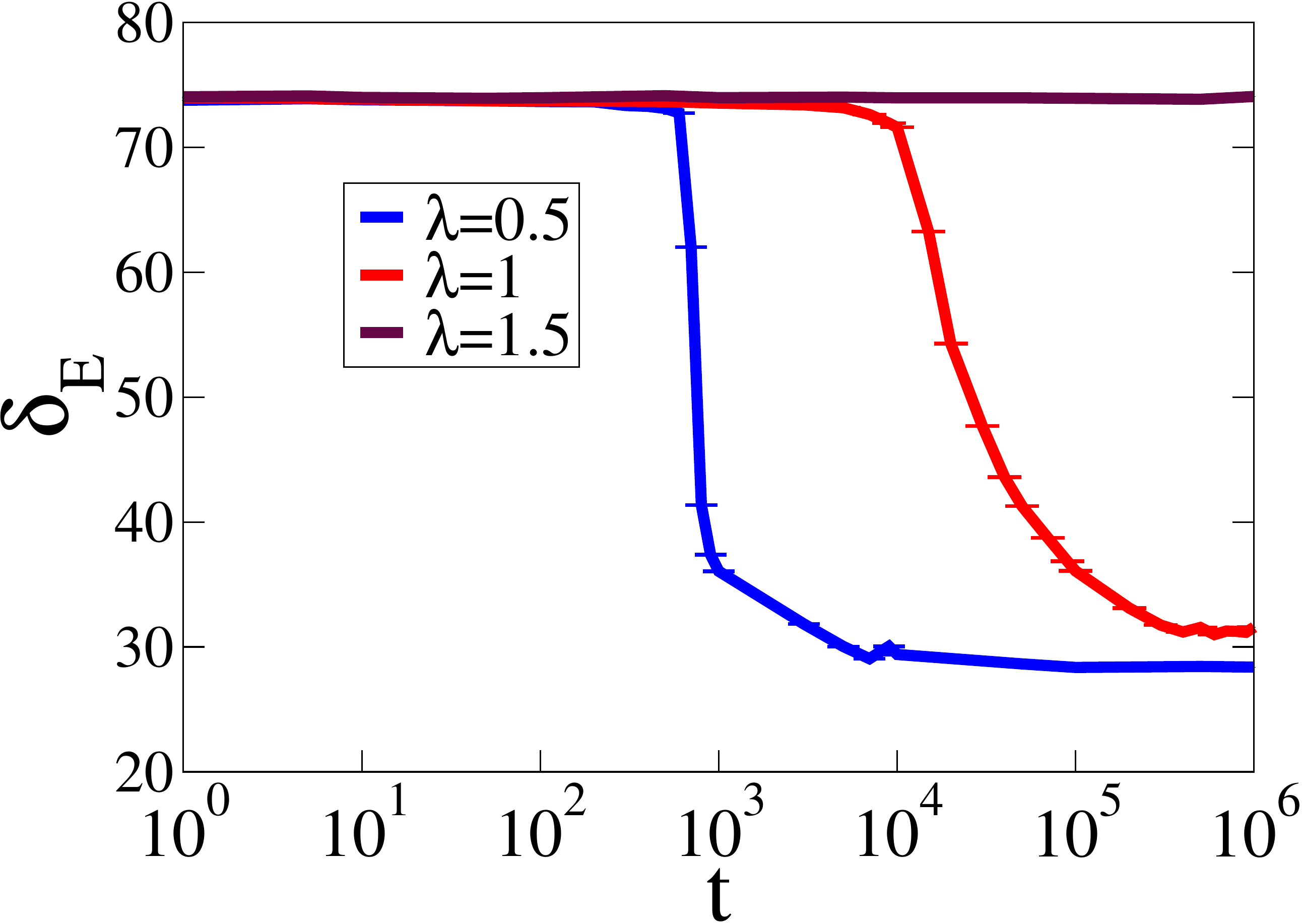}}
\caption{\label{fig1}Spectral properties of the EH in different phases
  under time evolution. Here $\lambda=0.5, 1, 1.5$ signify the
  delocalized phase, phase-transition point and the localized phase
  respectively.  Time evolution of (a) Nearest-neighbour spacing ratio
  $\left\langle r\right\rangle$, (b) Zeroth-order Renyi entropy $S_0$,
  (c) von Neumann entropy $S_1$ and (d) Entanglement Bandwidth
  $\delta_E$. Here the system size is $N=1920$ and subsystem size is
  $N_A=N/2$. The average has been taken over $100$ random values of
  $\theta_p$.}
\end{figure}
\begin{figure}
\vspace{3mm}
\centering
\includegraphics[height=4cm, width=6.0cm]{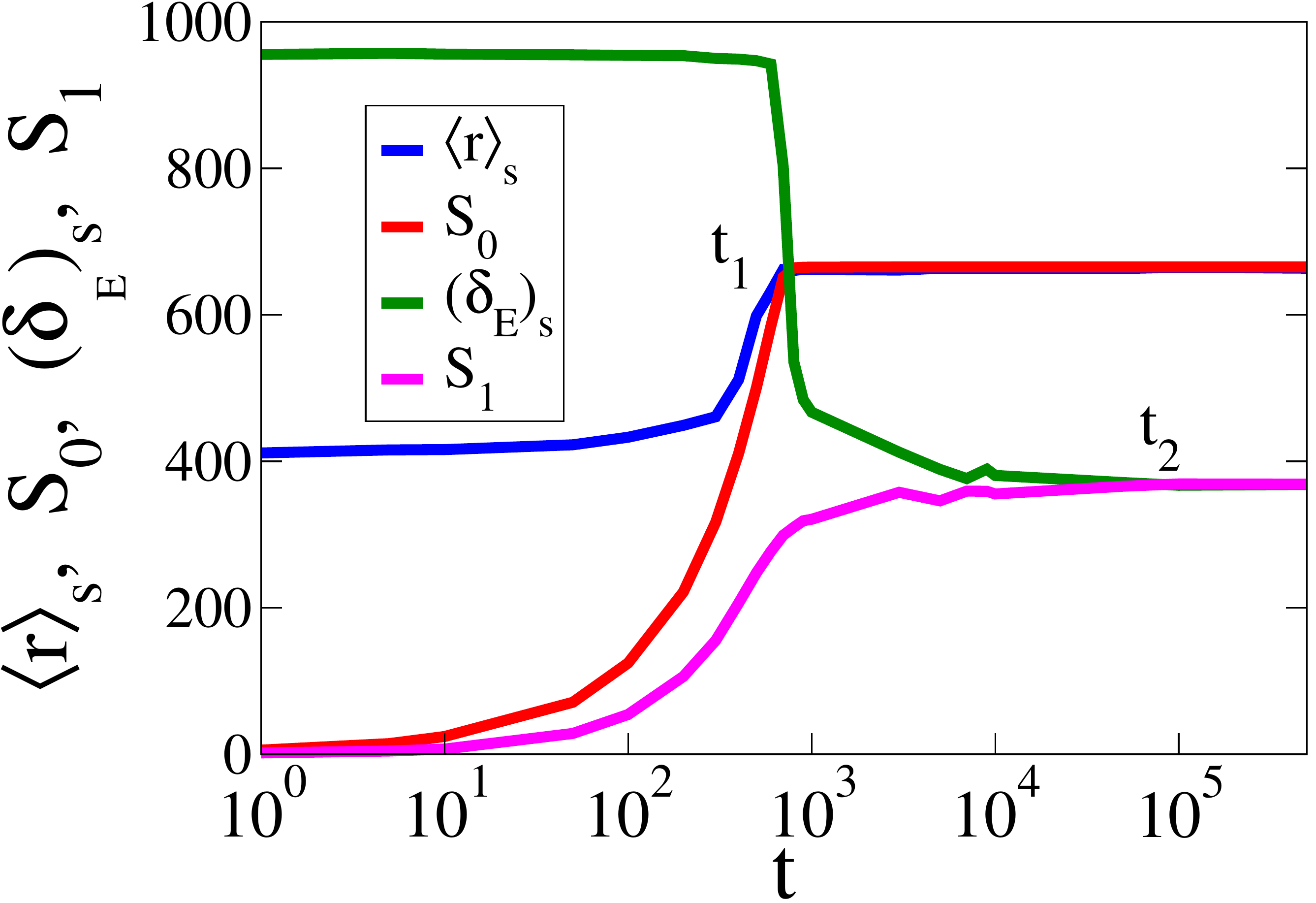}
\caption{\label{fig2}Scaled gap ratio $\left\langle
  r\right\rangle_s(=\eta_1\left\langle r\right\rangle)$ and zeroth
  order Renyi entropy $S_0$ saturating at time $t_1$ while scaled
  entanglement bandwidth $(\delta_E)_s (=\eta_2 \delta_E)$ and von
  Neumann entropy $S_1$ saturating at time $t_2$. The scaling factors
  are $\eta_1=1108.0$ and $\eta_2=12.9533$ in the delocalized phase
  ($\lambda = 0.5$). The system size is $N=1920$ and $N_A=N/2$. The
  average has been taken over $100$ random values of $\theta_p$.}
\end{figure}
\begin{figure}[b]
\centering
\stackunder{\hspace{-5.1cm}(a)}{\includegraphics[height=3.8cm, width=6cm]{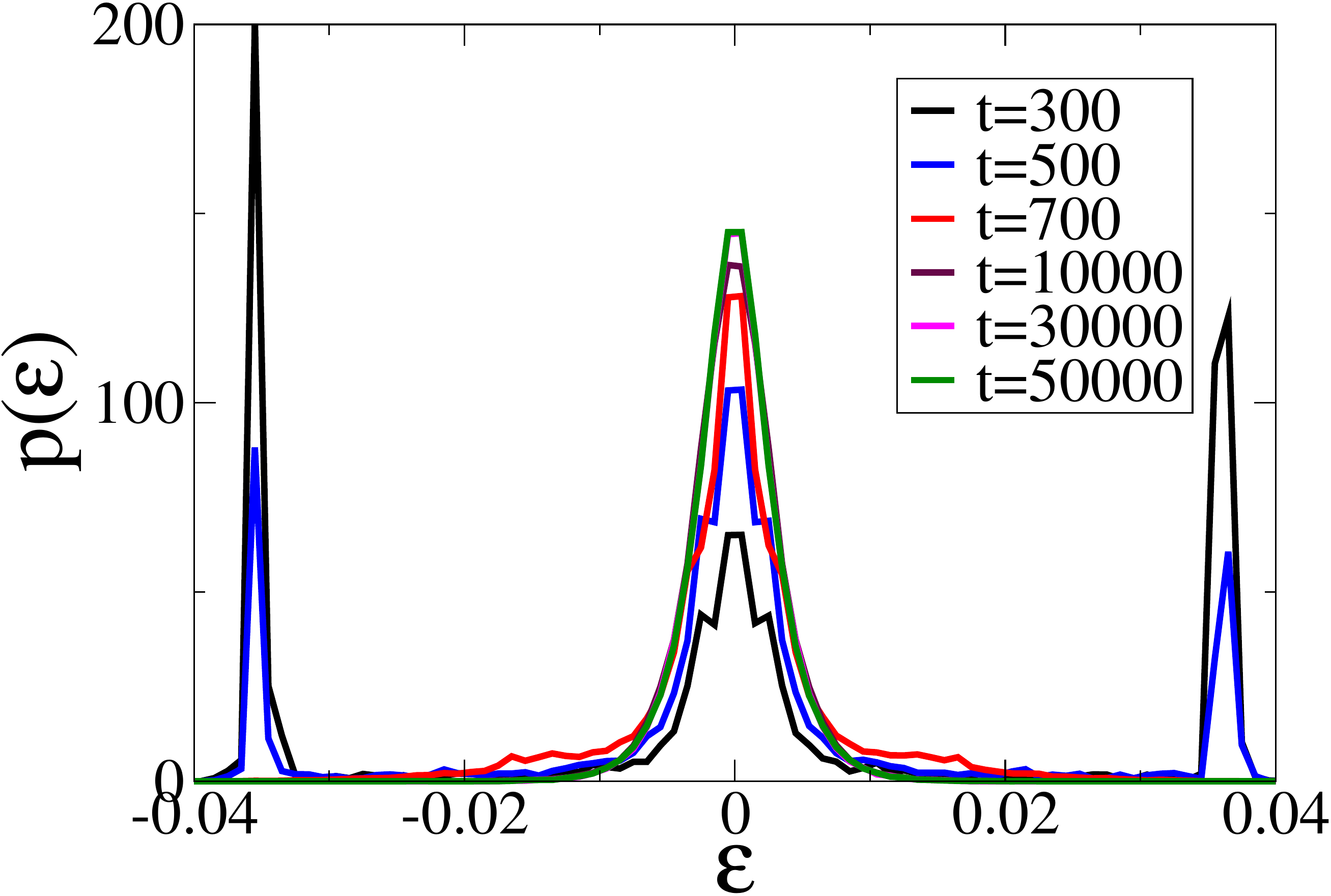}}\hspace{3mm}
\stackunder{\hspace{-5.1cm}(b)}{\includegraphics[height=3.8cm, width=6cm]{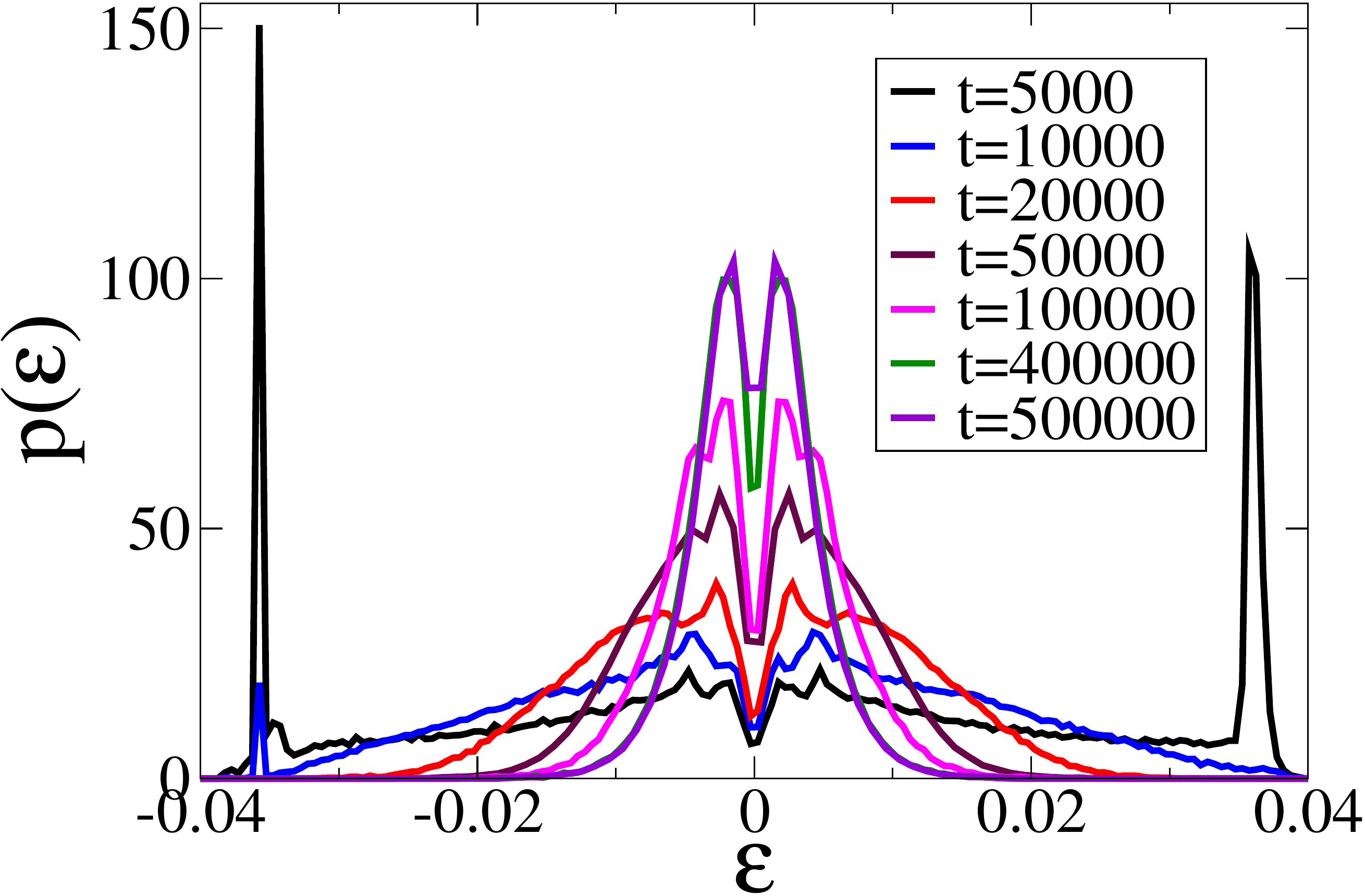}}\hspace{3mm}
\stackunder{\hspace{-5.1cm}(c)}{\includegraphics[height=3.8cm, width=6cm]{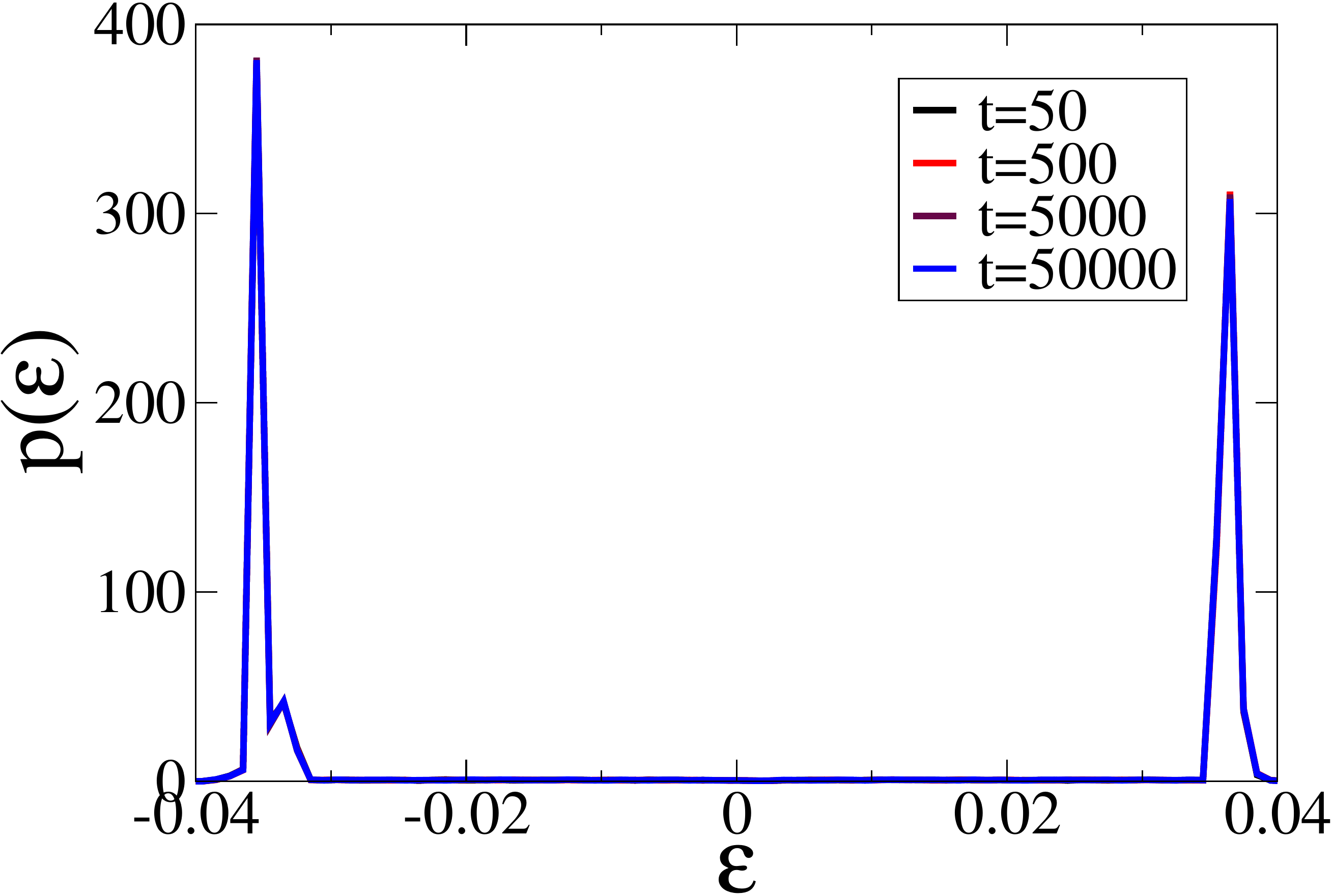}}
\caption{\label{fig3}Time evolution of spectral density of EH
  (scaled with subsystem size $N_A$) in different phases.
  (a) $\lambda=0.5$: delocalized phase, (b) $\lambda=1$: phase
  transition point and (c) $\lambda=1.5$: localized phase. Here the
  system size is $N=1920$ and $N_A=N/2$. The average has been taken
  over $100$ random values of $\theta_p$. }
\end{figure}
\section{Nonequilibrium dynamics}\label{sec:level3}
Here we intend to understand the dynamical evolution of the spectral
correlations of the entanglement Hamiltonian in different many-body
phases of the AAH model. The dynamics are studied in a nonequilibrium
setting starting from the initial product state $\ket
{\Phi_0}=\prod_{i=1}^{N/2} c^\dagger_{2i}\ket0$, where $N$ is an even number.
The unitary time evolution of the state is represented as 
\begin{equation}
\ket{\Phi(t)}=e^{-iHt}\ket{\Phi_0}.
\end{equation}
The time-dependent correlation matrix is then given by
\begin{equation}
C_{ij}(t) = \bra{\Phi(t)}c^{\dagger}_{i} c_{j}\ket{\Phi(t)}=\bra{\Phi_0}c^{\dagger}_{i}(t) c_{j}(t)\ket{\Phi_0},
\end{equation}
where the time evolution of the operator $c_i(t)$ is obtained in the Heisenberg picture~\cite{EE2}.
Thus the full time-dependent correlation matrix in the Heisenberg picture is given as
\begin{equation}
C(t)=e^{iHt}C(0)e^{-iHt},
\end{equation}
where $C(0)$ contains the initial correlations~\cite{Eisler_2012}. We
then obtain the correlation matrix and hence the time-dependent
entanglement Hamiltonian for the choosen subsystem using
Eq.~\ref{eq:ab}. We stick to this non-entangled initial product state
except in Subsection~\ref{subF} where we study the dynamics of
initially entangled states. Our non-equilibrium setting is similar to
studying a global quench in the hopping model. Using
Eq.~(\ref{eq:ab}), one can find the time-dependent entanglement
Hamiltonian, whose eigenvalues are used to study the dynamics of a
number of useful quantities such as the gap ratio, Renyi entropies,
and spectral form factor.

\subsection{Gap ratio and Renyi entropy}\label{subA}
The gap ratio \cite{gap-ratio} is defined as 
\begin{equation}
r_k=\frac{min(s_k,s_{k+1})}{max(s_k,s_{k+1})},
\label{eq:cd}
\end{equation}
where $s_k=e_{k}-e_{k-1}$ and $e_{k}$'s are the eigenvalues arranged
in ascending order. The mean gap ratio $\left\langle r\right\rangle$
is calculated by averaging $r_k$ over the energy spectrum.  In
Fig.~\ref{fig1}(a), $\langle r \rangle$ starts from a Poisson-like value
($\approx 0.386$). It evolves with time in a manner consistent with
growing nearest neighbour correlations until it reaches the Gaussian unitary ensemble (GUE) value
($\approx 0.599$~\cite{gap-ratio}). The dynamics of higher-order
gap ratios involving gaps between more than nearest energy levels is
also interesting and will be discussed later (see
Subsection~\ref{subD}). The time at which the first order gap ratio
hits its saturating value is denoted by $t_1$ and hence the first
timescale is defined by $t < t_1$.

The $\alpha$\textsuperscript{th} order time-dependent Renyi entropy in
terms of the eigenvalues $\lambda_k$'s of the subsystem correlation matrix
$C(t)$ is given as~\cite{Renyi}:
\begin{equation}
S_{\alpha}(t)=\frac{1}{1-\alpha} \sum\limits_{k=1}^{N_A}  \ln\big[(1-\lambda_k(t))^\alpha + (\lambda_k(t))^{\alpha} \big],
\end{equation} 
where $\alpha$ is the order of the Renyi entropy. The zeroth order 
Renyi entropy $S_0$, by definition determines the number of non-zero
eigenvalues or the rank of the reduced density matrix. The first order
Renyi entropy also called the von Neumann entropy can be defined as
$S_1=\lim\limits_{\alpha\rightarrow1}S_\alpha$ which is given by:
\begin{equation}
S_1=-\sum\limits_{k=1}^{N_A} \big[\lambda_k(t)\ln(\lambda_k(t)) +(1-\lambda_k(t))\ln((1-\lambda_k(t)) \big].
\label{eq:ab1}
\end{equation}
The above expression can also be derived via the relation in
Eqn.~\ref{eq:ab}~\cite{EE3}. For the initial product
state as considered here $S_{\alpha}=0$ at $t=0$, but
$S_{\alpha}(t)>0$ otherwise. It turns out that $S_0$ also hits
saturation along with the first-order gap ratio around
$t_1$. We have observed that only when the correlation
  matrix is obtained from a product state, $S_0$ saturates at time
  $t_1$. In all other cases, (see Subsection~\ref{subF}) $S_0$ remains
  saturated for times $t<t_1$ also. We are able to identify a second
timescale where the first order gap ratio (and $S_0$ as shown in
Fig.~\ref{fig1}(b)) is saturated whereas $S_1$ keeps on increasing.
At time $t_2$, $S_1$ also saturates [see Fig.~\ref{fig1}(c)], thus the
second timescale is defined by the interval $t_1 < t <
t_2$~\cite{model9}. We are also able to identify a third timescale
which is marked by $t > t_2$.

\begin{figure*}
\centering
\stackunder{\hspace{-5.0cm}(a)}{\includegraphics[height=3.6cm, width=5.1cm]{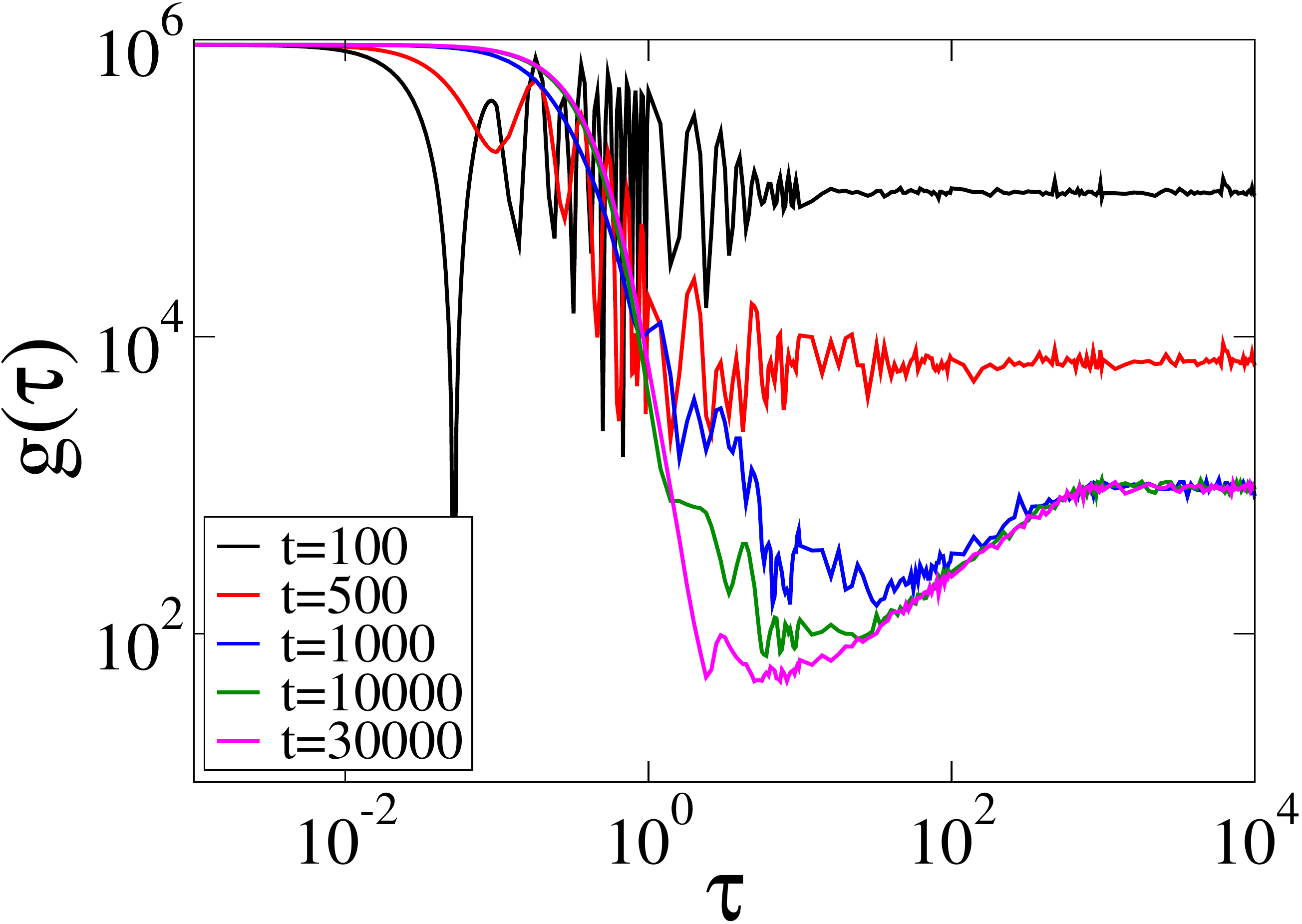}}\hspace{3mm}
\stackunder{\hspace{-5.0cm}(b)}{\includegraphics[height=3.6cm, width=5.1cm]{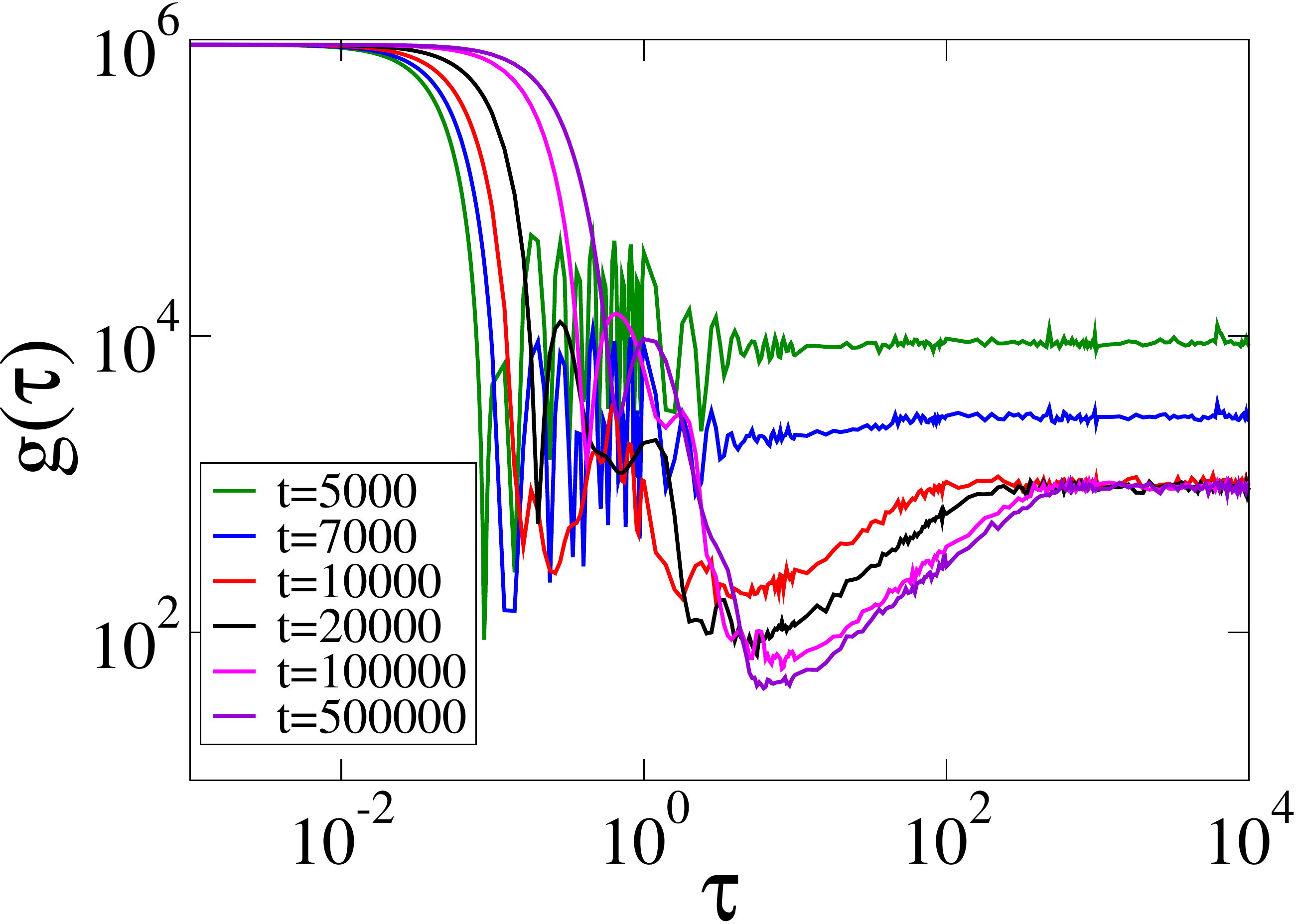}}\hspace{3mm}
\stackunder{\hspace{-5.0cm}(c)}{\includegraphics[height=3.6cm, width=5.1cm]{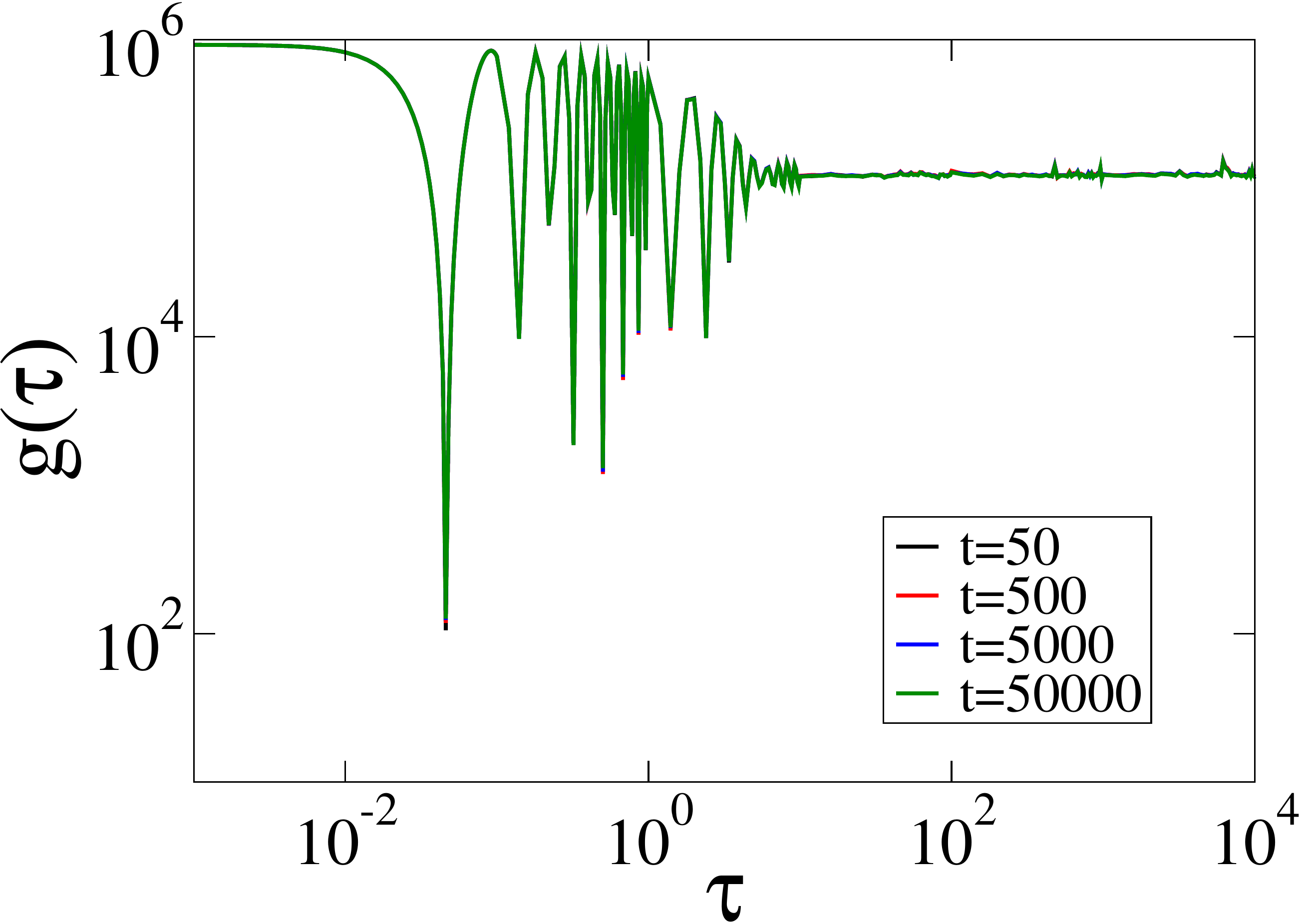}}

\stackunder{\hspace{-5.0cm}(d)}{\includegraphics[height=3.6cm, width=5.1cm]{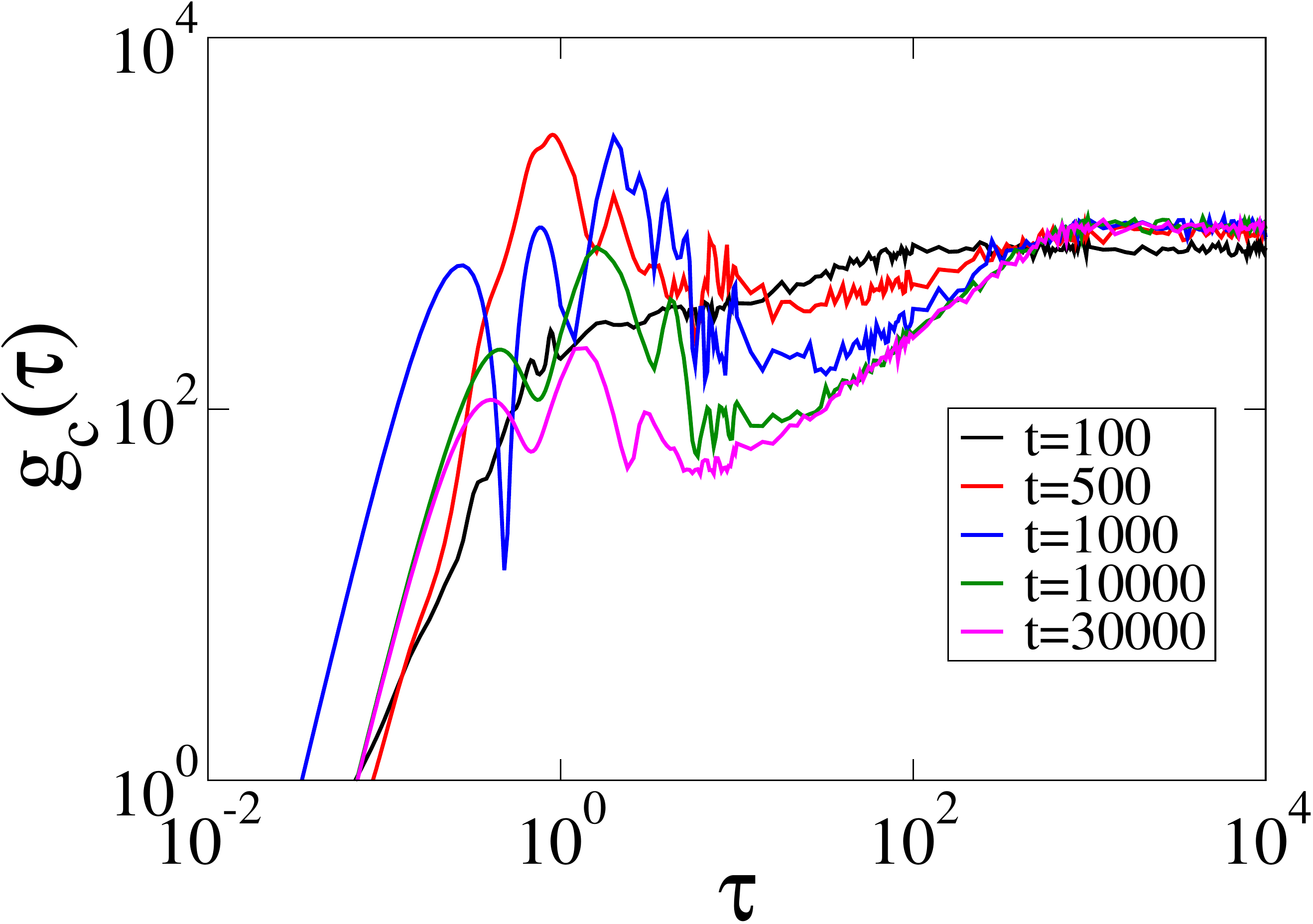}}\hspace{3mm}
\stackunder{\hspace{-5.0cm}(e)}{\includegraphics[height=3.6cm, width=5.1cm]{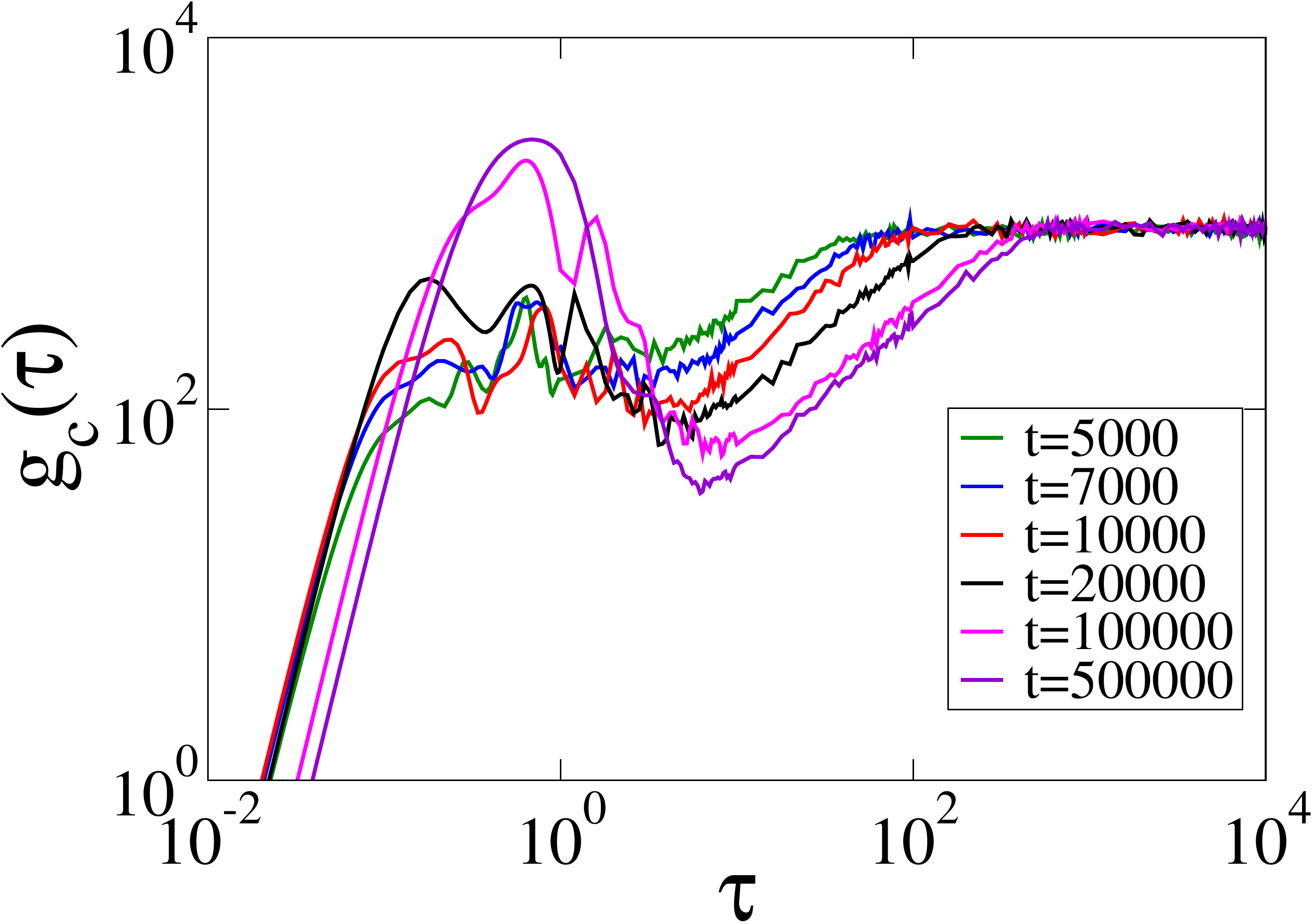}}\hspace{3mm}
\stackunder{\hspace{-5.0cm}(f)}{\includegraphics[height=3.6cm, width=5.1cm]{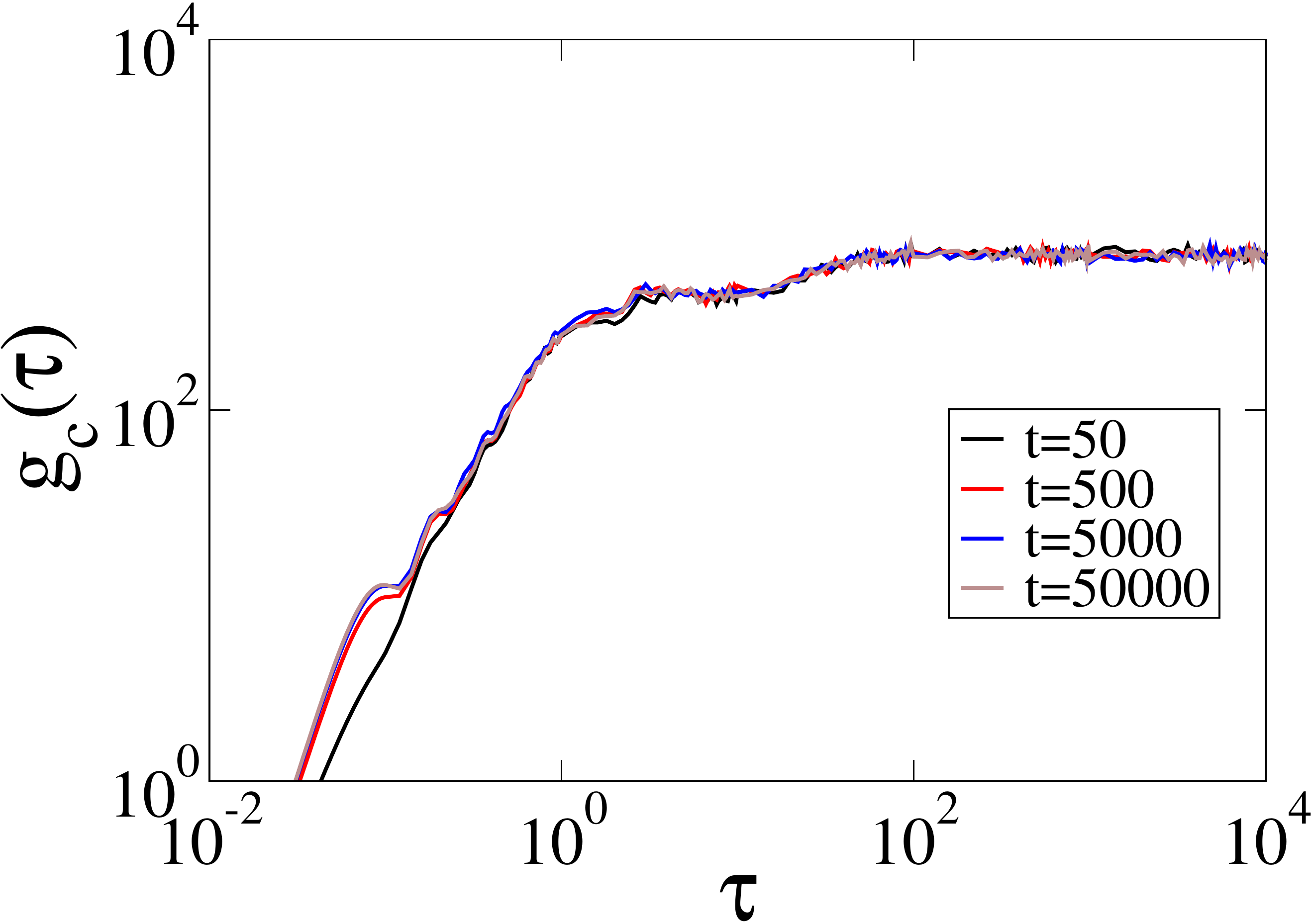}}
\caption{\label{fig4}Time evolution of spectral form factor. Here
  plots (a), (b) and (c) show evolution of $g(\tau)$ (SFF) and plots
  (d), (e) and (f) show the evolution of connected spectral form
  factor $g_{c}(\tau)$ corresponding to $\lambda= 0.5, 1$ and $1.5$
  respectively. The system size is $N=1920$ and $N_A=N/2$. The average
  has been taken over $500$ random values of $\theta_p$.}
\end{figure*}
Another useful quantity that can be constructed from the entanglement
spectrum is the entanglement bandwidth. It is defined as
\begin{equation}
\delta_E=E_{max}-E_{min},
\end{equation}
where $E_{max}$ and $E_{min}$ are the maximum and minimum eigenvalues
of the entanglement Hamiltonian. We find that for times beyond
$t>t_2$, it is not just $S_1$ that saturates, but in fact the entire
EH spectrum that saturates, and hence $\delta_E$ also saturates as
shown in Fig.~\ref{fig1}(d). Thus the onset of saturation in $\delta_E$
also determines the onset of the third timescale.
Also from Fig.~\ref{fig2} it can be observed, that on scaling the nearest-neighbour 
gap ratio $\left\langle r \right\rangle $ [see Fig.~\ref{fig1}(a)] by a factor $\eta_1$, 
it can be compared with the zeroth order Renyi entropy $S_0$ [see Fig.~\ref{fig1}(b)]
and hence it becomes apparent that they both reach saturation at time $t_1$. 
Similarly, it can be observed that the entanglement bandwidth $\delta_E$ can be scaled by a factor $\eta_2$ and at $t_2$ it reaches saturation together with the von Neumann entropy $S_1$. Since they both depend on the EH spectrum, they saturate as soon as the spectrum saturates i.e. at $t_2$.
This can be seen in both the delocalized phase and at the critical point.  

To understand the above observations better for different phases, it
is useful to study the probability distribution of the eigenvalues of
the entanglement Hamiltonian at different instants of
time\cite{EE4}. The evolution of the spectral density of EH is shown
in Fig.~\ref{fig3} for different phases. Referring to
Eq.~\ref{eq:ab} and Eq.~\ref{eq:ab1} we observe that the near-zero
eigenvalues of the EH contribute maximally to the entanglement entropy
whereas the eigenvalues with large magnitudes barely contribute to it.
As shown in Fig.~\ref{fig3}(a) and
Fig.~\ref{fig3}(b) the initial state corresponds to large
eigenvalues of the EH, and thus the entanglement entropy is also very
low. As EH evolves with time the eigenvalues become smaller thus
contributing more to the entanglement entropy. However,
Fig.~\ref{fig3}(c) shows that in the localized phase the
distribution does not change from its initial bimodal form with the
peaks remaining at large values. This implies almost no growth in
entanglement entropy in the localized phase [see Fig.~\ref{fig1}(c)].

We notice that in the delocalized phase during the first timescale ($t<t_1$) the
spectral distribution carries the bimodal peaks and at the end of this
timescale the distribution attains a single peak about zero with the
bimodal structure dismantled. In the second timescale ($t_1<t<t_2$), the single peak
continues to develop further and saturates on reaching the third
timescale. This is also observed from Fig.~\ref{fig2}
where at time $t_1$, a sharp transition can be seen in the scaled entanglement
bandwidth $(\delta_E)_s$, though it reaches saturation later only at time $t_2$. 
At the critical point $\lambda=1$ qualitatively the
scenario remains the same as in the delocalized phase. However, one
obtains bimodal peaks close to zero in the second timescale in this
case instead of a single peak about zero in the delocalized phase. In
the localized phase, the spectrum does not evolve with time and hence
the distribution does not change. This analysis of timescales as
discussed here is consistent with the analysis obtained from the other
quantities studied in Fig.~\ref{fig1}. 
\begin{figure}[b]
\vspace{3mm}
\centering
\stackunder{\hspace{-3.5cm}(a)}{\includegraphics[height=3.3cm, width=4cm]{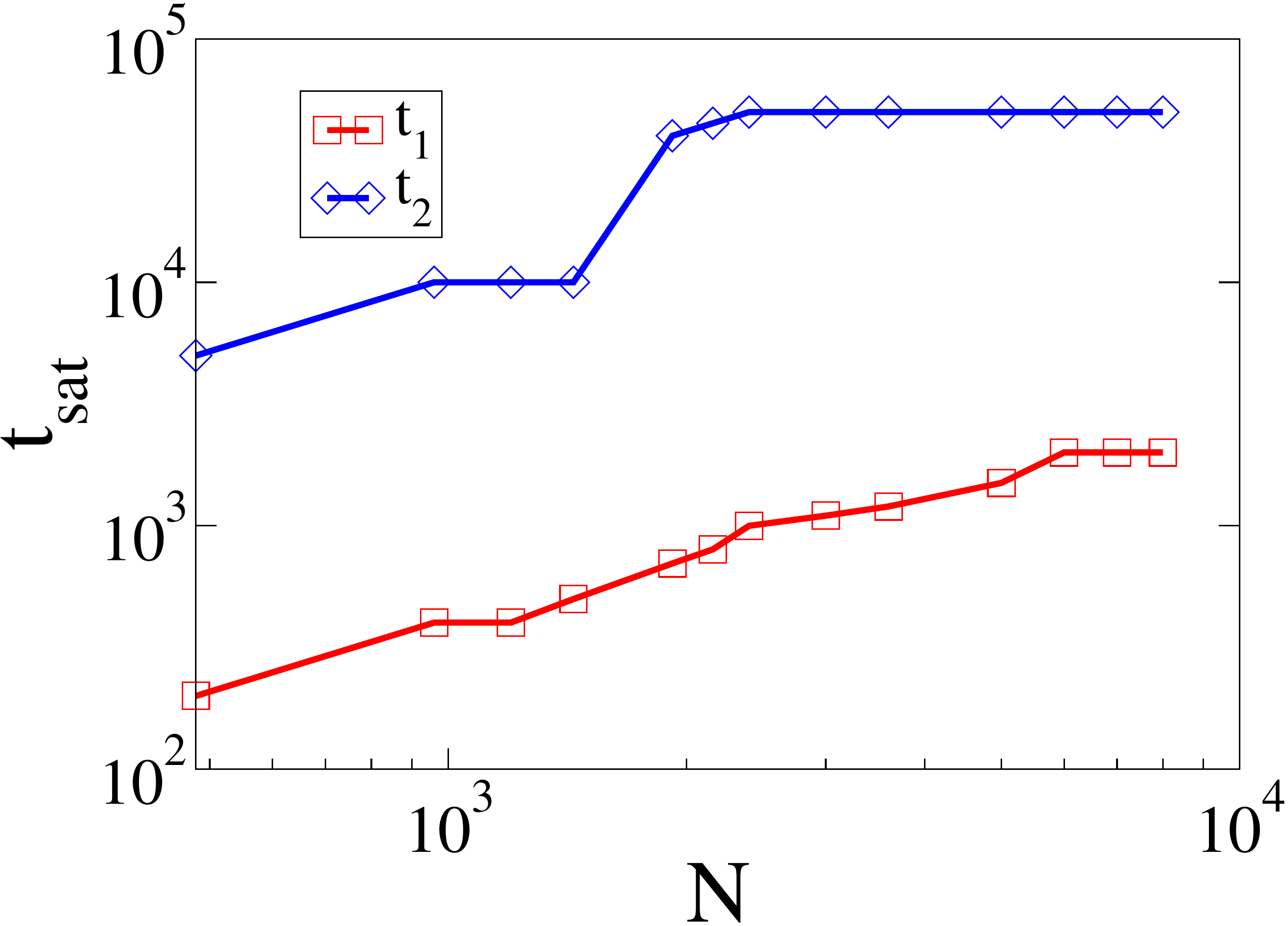}}\hspace{1mm}
\stackunder{\hspace{-3.5cm}(b)}{\includegraphics[height=3.3cm, width=4cm]{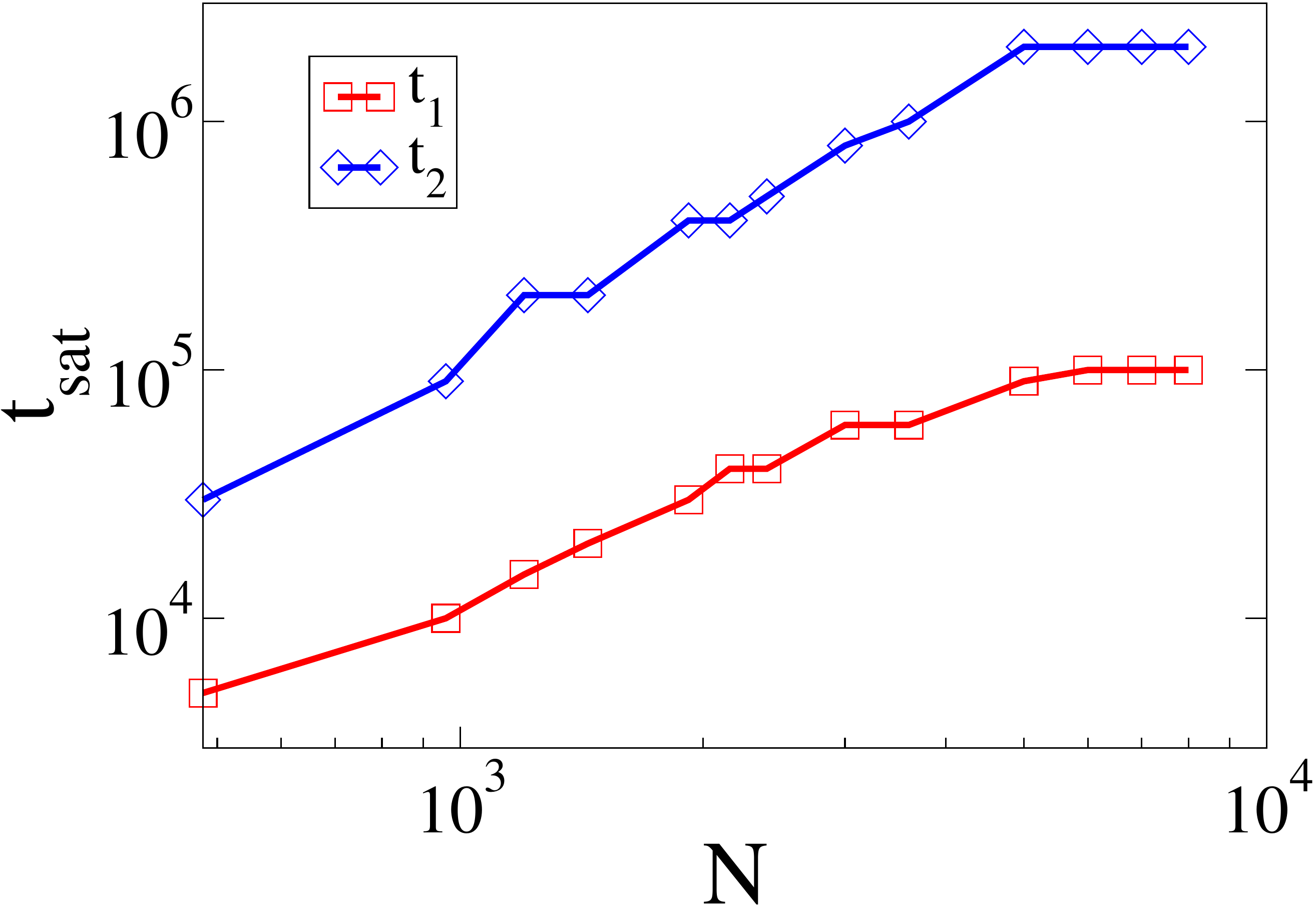}}
\caption{\label{fig5}System size dependence of times $t_1$ and $t_2$
  for the initial product state (a) $\lambda=0.5$ and (b)
  $\lambda=1$. In all cases $N_A=N/2$ and average has been taken over
  a range of $100-30$ random values of $\theta_p$ (depending on system
  size).}
\end{figure}
\begin{figure*}
\centering
\stackunder{\hspace{-5.0cm}(a)}{\includegraphics[height=3.8cm, width=5.6cm]{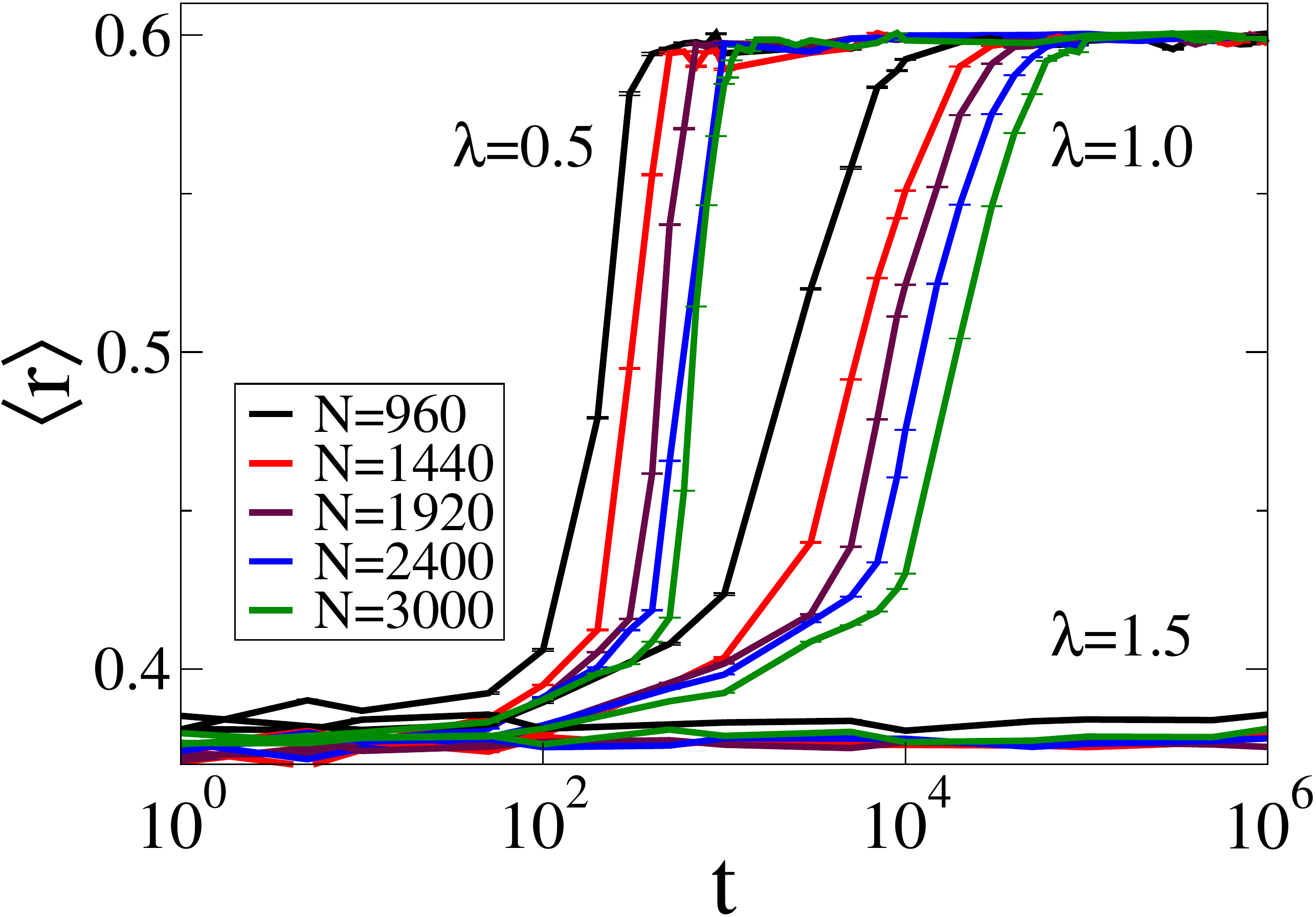}}\hspace{2mm}
\stackunder{\hspace{-5.0cm}(b)}{\includegraphics[height=3.8cm, width=5.6cm]{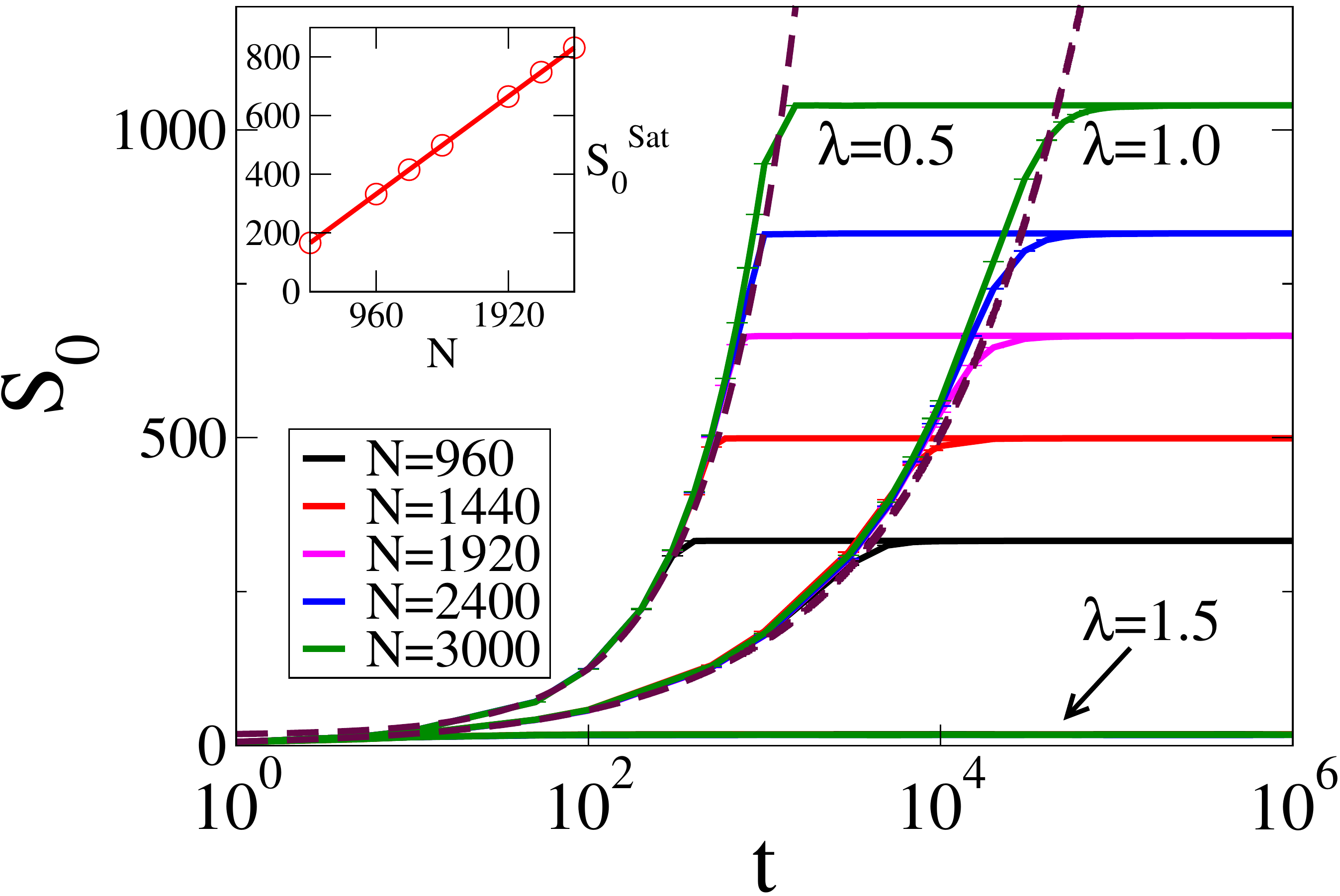}}\hspace{2mm}
\stackunder{\hspace{-5.0cm}(c)}{\includegraphics[height=3.8cm, width=5.6cm]{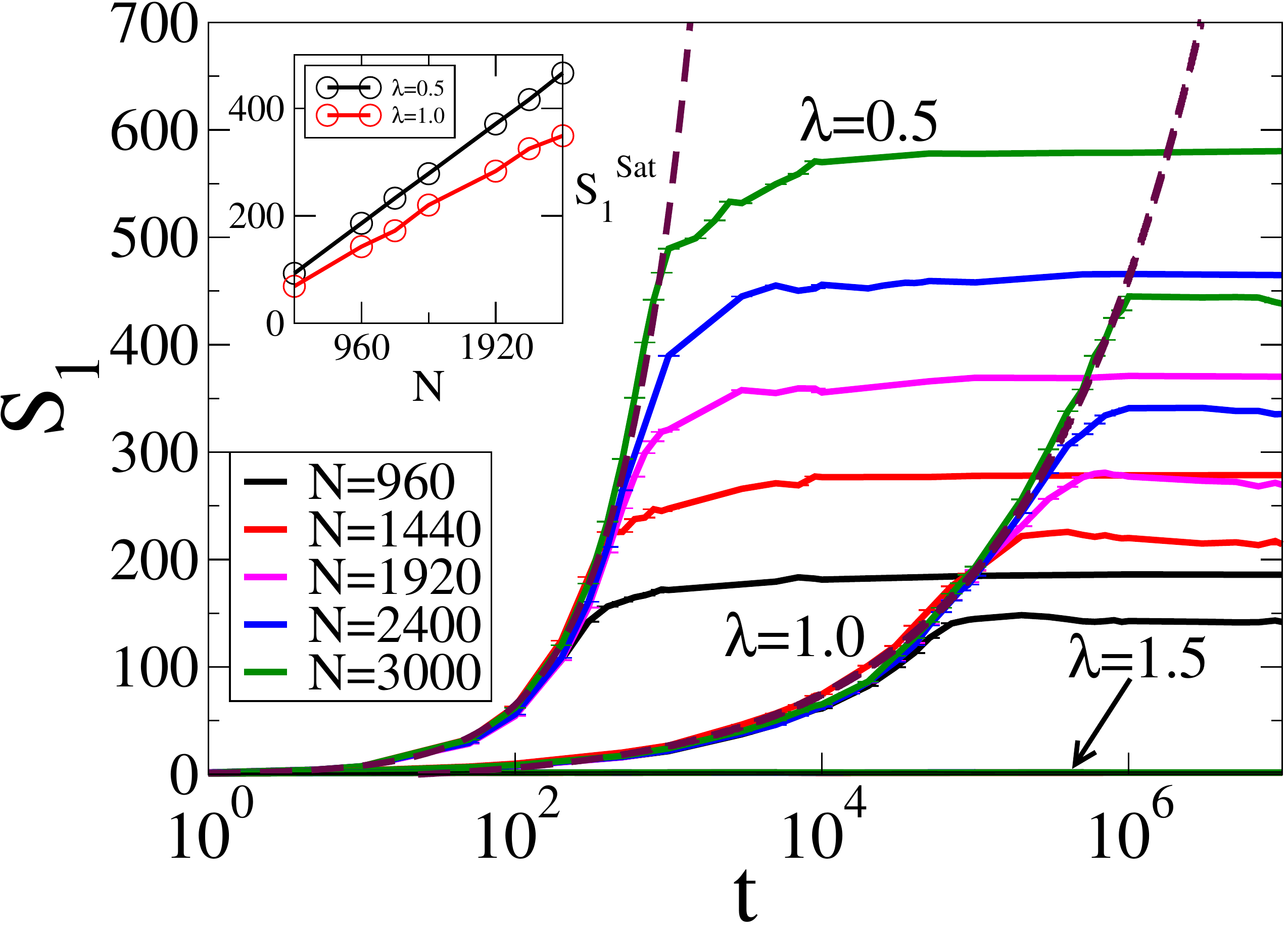}}
\caption{\label{fig6}System size effects on (a) Gap ratio, (b)
  Zeroth-order Renyi Entropy and (c) von Neumann Entropy for the
  different phases corresponding to $\lambda=0.5, 1.0, 1.5$. Here the
  power law dependence is shown by fitting non-linear maroon curves in
  (b) and (c) with the expression
  $S_{\alpha}^{\lambda}(t)=ct^{\gamma}+d$:
  $S_{0}^{0.5}(t)=1.86t^{0.88}+17.20$, $S_{0}^{1}(t)=6.57t^{0.47}$,
  $S_{1}^{0.5}(t)=t^{0.90}$ and $S_{1}^{1}(t)=5.15t^{0.31}-17.02$. The
  insets in (b) and (c) show dependence of saturated magnitude of
  $S_0$ and $S_1$ on system size. Here the subsystem size is $N_A=N/2$
  for all cases. The average has been taken over $100$ random values
  of $\theta_p$.}
\end{figure*}
\subsection{Spectral form factor}\label{subB}

Through the above analysis based on gap ratio, the spread of the
nearest-neighbor (NN) spectral correlations in the entanglement
spectrum becomes clear. However, we are interested in understanding
the global spread of correlations in the spectrum. Hence the spectral
form factor (SFF) defined in Subsection~\ref{SFF} becomes an
appropriate measure for it. It is sensitive not only to long-range
correlations but also to the density of states, unlike the gap
ratio. Figures ~\ref{fig4}(a)--\ref{fig4}(c) show the evolution
of the SFF and Figs.~\ref{fig4}(d)--\ref{fig4}(f)
show the evolution of the CSFF of the EH in
different phases.

Here we associate the time evolution of the features of SFF with times
$t_1$ and $t_2$ as discussed in the previous section. The
nearest-neighbor repulsion in the spectrum develops till $t<t_1$.
However as the spectral form factor also quantifies
longer-correlations, it keeps developing between $t_1<t<t_2$.  For
time $t>t_2$ the entanglement spectrum does not evolve and
consequently the SFF also becomes time-independent.

The time taken for the SFF to reach saturation in the delocalized
phase is less than that at the critical point [see
  Figs.~\ref{fig4}(a)--\ref{fig4}(b)]. The short-time growth of the
entanglement entropy is seen to be slower at the critical point [see
  Fig.~\ref{fig1}(c)] in comparison with the delocalized phase because
the eigenstates at the critical point are barely extended in
nature. The length of the ramp (in units of $\tau$) is nearly the same
in both cases which shows that the number of eigenvalues that exhibit
universal spectral correlations is also approximately equal. Also
there are initial fluctuations seen at earlier values of $\tau$ in all
the cases of $g(\tau)$ which are absent in $g_c(\tau)$. These are
non-universal and depend on the spectral density of the Entanglement
Hamiltonian of the AAH model. In the localized phase, one can observe
that irrespective of the time of evolution, $g(\tau)$ does not develop
any ramp-like structure. For $\tau \rightarrow
  \infty$, it can be concluded from Eq.~\ref{ab5} that the only terms
  which will survive are those where $\lambda_i=\lambda_j$. Since the
  subsytem size is $N/2$ here, hence $g(\tau)=N/2$. Also as $\tau
  \rightarrow \infty$, the disconnected part $\left\langle
  Z(\tau)\right\rangle\left\langle Z^*(\tau)\right\rangle=0$ (see
  Eq.~\ref{ab6}), which in turn implies that $g_c(\tau)= g(\tau) =
  N/2$. The same can be observed from Fig.~\ref{fig4}. Hence the
spectral form factor serves as a probe to distinguish between the
different phases of the spectrum.

\subsection{System and Subsystem size dependence}\label{subC}
Starting from the initial product state, the system size and subsystem
size dependences are studied here in different phases. The system size
dependences on $t_1$ and $t_2$ in the delocalized phase and at the
phase-transition point are shown in Fig.~\ref{fig5}(a) and
Fig.~\ref{fig5}(b) respectively. We define $t_1$ as the time beyond
which the (numerically determined) time derivative of the NN gap ratio
i.e.  $\frac{d\left\langle r\right\rangle}{dt}$ remains consistently
below $10^{-5}$. Similarly, $t_2$ is defined as the time beyond which
the time-derivative of the entanglement entropy ($\frac{dS_1}{dt}$)
remains consistently below $10^{-5}$. For a clean
  system (results not shown here), we find that there is a systematic
  increase in the time $t_2$ with increasing system sizes. A similar
  trend is also expected in the case of disordered systems, which is
  unclear here from Fig.~\ref{fig5} on account of the high cost of the
  numerical probe. However, we are guaranteed that the time $t_1$ will
  not cross time $t_2$. This is due to the fact that the the butterfly
  velocity $v_B$ is generically larger than the entanglement velocity
  $v_E$ and hence the spectrum develops level repulsion before the von
  Neumann entropy reaches its saturation value.\cite{model9}.
\begin{figure}[b]
\vspace{3mm}
\centering
\stackunder{\hspace{-3.5cm}(a)}{\includegraphics[height=3.3cm, width=4cm]{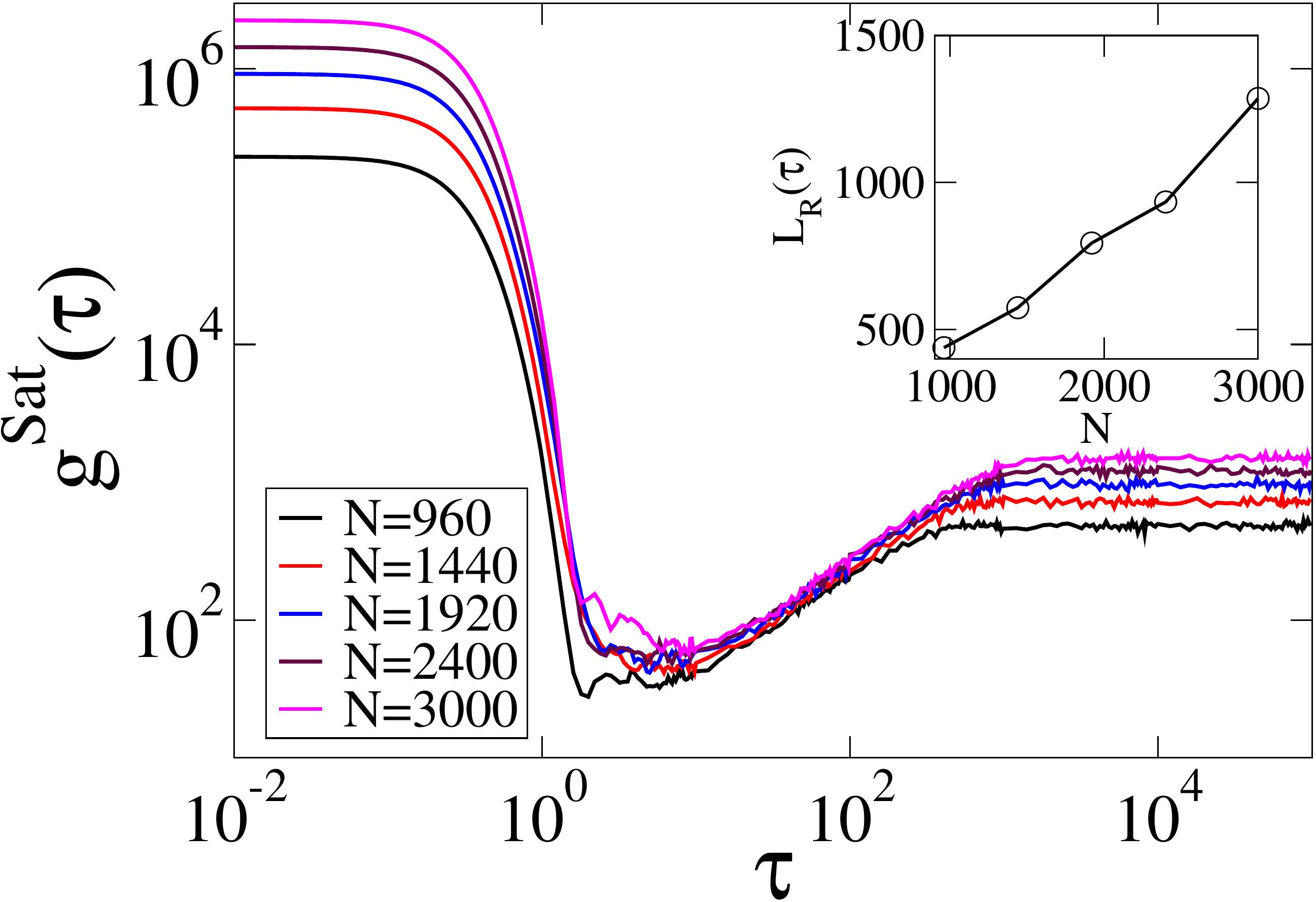}}\hspace{1mm}
\stackunder{\hspace{-3.5cm}(b)}{\includegraphics[height=3.3cm, width=4cm]{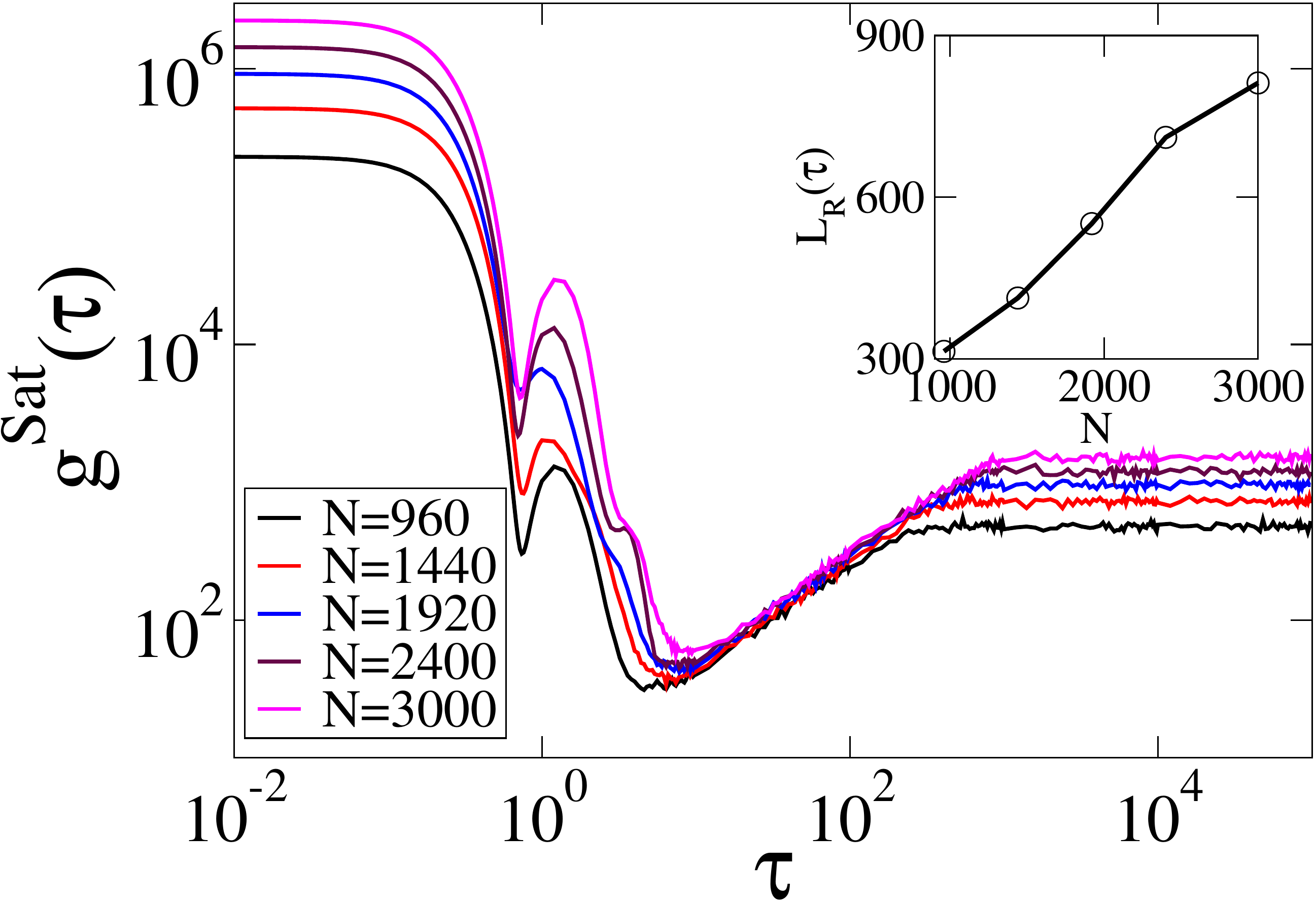}}
\caption{\label{fig7}SFF at time $t=10^6$ corresponding to various
  system sizes for (a) the delocalized phase: $\lambda=0.5$ and (b) at
  the critical point : $\lambda=1.0$. Here subsystem size is $N_A=N/2$
  in all cases. Inset shows the length of the ramp $L_R$ (in units of
  $\tau$) corresponding to various system sizes.  The length of the
  ramp here is determined numerically by subtracting the Thouless time
  (local minimum of the CSFF) from the plateau time, which is taken to
  be the $\tau$ at which $g_c(\tau)$ reaches $98\%$ of its maximum
  value (i.e. $N_A$). The average has been taken over $500$ random
  values of $\theta_p$.}
\end{figure} 
The system size dependences of the gap ratio, zeroth-order Renyi
entropy and entanglement entropy are also discussed here [see
Figs.~\ref{fig6}(a)--\ref{fig6}(c)]. It can be seen from
Fig.~\ref{fig6}(c), that the EE in the localized phase is
independent of the system size and hence it exhibits an area-law-like
behavior i.e. $S_1$ is constant, despite the state itself involving
all the excited eigenstates as well. Also, it can be observed that the
zeroth and first-order Renyi entropy at short times shows a power-law
dependence on time $t$ in the delocalized phase and at the critical
point i.e. $S \propto t^{\gamma}$ where $\gamma$ is some exponent
which is system size independent~\cite{EE5}. The span of growth of
entanglement entropy is larger at the critical point with smaller
saturation values as compared to the delocalized phase. The EH
eigenvalues in the delocalized phase have smaller magnitude than those
at the phase transition point [see
Figs.~\ref{fig3}(a)--\ref{fig3}(b)], therefore the EE
tends to be larger in the delocalized phase. It can also be observed
that when entropy reaches its saturation value it obeys volume law,
i.e. both $S_0$ and $S_1$ at saturation increase linearly with $N$ [see
insets of Figs.~\ref{fig6}(b)--\ref{fig6}(c)].
 
Since $S_0$ signifies the number of non-zero eigenvalues of the RDM of
the subsystem ${\rho_m}$,  and eigenvalues of EH are related to it as
$\left\lbrace E_m \right\rbrace =\left\lbrace-\log\rho_m
\right\rbrace$, it can be concluded that though the number of non-zero
eigenvalues of the RDM is the same in both phases, their magnitudes
are different as observed from $S_1$. In general for any $i>j$,
$S_i\leq S_j$~\cite{Beck}; in particular, $S_1$ the von Neumann entropy is a lower
bound for $S_0$ as observed from Fig.~\ref{fig6}(b) and
Fig.~\ref{fig6}(c).

We have seen that at long times, the system saturates to a state whose
properties depend on the phase in which the Hamiltonian is tuned 
(see Fig.~\ref{fig1}). Here we study the system-size dependence of the
SFF of the states attained at long times. Figure~\ref{fig7}(a)
and Fig.~\ref{fig7}(b) show the system size dependence of the
SFF calculated from the states at time $t=10^6$ for the delocalized
phase and at the phase transition point respectively. It can be
observed, especially from the linear scale (see inset) that the length
of the ramp `$L_R$' (in units of $\tau$) increases as the system size
becomes larger. This signifies that with a change in the system size,
the number of eigenvalues that exhibit universal spectral correlations
also increases which in turn leads to a longer ramp. Also, we have
studied the subsystem size dependence on the length of the ramp, which
is not shown here.  Our main observation from this study is that for a
given system size, as the subsystem size is increased (even for sizes
much smaller than $\frac{N}{2}$), the length of the ramp tends to
saturate, thus signaling a characteristic length of the ramp for a
given system size.
 
\subsection{Higher-order gap ratio}\label{subD}

The nearest-neighbor level spacing ratio as defined in Eq.~\ref{eq:cd}
can be generalized to the $n^{th}$ order gap ratio which examines the
variations in the higher-order gaps of the spectrum. These
higher-order gap ratios are focussed on larger spectral intervals and can
be compared with the SFF, which is also a probe of global correlations
in the spectrum. The $n^{th}$-order gap ratio is defined as:
\begin{equation}
r_k^{(n)}=\frac{min(s_k^{(n)},s_{k+n}^{(n)})}{max(s_k^{(n)},s_{k+n}^{(n)})},
\label{ratio}
\end{equation}
where $s_k^{(n)}=e_{k}-e_{k-n}$. We study the evolution of this
quantity for the dynamics starting with the initial product state,
when the system is tuned in the delocalized phase and at the critical
point.

\begin{figure}
\centering
\stackunder{\hspace{-3.5cm}(a)}{\includegraphics[height=3.3cm, width=4cm]{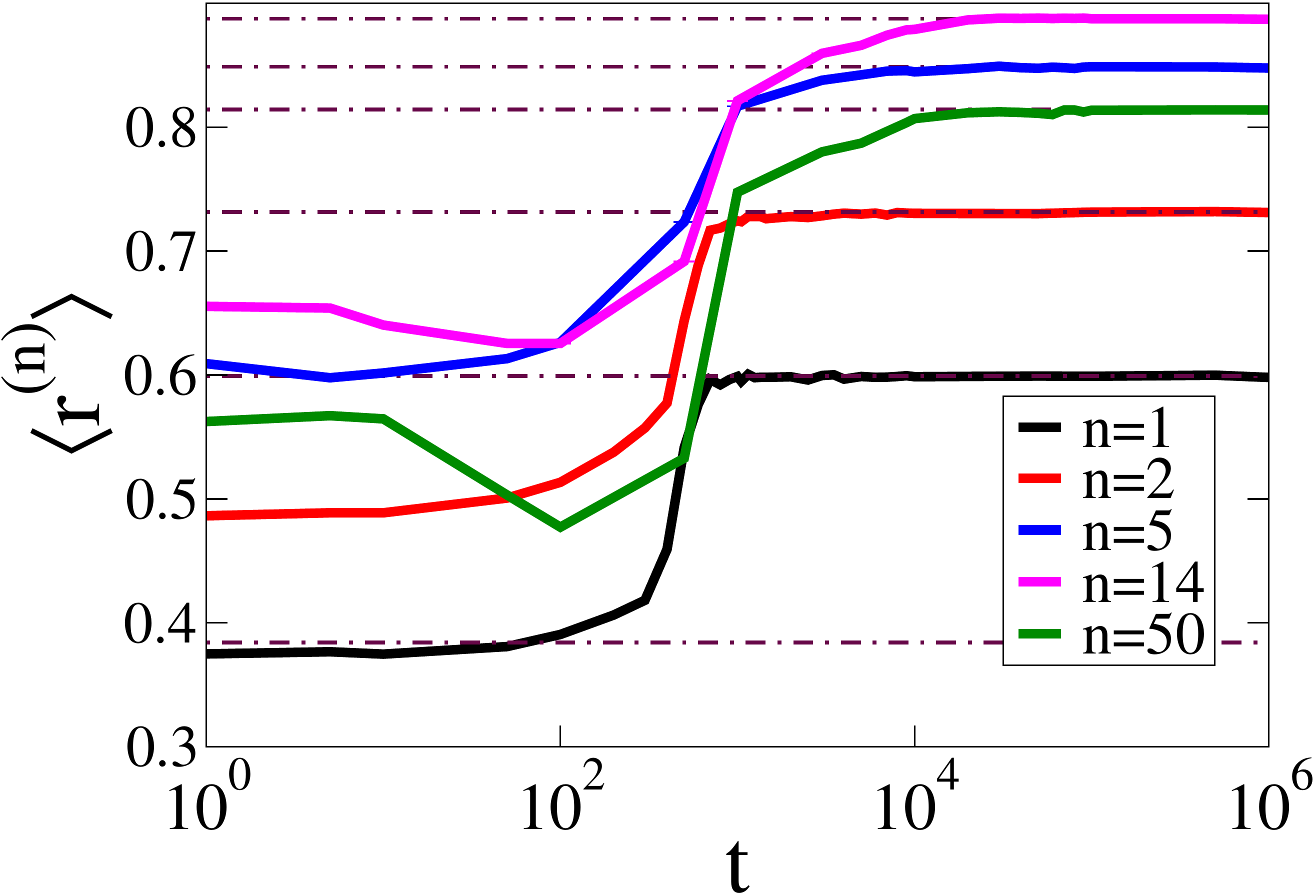}}\hspace{1mm}
\stackunder{\hspace{-3.5cm}(b)}{\includegraphics[height=3.3cm, width=4cm]{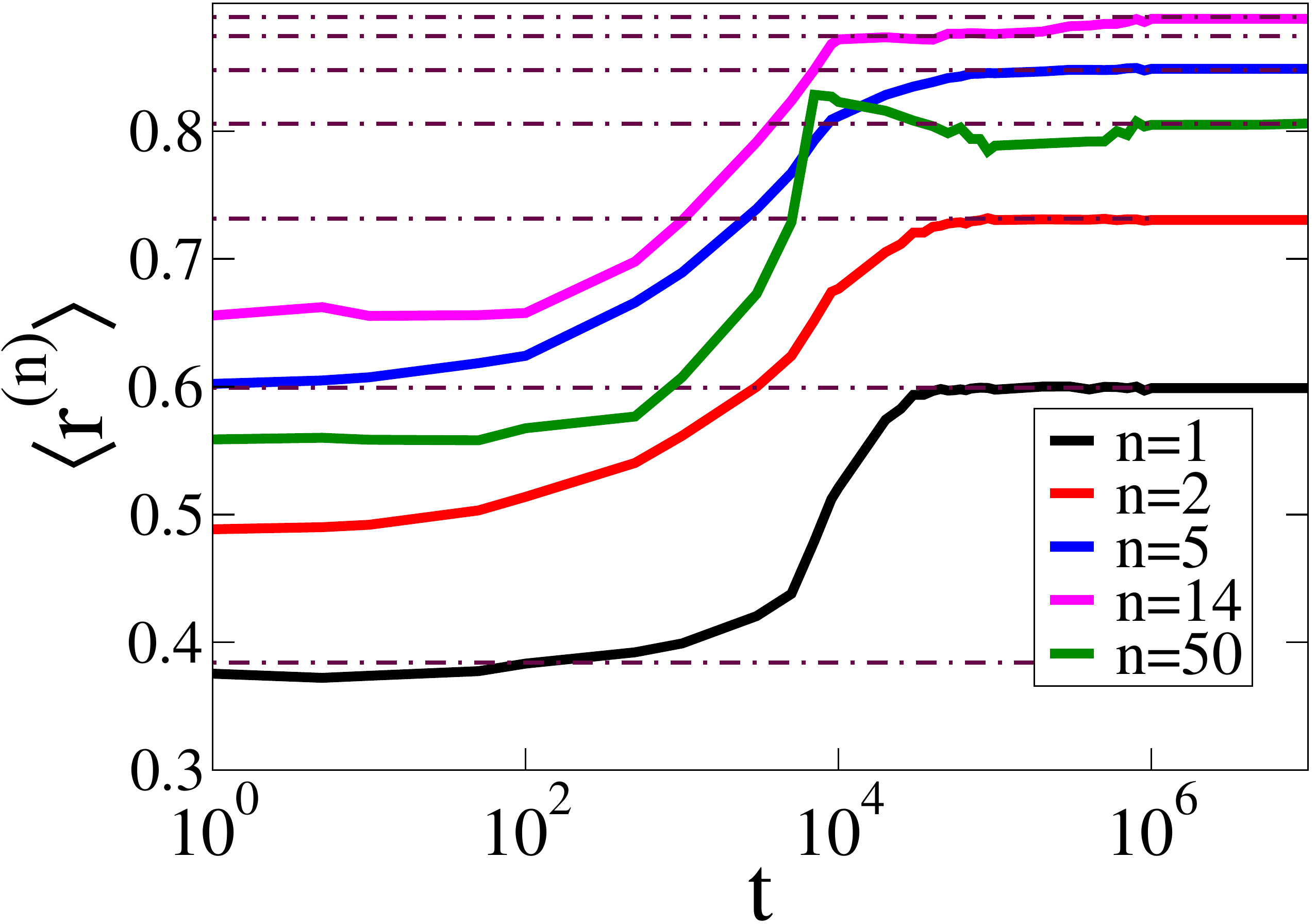}}
\caption{\label{fig8}Time evolution of $n\textsuperscript{th}$ order gap ratio of EH spectrum for (a) $\lambda=0.5$ and (b) $\lambda=1$. The horizontal dot-dashed lines are used to show time of satuartion. Figures correspond to $N=1920$ and $N_A=N/2$. The average has been taken over $100$ random values of $\theta_p$.}
\vspace{3mm}
\end{figure}

Figure~\ref{fig8} shows the time evolution of the higher-order gap
ratio for a number of different orders. In general, it can be observed
that as the order `$n$' increases, the time taken for the gap ratio
$\left\langle r^{(n)}\right\rangle$ to saturate also increases.  As
discussed earlier $\left\langle r^{(1)}\right\rangle$ saturates at
time $t_1$. In the delocalized phase, it can be seen from
Fig.~\ref{fig8}(a) that the time taken for the gap ratio of order
$n=2$ to saturate is around $t=10^3$ at which, it can be observed [see
  Fig.~\ref{fig4}(a)] that the SFF is still evolving. Tracking the
higher-order gap ratios and their saturation, we observe that
$\left\langle r^{(14)} \right\rangle$ reaches saturation at $t\approx
30000$ (which corresponds to time $t_2$), at which point the SFF is
also saturated [see Fig.~\ref{fig4}(a)]. All higher-order gap ratios
beyond $n=14$, saturate at time $t_2$ which can also be seen from
$\left\langle r^{(n)} \right\rangle$ with $n=50$ here. We conclude
that the development of higher-order correlations are encapsulated in
the SFF. Similarly at the critical point, from Fig.~\ref{fig8}(b) and
Fig.~\ref{fig4}(b), all $\left\langle r^{(n)} \right\rangle$ with
orders higher than $n=14$ saturate approximately at time $t_2$ along
with the SFF.  Thus there exists a correspondence between the
higher-order gap ratios and SFF, both of which are measures of
long-range spectral correlations.

\subsection{Spectrum resolved SFF}\label{subE}

\begin{figure}
\vspace{3mm}
\centering
\stackunder{\hspace{-3.5cm}(a)}{\includegraphics[height=3.3cm, width=4cm]{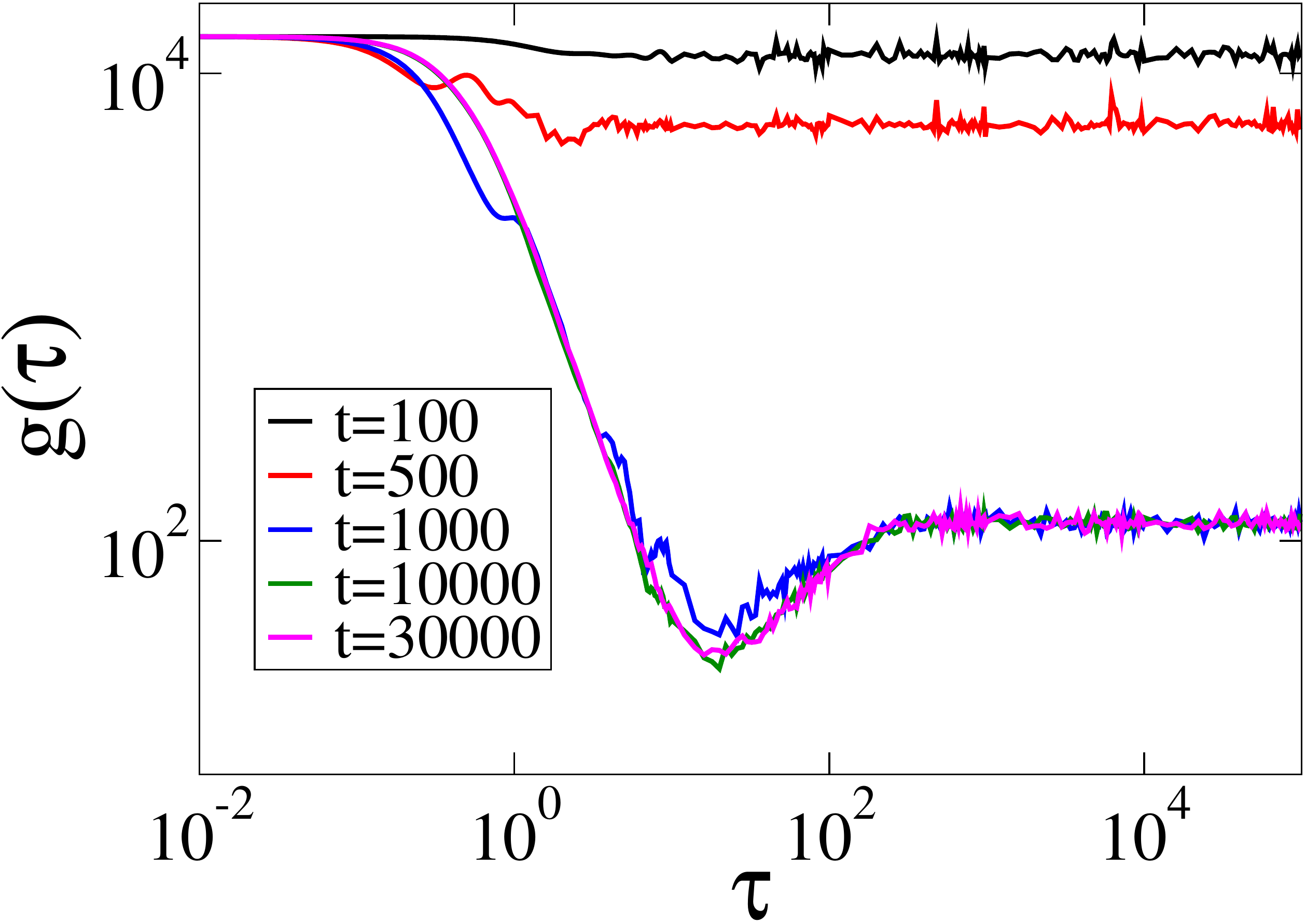}}\hspace{1mm}
\stackunder{\hspace{-3.5cm}(b)}{\includegraphics[height=3.3cm, width=4cm]{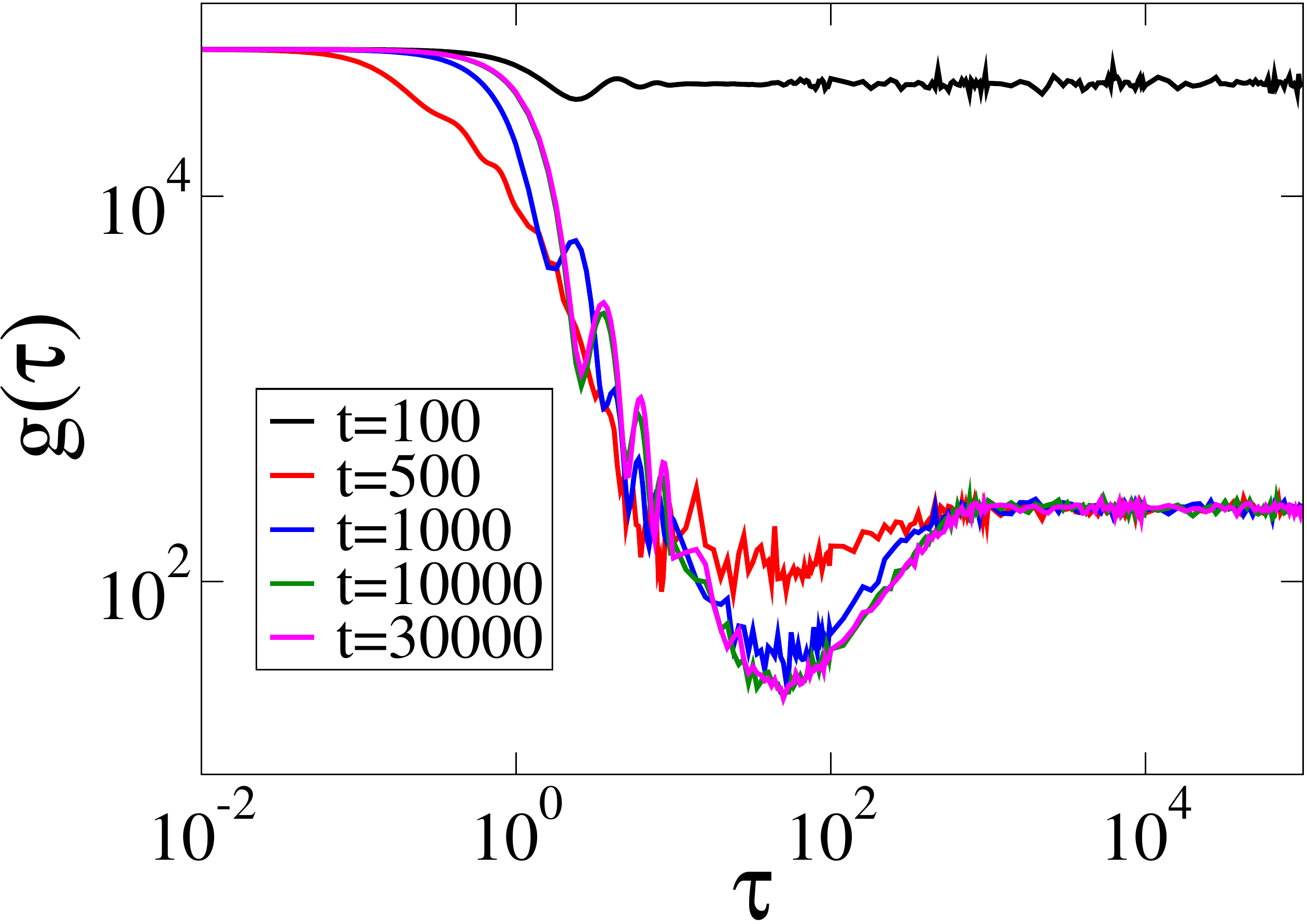}}
\stackunder{\hspace{-3.5cm}(c)}{\includegraphics[height=3.3cm, width=4cm]{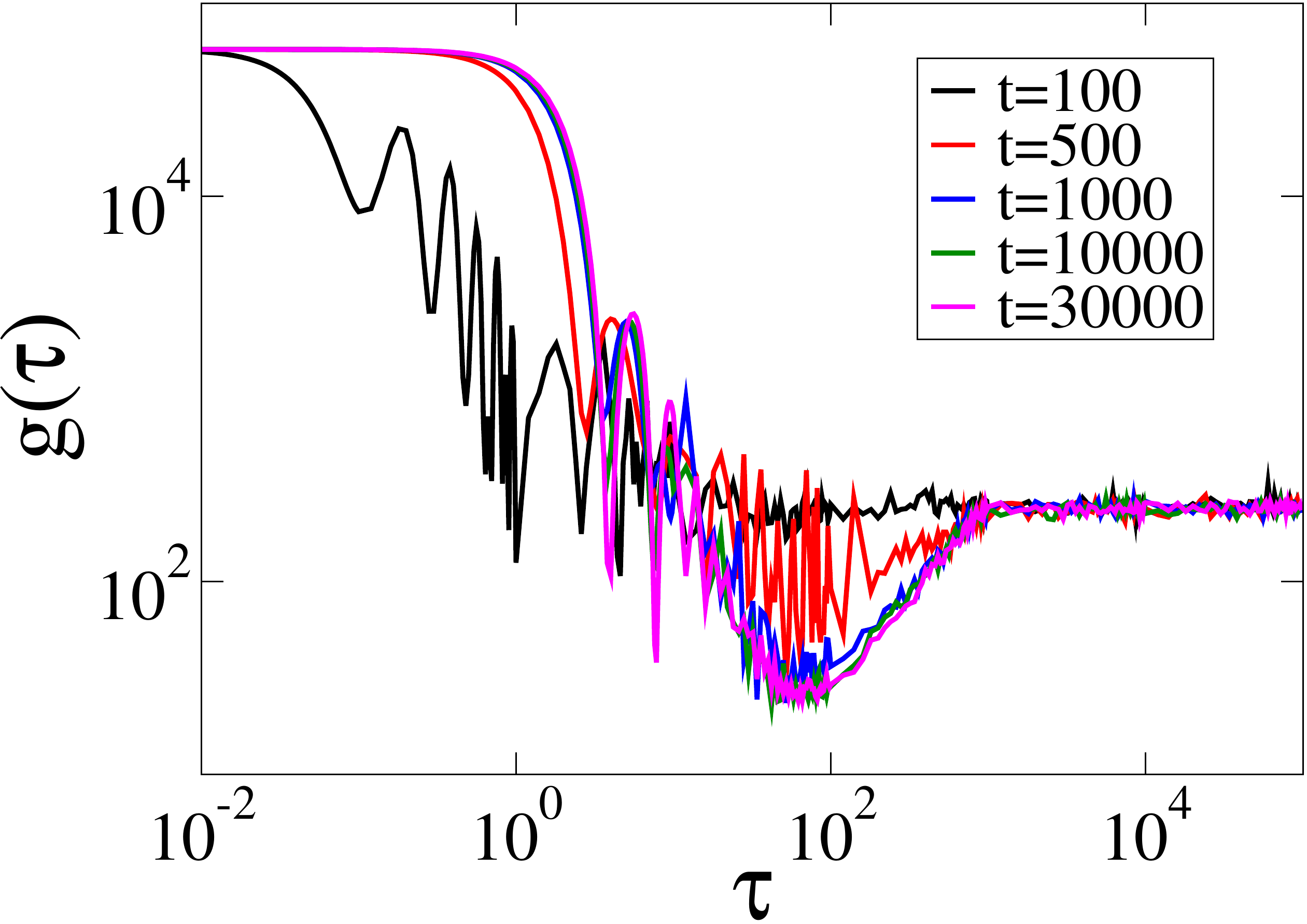}}\hspace{1mm}
\stackunder{\hspace{-3.5cm}(d)}{\includegraphics[height=3.3cm, width=4cm]{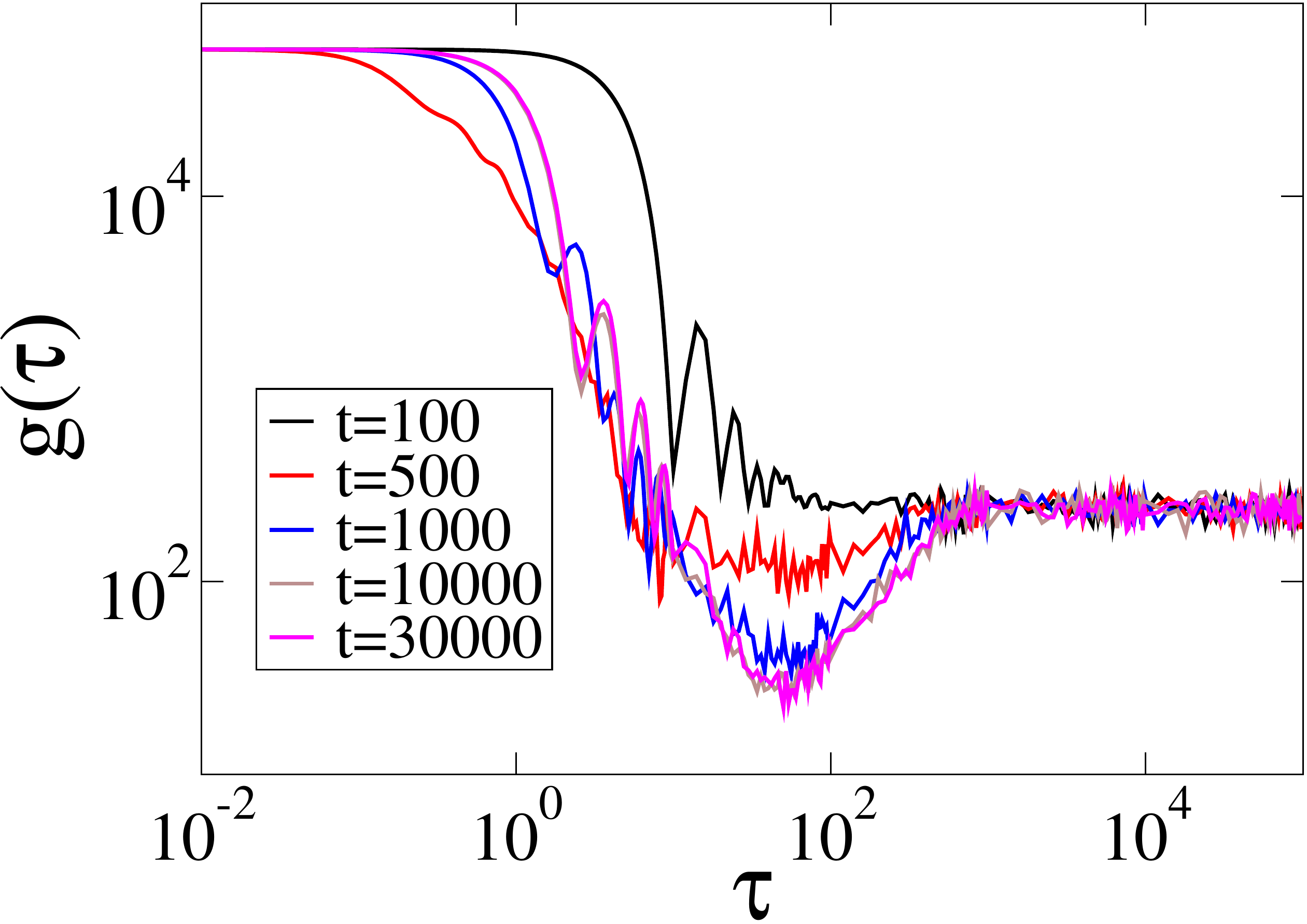}}
\caption{\label{fig9} SFF for subset of eigenvalues (arranged in
  ascending order) at different time steps $t$ of the EH spectrum for
  $\lambda=0.5$, $N=1920$ and $N_A=N/2$ over $100$
  realizations. Subset of eigenvalues (a) $1-120$ (bottom
  $1/8\textsuperscript{th}$), (b) $121-360$ ($1/4\textsuperscript{th}$
  above (a)), (c) $361-600$ (central $1/4\textsuperscript{th}$) and
  (d) $601-840$ ($1/4\textsuperscript{th}$ above (c)).}
\end{figure}

Here the spectrum of the EH is divided into subsets~\cite{SFF3}(with
eigenvalues arranged in ascending order) and time evolution of the SFF
corresponding to each subset has been plotted in
Fig.~\ref{fig9}. This is used to understand the time evolution
as well as the spread of the spectral correlations in the eigenvalues
present in a particular span of the spectrum.

In Fig.~\ref{fig9}(a), the spectrum-resolved SFF considering the
bottom $1/8\textsuperscript{th}$ of the eigenvalues of the EH spectrum
is plotted. Since the spectrum is symmetric about zero, the same
behavior can also be observed with the top $1/8\textsuperscript{th}$
of the eigenvalues of the spectrum. The ramp here is smaller as
compared to subsets of eigenvalues that lie at the center of the
spectrum. For example, we see a larger ramp in Fig.~\ref{fig9}(b)
where the SFF is computed with $1/4\textsuperscript{th}$ of the total
number of eigenvalues above the bottom set. We also observe an earlier
dephasing in Fig.~\ref{fig9}(a) on account of the larger magnitude of
the eigenvalues involved. As one moves towards the center of the
spectrum, the ramp becomes longer and dephases later. At saturation,
the length of the ramp is longest for the middle of the spectrum as
shown in Fig.~\ref{fig9}(c) (where the central
$1/4\textsuperscript{th}$ of the eigenvalues are used to compute the
SFF). Figure~\ref{fig9}(d) shows the evolution of the SFF considering
the next $1/4\textsuperscript{th}$ of the total number of
eigenvalues. Due to the symmetry of the eigenvalues on either side of
the central eigenvalue, the features of the SFF at saturation in
Fig.~\ref{fig9}(b) and Fig.~\ref{fig9}(d) are almost identical.
 
\begin{figure*}
\centering
\stackunder{\hspace{-5.0cm}(a)}{\includegraphics[height=3.7cm, width=5.1cm]{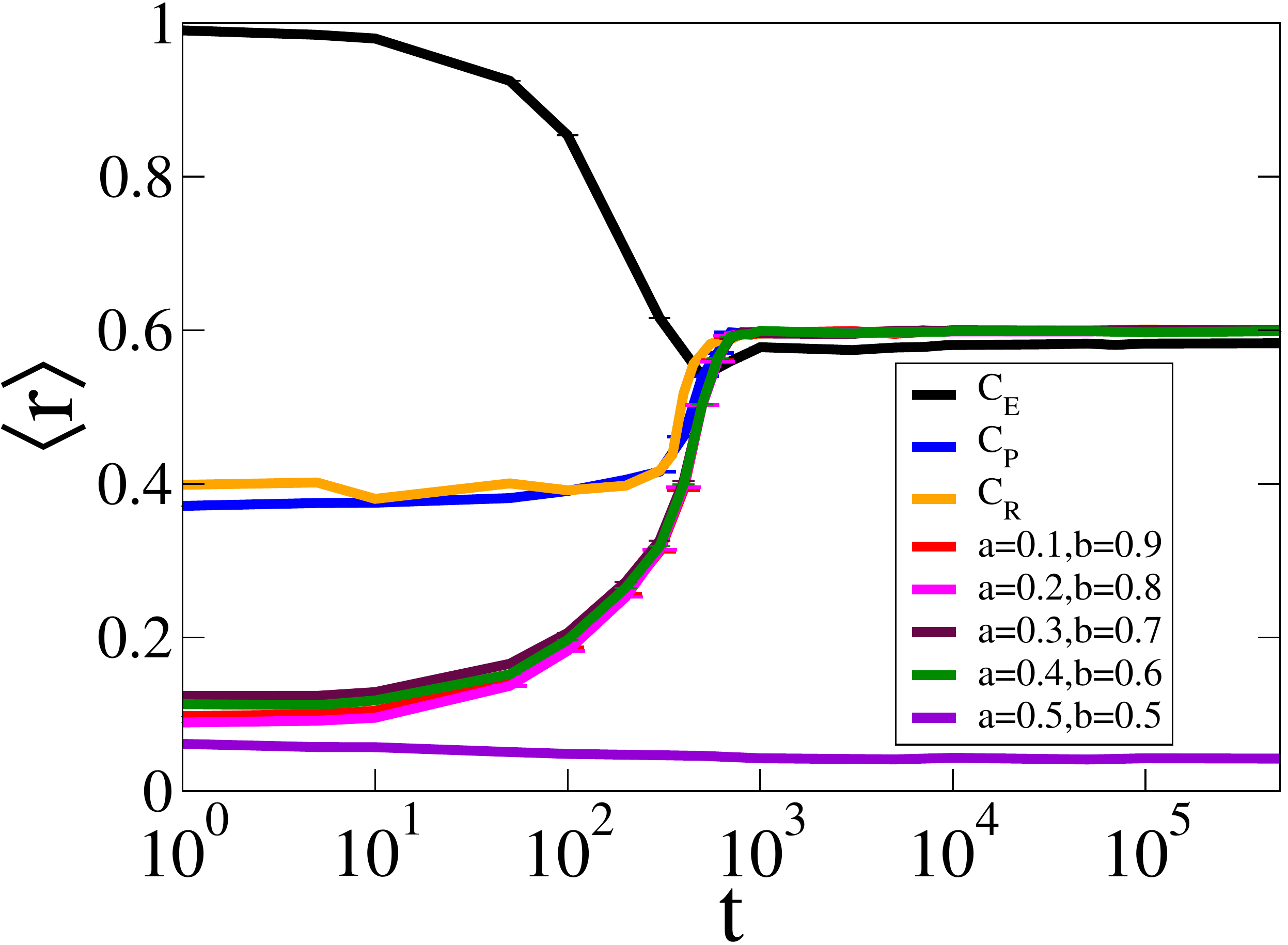}}\hspace{3mm}
\stackunder{\hspace{-5.0cm}(b)}{\includegraphics[height=3.7cm, width=5.1cm]{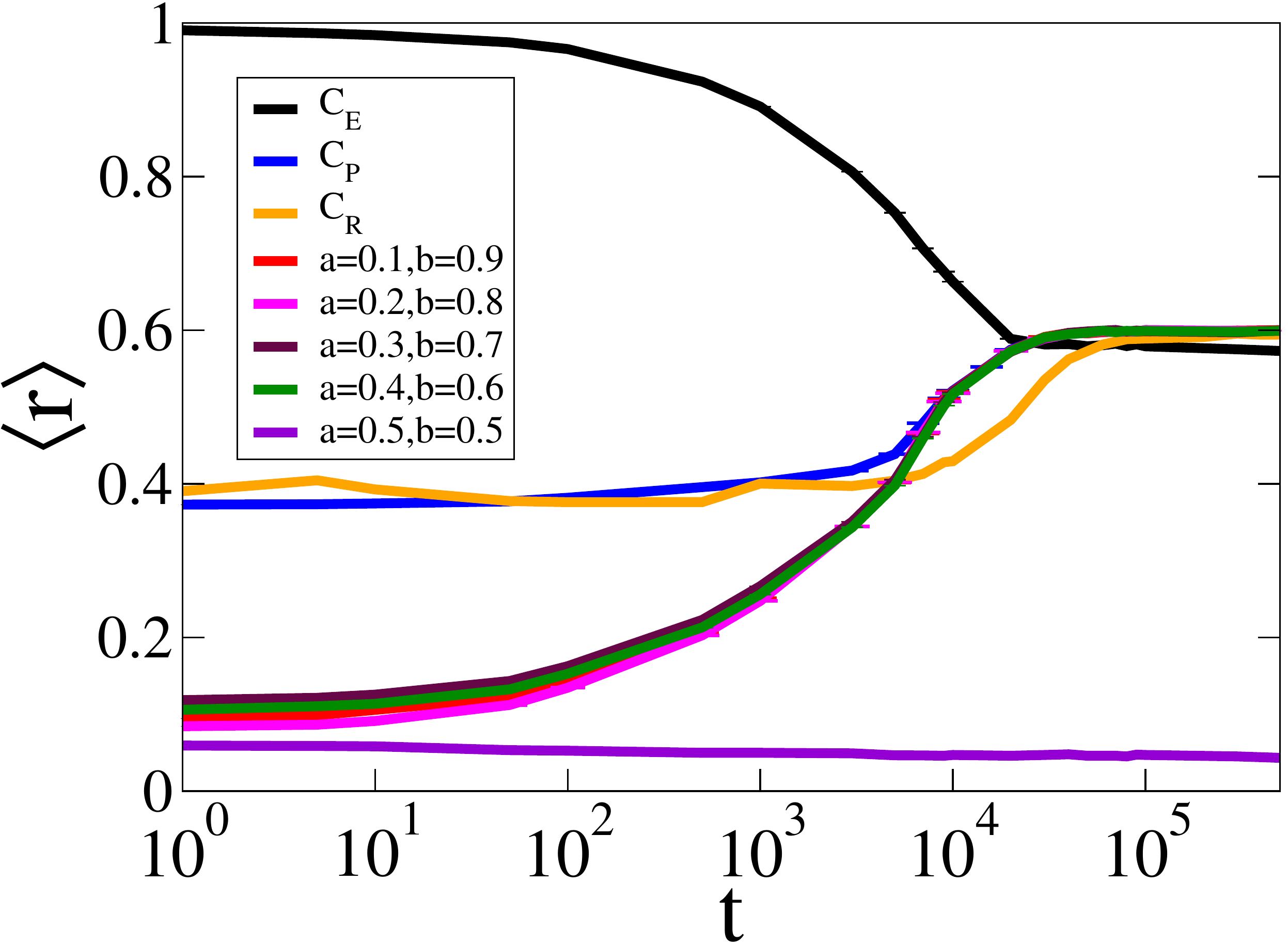}}\hspace{3mm}
\stackunder{\hspace{-5.0cm}(c)}{\includegraphics[height=3.7cm, width=5.1cm]{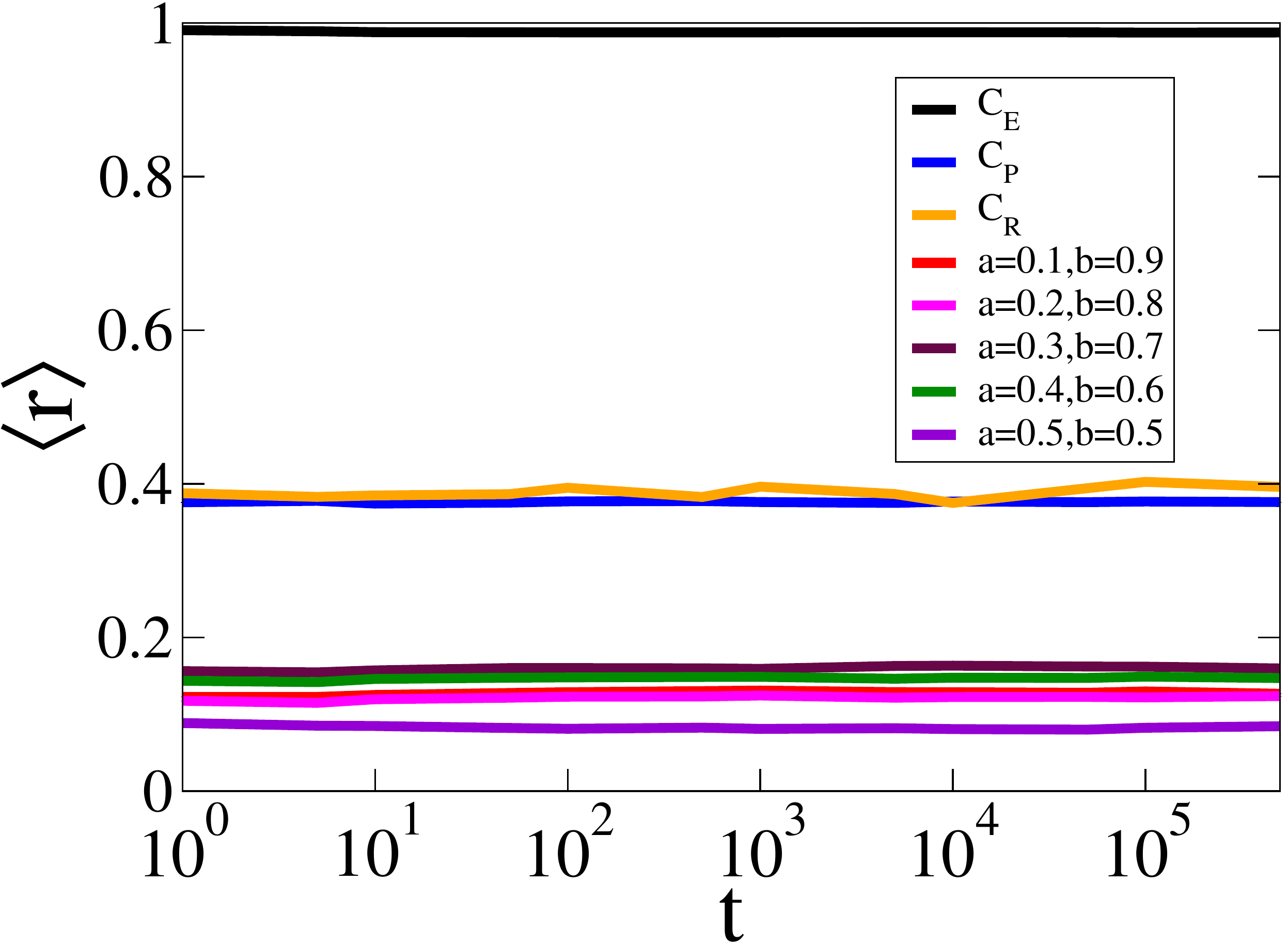}}

\stackunder{\hspace{-5.0cm}(d)}{\includegraphics[height=3.7cm, width=5.1cm]{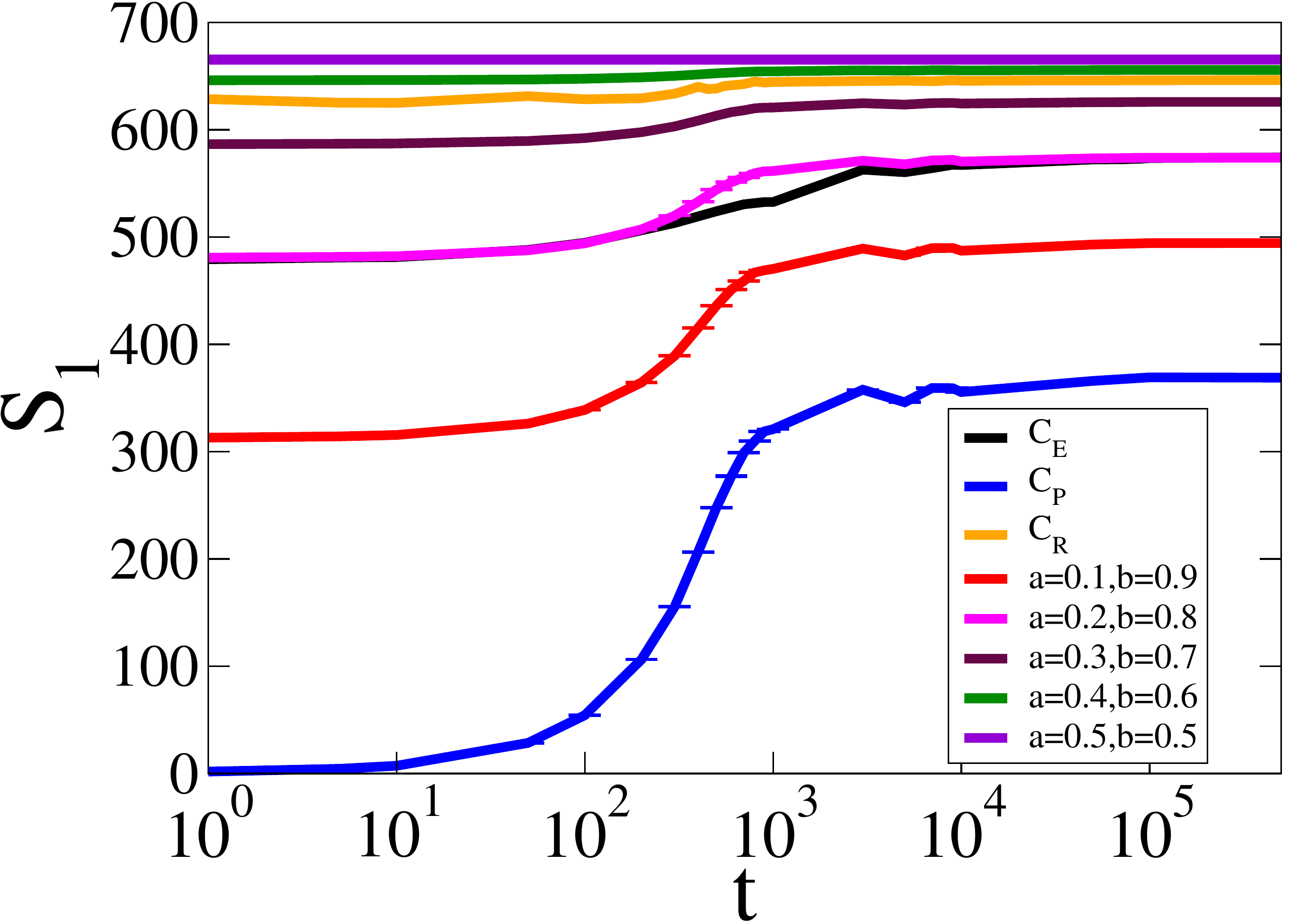}}\hspace{3mm}
\stackunder{\hspace{-5.0cm}(e)}{\includegraphics[height=3.7cm, width=5.1cm]{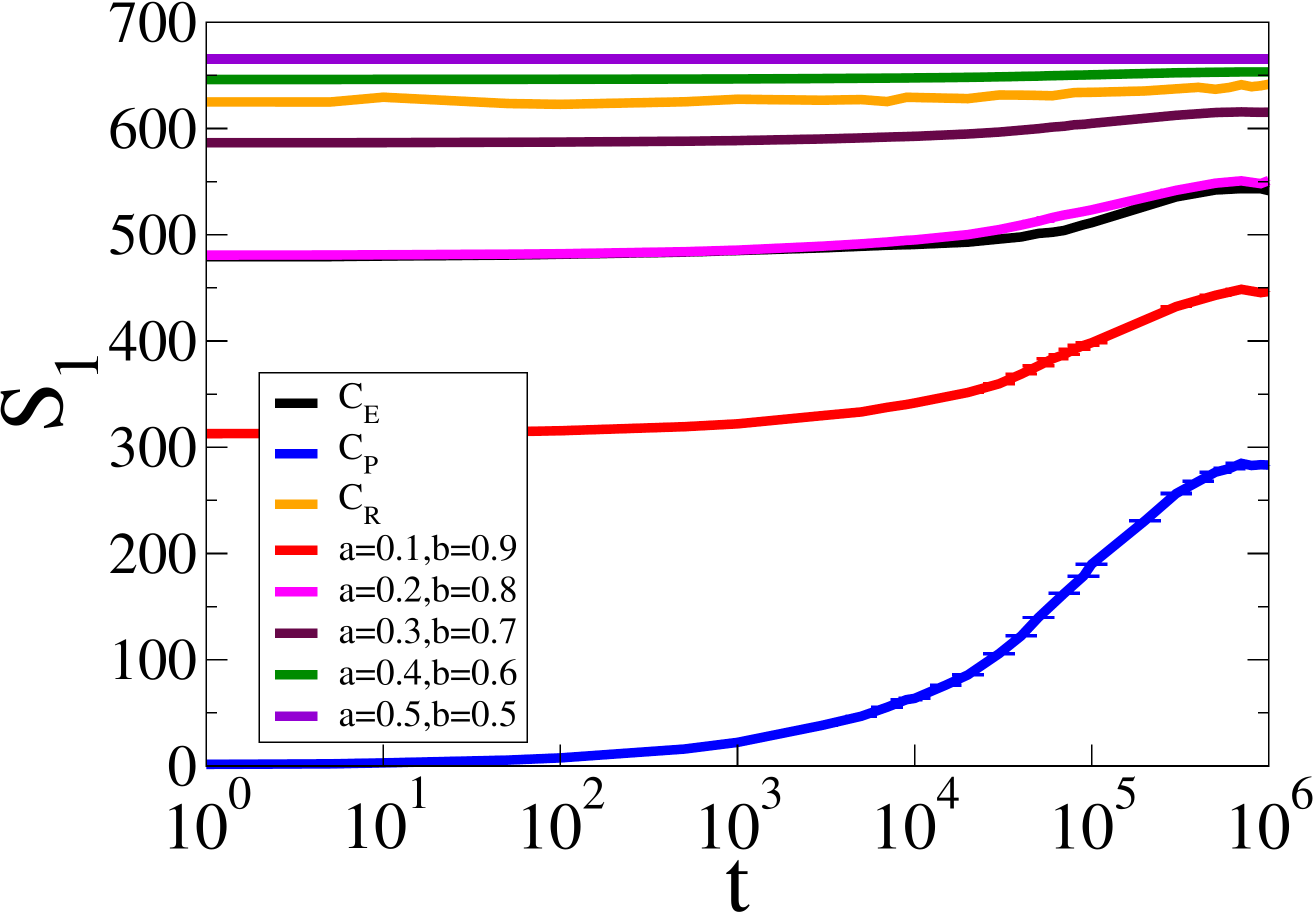}}\hspace{3mm}
\stackunder{\hspace{-5.0cm}(f)}{\includegraphics[height=3.7cm, width=5.1cm]{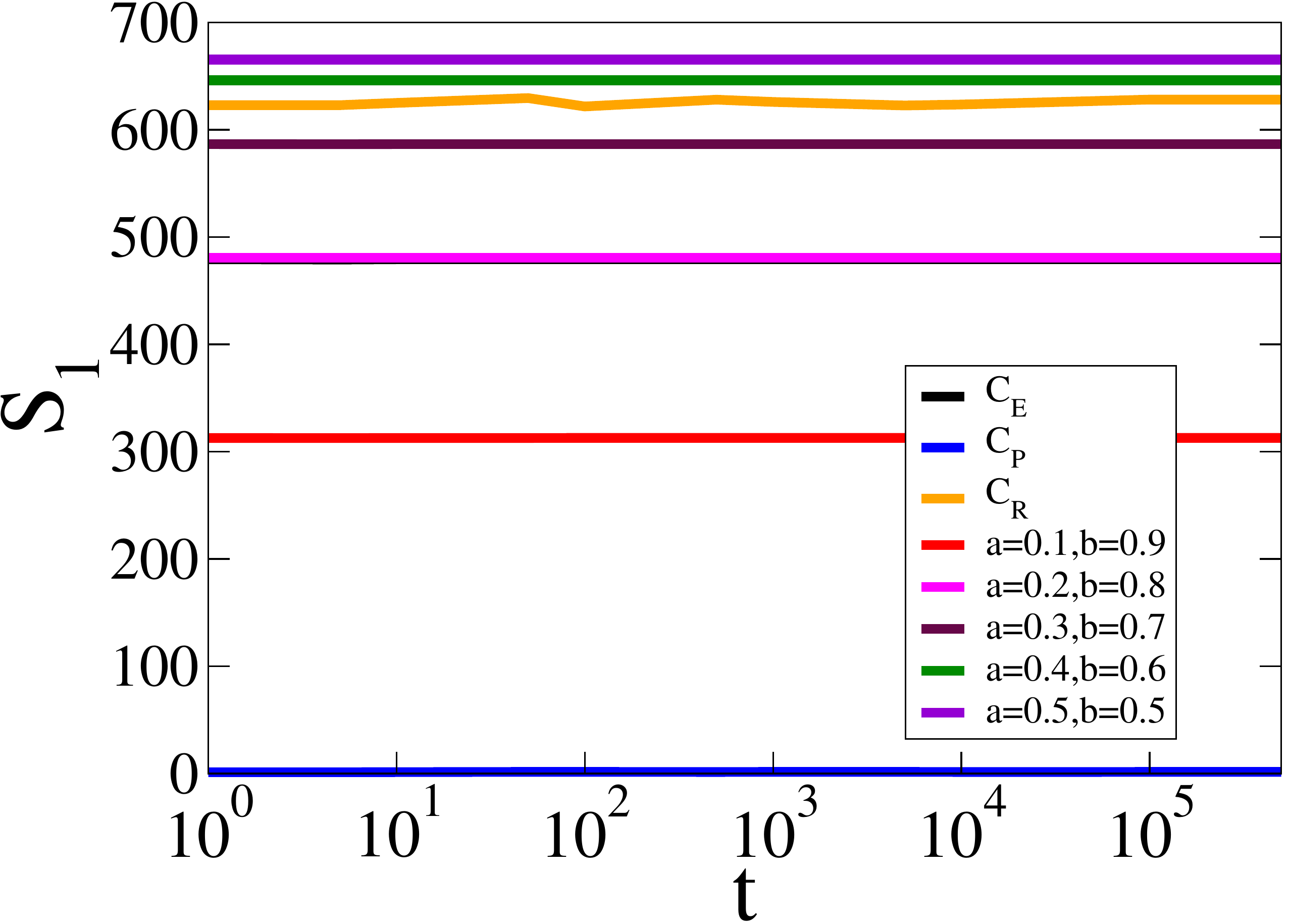}}
\caption{\label{fig10}Evolution of the spectral properties of the
  EH obtained from various initial correlation matrices: $C_E$ matrix,
  $C_P$ matrix, $C_R$ matrix and $C_S$ matrix for $N=1920$ and
  $N_A=N/2$. Gap ratio for (a) $\lambda=0.5$, (b) $\lambda=1.0$ and
  (c) $\lambda=1.5$. Also von Neumann entropy for (d) $\lambda=0.5$,
  (e) $\lambda=1.0$ and (f) $\lambda=1.5$. The average has been taken
  over $100$ random values of $\theta_p$.}
\end{figure*}

It can be concluded that in the delocalized phase, the correlations
are concentrated more at the center of the spectrum and as one moves
towards the edges, these correlations decrease. This happens as the
contribution to entanglement come mostly from smaller magnitude
eigenvalues. We have checked that at the phase-transition point
$\lambda=1$, the length of the ramp at saturation corresponding to the
top(or bottom) $1/8\textsuperscript{th}$ of the eigenvalues is
smallest (in units of $\tau$). However, the other subsets, namely the
next $1/4\textsuperscript{th}$ and the central
$1/4\textsuperscript{th}$ of the eigenvalues show the same length of
the ramp at saturation. Our main observation from the study of the
length of the ramp at saturation is that the correlations at the
critical point are more uniformly spread out than in the delocalized
phase.  We have also calculated the first order gap ratio $\langle
r\rangle$ for various subsets of the EH spectrum. We observe that all
subsets eventually reach close to the GUE value $\approx 0.599$ at
nearly the same time i.e. the time taken for all the subsets to
develop NN level repulsion is the same.

\subsection{Dynamics of initially entangled system}\label{subF}

So far, we have considered the dynamics of the system starting from an
initial product state with a non-entangled subsystem correlation
matrix. As time evolves, we have seen how entanglement tends to grow
and saturate to a characteristic value at long times. In the current
subsection, we will study the dynamics of the system considering a
variety of initial (subsystem) correlation matrices with non-zero
entanglement entropy at time $t=0$. The gap ratio of the EH related to
the product state we have considered so far corresponds to the
Poissonian value of $\langle r \rangle = 0.38$. We now consider three
different kinds of initial states whose gap ratios are different from
the Poissonian value and track the evolution of these states.
 
Firstly, we take the correlation matrix of the full system to be a
diagonal matrix whose elements are drawn from a Gaussian distribution,
with the constraint that the sum of its diagonal terms is equal to
the number of fermions. The corresponding correlation matrix denoted
by `$C_R$' is given by:
\begin{equation}
C_{R}=\xi\begin{bmatrix}
c_{11} & 0 & 0 & 0 & \cdot\\
0 & c_{22} & 0 & 0 & \cdot\\
0 & 0 & c_{33} & 0 & \cdot\\
0 & 0 & 0 & c_{44} & \cdot\\
\cdot & \cdot & \cdot & \cdot & \cdot\\
\end{bmatrix},
\label{eq:df}
\end{equation}
where $c_{11},c_{22},c_{33},..$ are numbers drawn randomly from a
Gaussian distribution $\mathcal{N}(0.5,0.02)$, and the normalization
$\xi$ is set according to the number of fermions. For implementing the
constraint, we fix $\xi$ such that the trace of the resulting matrix
is equal to the number of fermions which in our case is taken to be
$N/2$. The results shown here are qualitatively independent of the
type of distribution from which the diagonal elements are drawn (we
have checked the same for a uniform distribution $\mathcal{U}(0,1)$.)

Using the above correlation matrix, quench dynamics is studied for the
AAH model. The NN gap ratio evolves starting from the Poissonian value
and saturates to the GUE value at late times except in the case of the
localized phase which does not show any evolution [see
  Figs.~\ref{fig10}(a)--\ref{fig10}(c)]. Here since the initial
correlation matrix eigenvalues are engineered to lie close to $0.5$,
the corresponding EH eigenvalues have very small magnitude, and thus
it can be seen that the EE is close to the maximum entropy. Also its
evolution is similar to that of the non-entangled initial product
state as discussed in the previous subsections whose correlation
matrix is denoted here as `$C_P$' [see
  Figs.~\ref{fig10}(d)--\ref{fig10}(f)].  We have also checked the
effect of the inclusion of small off-diagonal terms into Eq.~
\ref{eq:df}. We find that the resulting time-evolved state's spectral
properties are similar to those evolving from the `$C_R$' matrix,
though the magnitude of EE becomes higher as the number of non-zero
off-diagonal elements of the matrix increase.

Secondly we have considered an initial (system) correlation matrix
obtained from the superposition of two DW-type pure states, $\ket
{\Phi_0}=\prod_{i=1}^{N/2} c^\dagger_{2i}\ket0$ with probablilty
amplitute $\sqrt{a}$ and $\ket {\psi_0}=\prod_{i=1}^{N/2}
c^\dagger_{(2i-1)}\ket0$ with probabilty amplitute $\sqrt{b}$. This
correlation matrix is denoted by `$C_{S}$':
\begin{equation}
C_{S}=\begin{bmatrix}
c_{11} & 0 & 0 & 0 & \cdot\\
0 & c_{22} & 0 & 0 & \cdot\\
0 & 0 & c_{33} & 0 & \cdot\\
0 & 0 & 0 & c_{44} & \cdot\\
\cdot & \cdot & \cdot & \cdot & \cdot\\
\end{bmatrix}.
\end{equation}

Here for all odd values of `$i$', $c_{ii}=b$ while for all even values
of `$i$', $c_{ii}=a$. Tuning the parameters $a=1$ and $b=0$, we get
the initial product state denoted by `$C_P$', for which the results
have already been discussed. For $a=0.5$ and $b=0.5$, we get the
maximally entangled state which does not show any evolution, and hence
there is no timescale associated with it. For all other values of $a$
and $b$, the state is entangled. It can be observed from
Figs.~\ref{fig10}(a)--\ref{fig10}(c) that the magnitude of the NN gap ratios
corresponding to various values of $a$ and $b$ are initially much
below the Poissonian value (i.e. 0.38). However, they also evolve with
time to saturate at the GUE value except in the localized phase, where
completely flat time evolution is seen.  Also, it can be noted from
Figs.~\ref{fig10}(d)--\ref{fig10}(f), that with the change in parameters the
initial entanglement entropy of the subsystem increases until it
reaches the value $N_A\ln2$ for the maximally entangled state. As a
result of the subsystem having non-zero entanglement initially, the
time taken for the EE to reach saturation decreases which reduces the
second timescale in both the delocalized phase and at the
phase-transition point in comparison to $C_P$.
  
Next, we study another kind of initially entangled state with a
tridiagonal correlation matrix denoted by `$C_E$'. The main aim here
is to understand the evolution of a state which initially has nearly
equally distant eigenvalues and hence shows a higher value of the NN
gap ratio, i.e. whose eigenvalues are more correlated than GUE. Here
the diagonal terms are determined in such a way, that they mimic the
energy levels of a harmonic oscillator ~\cite{Harmonic} and the
off-diagonal terms are added keeping in mind the positive
semi-definite nature of the correlation matrix. The matrix
corresponding to it is given by:
\begin{equation}
C_{E}=\begin{bmatrix}
c_{11} & c_{12} & 0 & 0 & \cdot\\
c_{12} & 2c_{11} & c_{23} & 0 & \cdot\\
0 & c_{23} & 3c_{11} & c_{34} & \cdot\\
0 & 0 & c_{34} & 4c_{11} & \cdot\\
\cdot & \cdot & \cdot & \cdot & \cdot\\
\end{bmatrix}.
\end{equation}
and has nearly equidistant eigenvalues. Here the value of the term
$c_{11}=\frac{2N_P}{N(N+1)}$, where `$N_P$' is the number of fermions
(which is equal to $N/2$ here) and `$N$' is the total number of sites
(i.e. the dimension of the system). The off-diagonal elements here are
chosen such that for any value of `$i$', $\sum_{i\neq j}c_{ij}<c_{ii}$
to satisfy the positive semi-definite nature of the matrix. We also
observe that evidently, the trace of the matrix $C_E$ is equal to
$N/2$, the number of fermions. The resulting EH of the subsystem also
has nearly equally distant eigenvalues, thus the gap ratio is
initially $\approx1$. We notice from
Figs.~\ref{fig10}(a)--\ref{fig10}(b) that it evolves with time
eventually to nearly the GUE value.

\begin{figure}[b]
\vspace{3mm}
\centering
\stackunder{\hspace{-3.5cm}(a)}{\includegraphics[height=3.3cm, width=4cm]{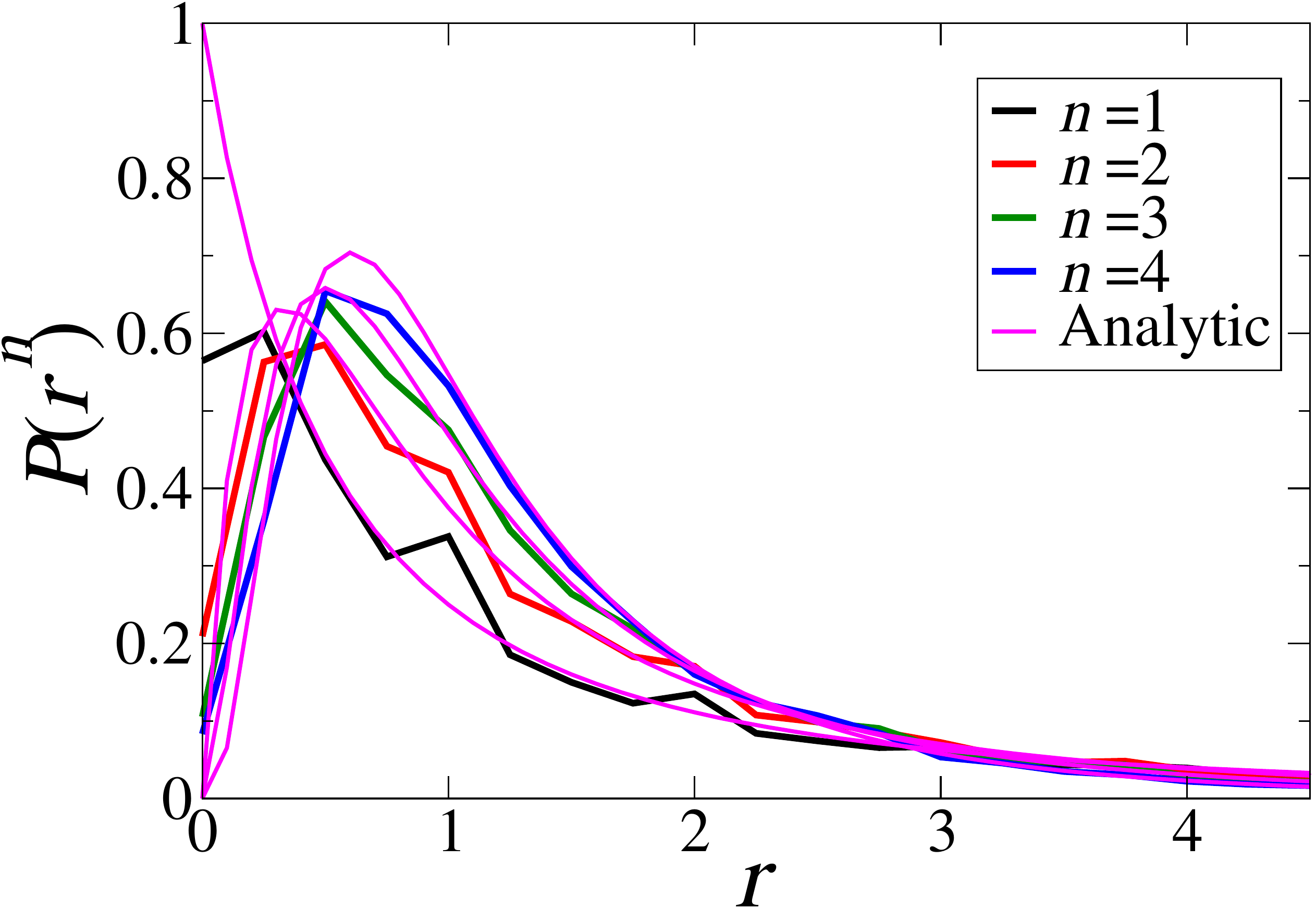}}\hspace{1mm}
\stackunder{\hspace{-3.5cm}(b)}{\includegraphics[height=3.3cm, width=4cm]{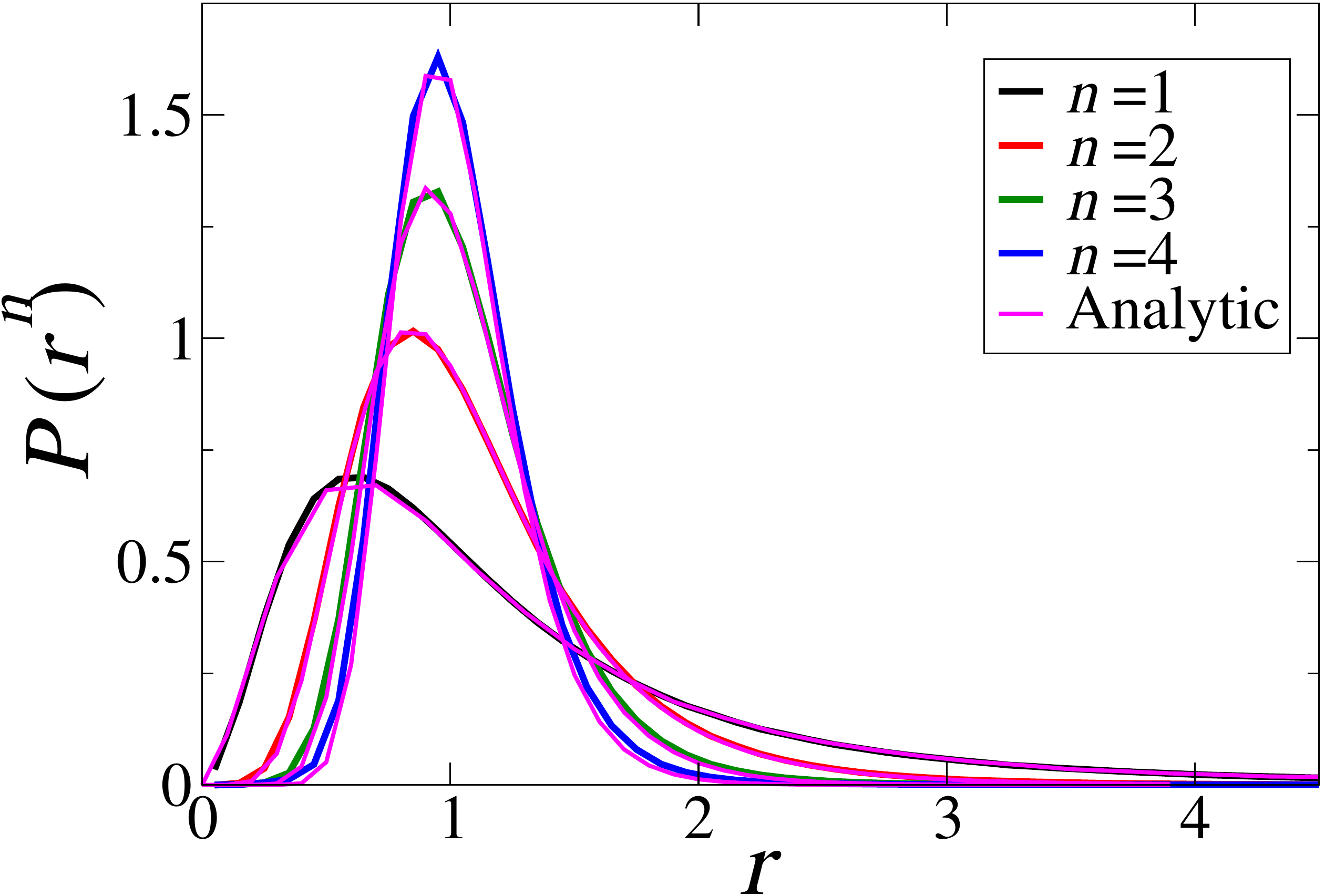}}
\caption{\label{fig11}Distribution of $n\textsuperscript{th}$
  order spacing ratios of EH spectrum for $\lambda=0.5$, $N=1920$ and
  $N_A=N/2$ with corresponding analytic curves. (a) At an early time
  $t=1$ and (b) at a late time $t=10^5$. The average has been taken
  over $5000$ random values of $\theta_p$. }
\end{figure}

It can be concluded that irrespective of the initial state, the
nearest neighbor gap ratio of the system evolves and saturates to the
GUE value in the delocalized phase and at the phase transition point
[see Figs.~\ref{fig10}(a)--\ref{fig10}(b)]. However for the localized
phase, the system does not show any evolution irrespective of the
initial state of the system [see Fig.~\ref{fig10}(c)]. Also, the
evolution of the EE for the delocalized phase and at the critical
point, are plotted in Fig.~\ref{fig10}(d) and Fig.~\ref{fig10}(e)
respectively.

In order to get a finer understanding of the nature of the initial and
final states, we also study the $n\textsuperscript{th}$ order
gap ratio given by:
\begin{equation}
r_i^{(n)}=\frac{s^{(n)}_{i+n}}{s^{(n)}_{i}}.
\label{ratio2}
\end{equation}
Note that this definition is a little different from the higher-order
gap ratio definition in Eq.~\ref{ratio}. We have considered this
variant (Eq.~\ref{ratio2}) here, since benchmark analytic results for
the probability distribution of this $n\textsuperscript{th}$ order
spacing ratios have been obtained for Gaussian
ensembles~\cite{Higher-order} and for uncorrelated
eigenvalues~\cite{Higher-order-P}. For the initial product state
($C_P$) in the delocalized phase i.e. $\lambda=0.5$, the distribution
for NN spacing ratio $n=1$ as well as higher-orders $n=2, 3, 4$ have
been plotted at the initial time, Fig.~\ref{fig11}(a) and in the
long-time limit, Fig.~\ref{fig11}(b).  The analytic
distributions~\cite{Higher-order,Higher-order-P} $P(r^{n})$ are found
to match with our numerically determined distributions. We observe
that at early times when the NN gap ratio is $\approx0.38$, the
distributions of $n\textsuperscript{th}$ order spacing ratios are also
found to match with those obtained from a sequence of uncorrelated
eigenvalues whereas at late times when $\left\langle r\right\rangle
\approx 0.599$ the distributions match with those obtained from the
spectrum of random matrices drawn from a GUE. Also in
Figs.~\ref{fig8}(a)--\ref{fig8}(b) at saturation the average value of
$r$, was found to match with the GUE values\cite{Higher-order}.

\begin{figure}
\centering
\stackunder{\hspace{-3.5cm}(a)}{\includegraphics[height=3.3cm, width=4cm]{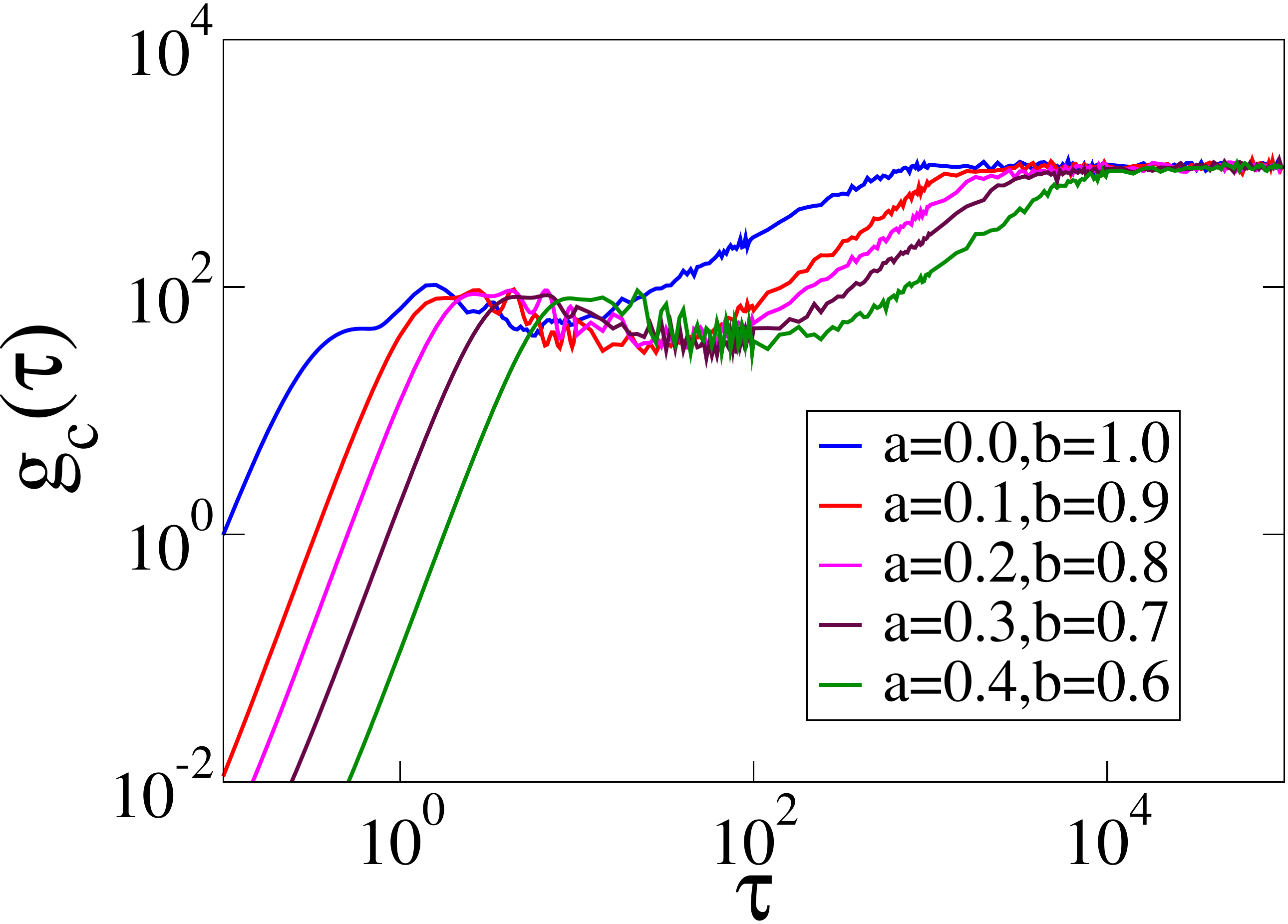}}\hspace{1mm}
\stackunder{\hspace{-3.5cm}(b)}{\includegraphics[height=3.3cm, width=4cm]{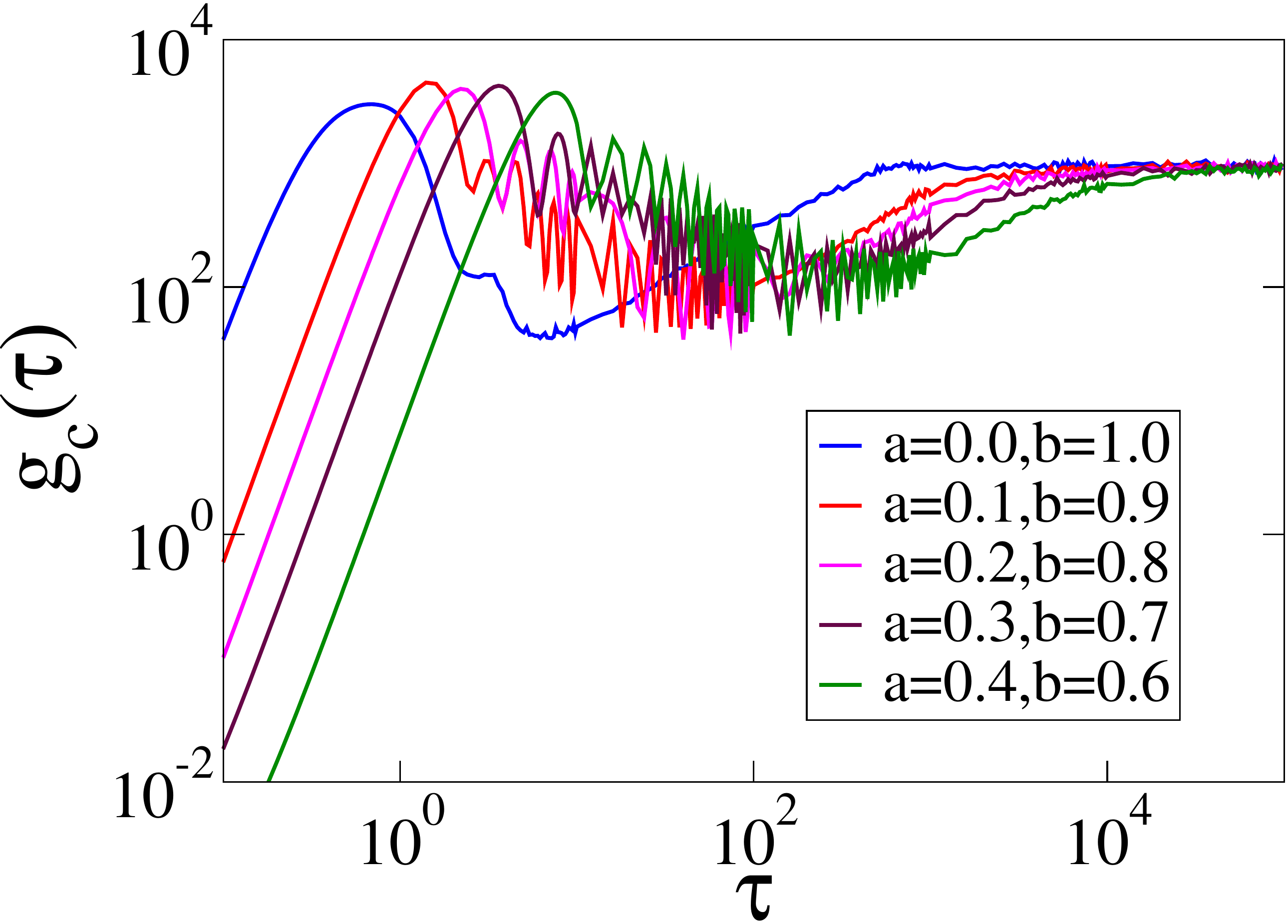}}
 \caption{\label{fig12}Connected spectral form factor (at saturation) $t=5\times 10^5$, considering various values of $a$ and $b$ of the correlation matrix `$C_S$'. Here (a) $\lambda=0.5$ and (b) $\lambda=1.0$ for $N=1920$ and $N_A=N/2$. The average has been taken over $500$ random values of $\theta_p$.}
 \vspace{4mm}
\end{figure}   

Similar results were observed for the higher-order spacing
distributions at both early and late times in the case of the
the initially entangled state obtained from the correlation matrix
$C_R$. We have also studied the same for the case of the entangled
initial state obtained from the correlation matrix $C_S$. We find that
$P(r^{n})$ here does not match with the analytic distributions
corresponding to uncorrelated eigenvalues at early times, but at late
times they show distributions obtained from the spectrum of random
matrices drawn from a GUE. In the case when the EH of the subsystem has
nearly equally distant eigenvalues (corresponding to $C_E$), at early
times the higher-order distributions of the gap ratios are peaked at
$\approx 1$ as the eigenvalues are highly correlated. At late times
only the distribution for the order $n=1$ was found to match with that
proposed by Ref.~\onlinecite{Higher-order}.

The SFF-related results are next discussed for the various initially
entangled subsystems described above. Figure~\ref{fig12}(a) and
Fig.~\ref{fig12}(b) show the connected spectral form factor
$g_c(\tau)$ at saturation (long-time limit) for different values of
$a$ and $b$ for the delocalized phase and at the critical point
respectively. We have determined the length of the ramp as
follows. The time ($\tau$) at which the ramp starts (i.e. the Thouless
time) is also a local minimum of the connected SFF which can be found
by zooming in the area close to the local minimum and then determining
it. For the plateau time (i.e the value of $\tau$ at which $g(\tau)$
or $g_c(\tau)$ saturates), we consider the time taken for $g_c(\tau)$
to reach $90\%$ (allowing some room for fluctuations) of its
saturation value (i.e.  $\lim_{\tau\rightarrow \infty}
g_c(\tau)=N_A$). The length of the ramp can be found by deducting the
Thouless time from the plateau time.
\begin{figure}
\vspace{3mm}
\centering
\stackunder{\hspace{-3.5cm}(a)}{\includegraphics[height=3.3cm, width=4cm]{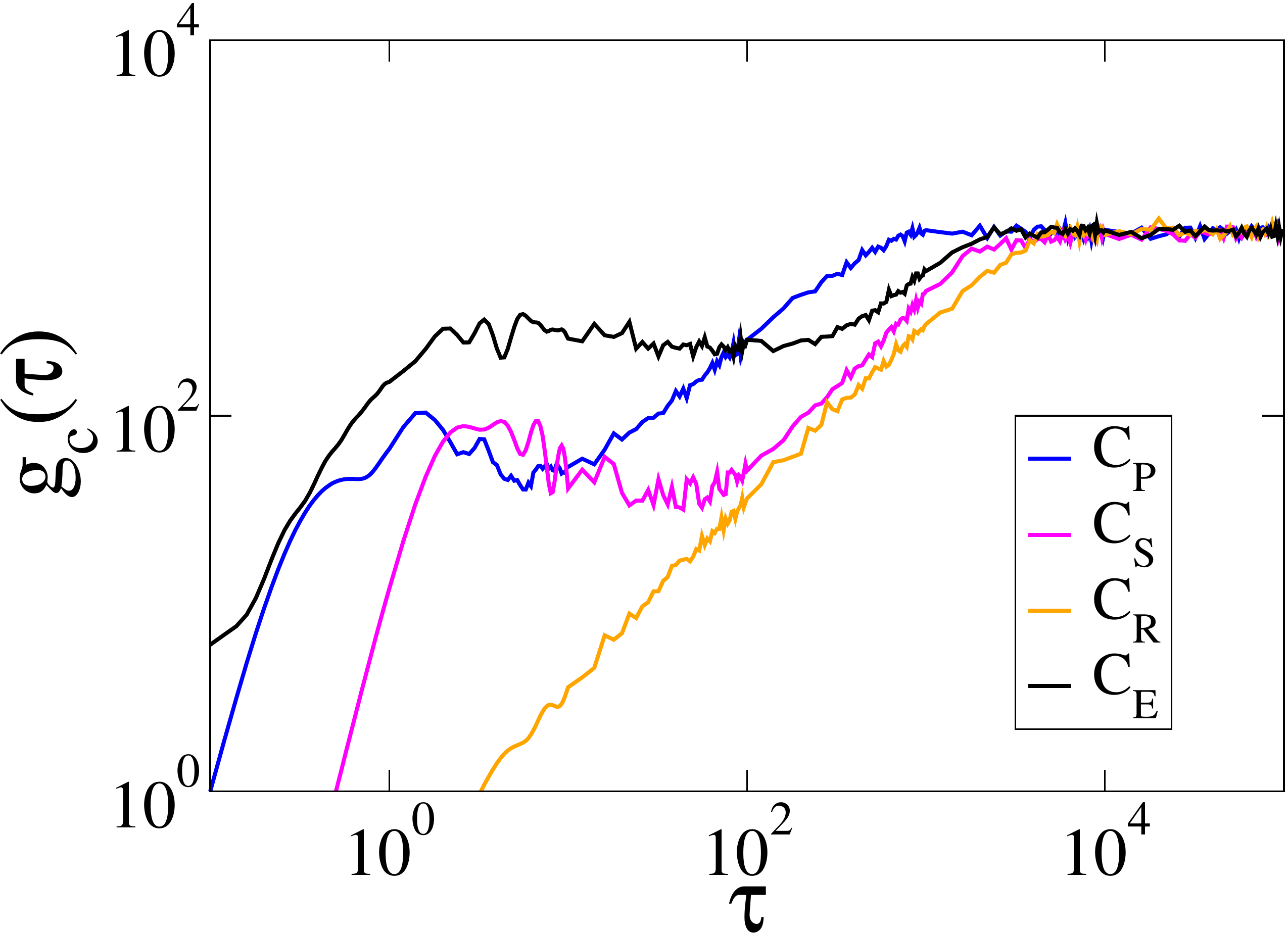}}\hspace{1mm}
\stackunder{\hspace{-3.5cm}(b)}{\includegraphics[height=3.3cm, width=4cm]{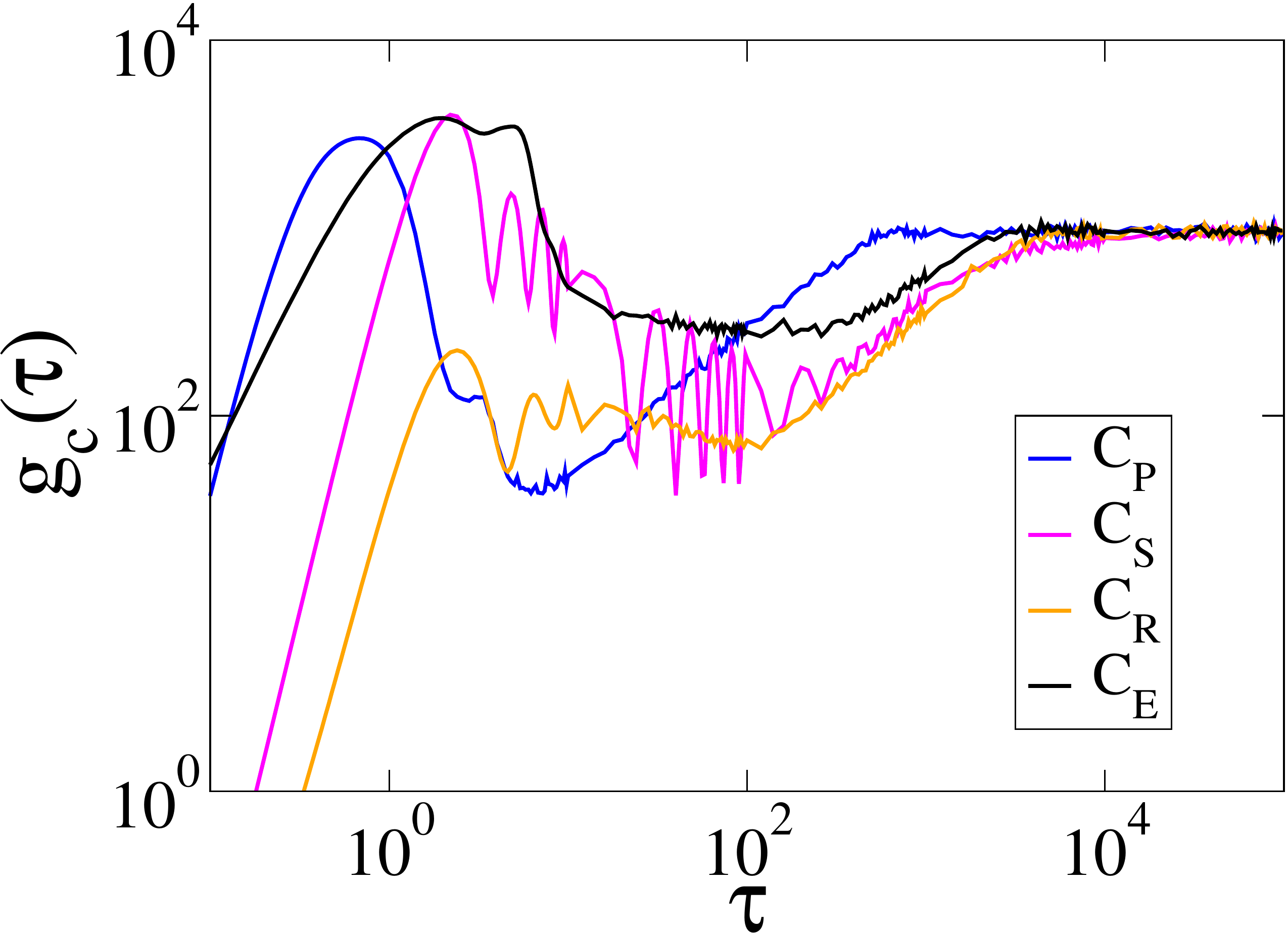}}
\caption{\label{fig13}Connected spectral form factor (at saturation) $t=5\times 10^5$, for various initial correlation matrices: $C_E$ matrix, $C_P$ matrix, $C_R$ matrix and $C_S$ matrix (here $a=0.2$, $b=0.8$) for $N=1920$ and $N_A=N/2$. Here (a) $\lambda=0.5$ and (b) $\lambda=1.0$ for $N=1920$ and $N_A=N/2$. The average has been taken over $500$ random values of $\theta_p$.}
\end{figure} 

\begin{figure}[b]
\centering
\stackunder{\hspace{-3.5cm}(a)}{\includegraphics[height=3.3cm, width=4cm]{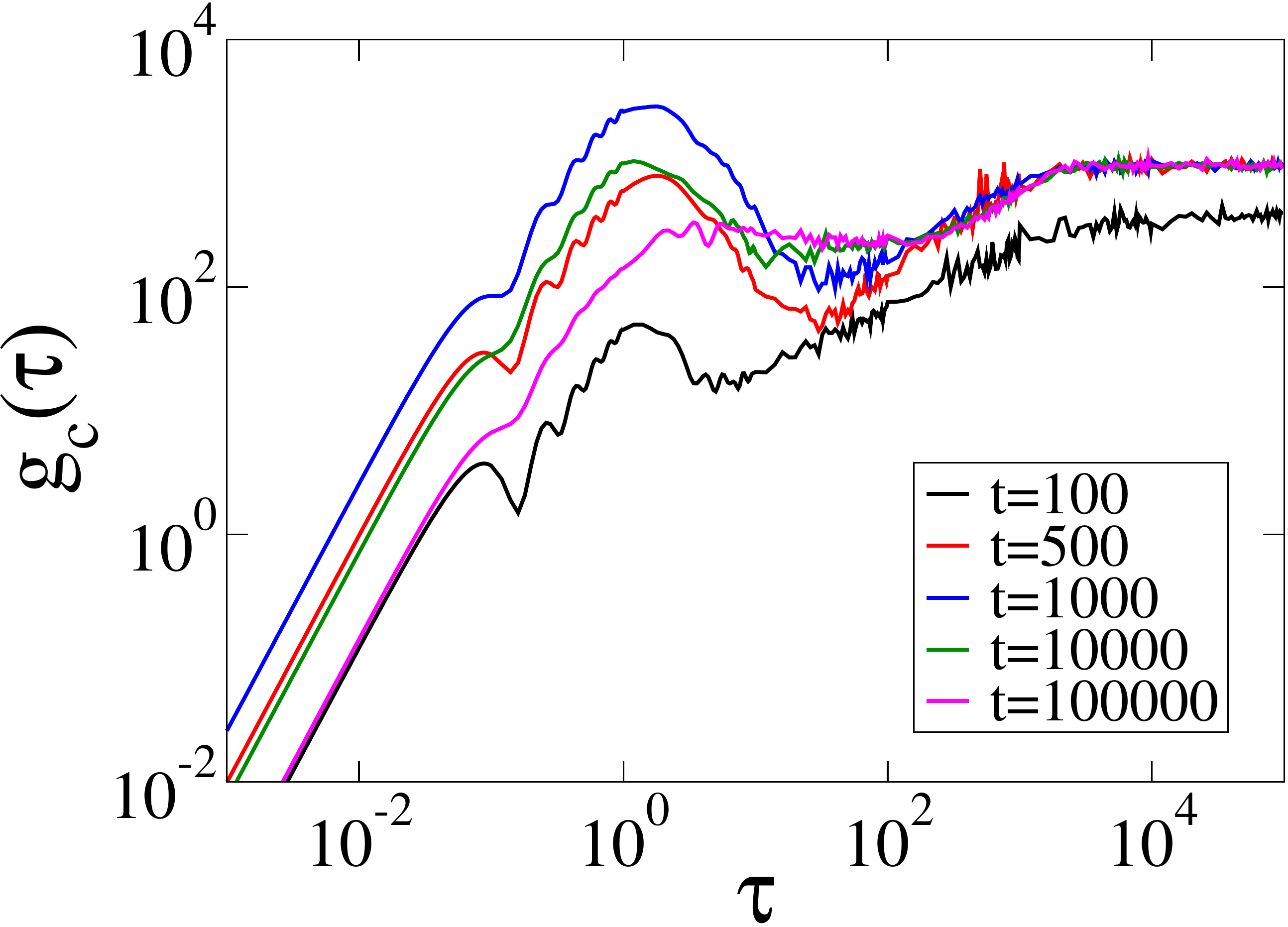}}\hspace{1mm}
\stackunder{\hspace{-3.5cm}(b)}{\includegraphics[height=3.3cm, width=4cm]{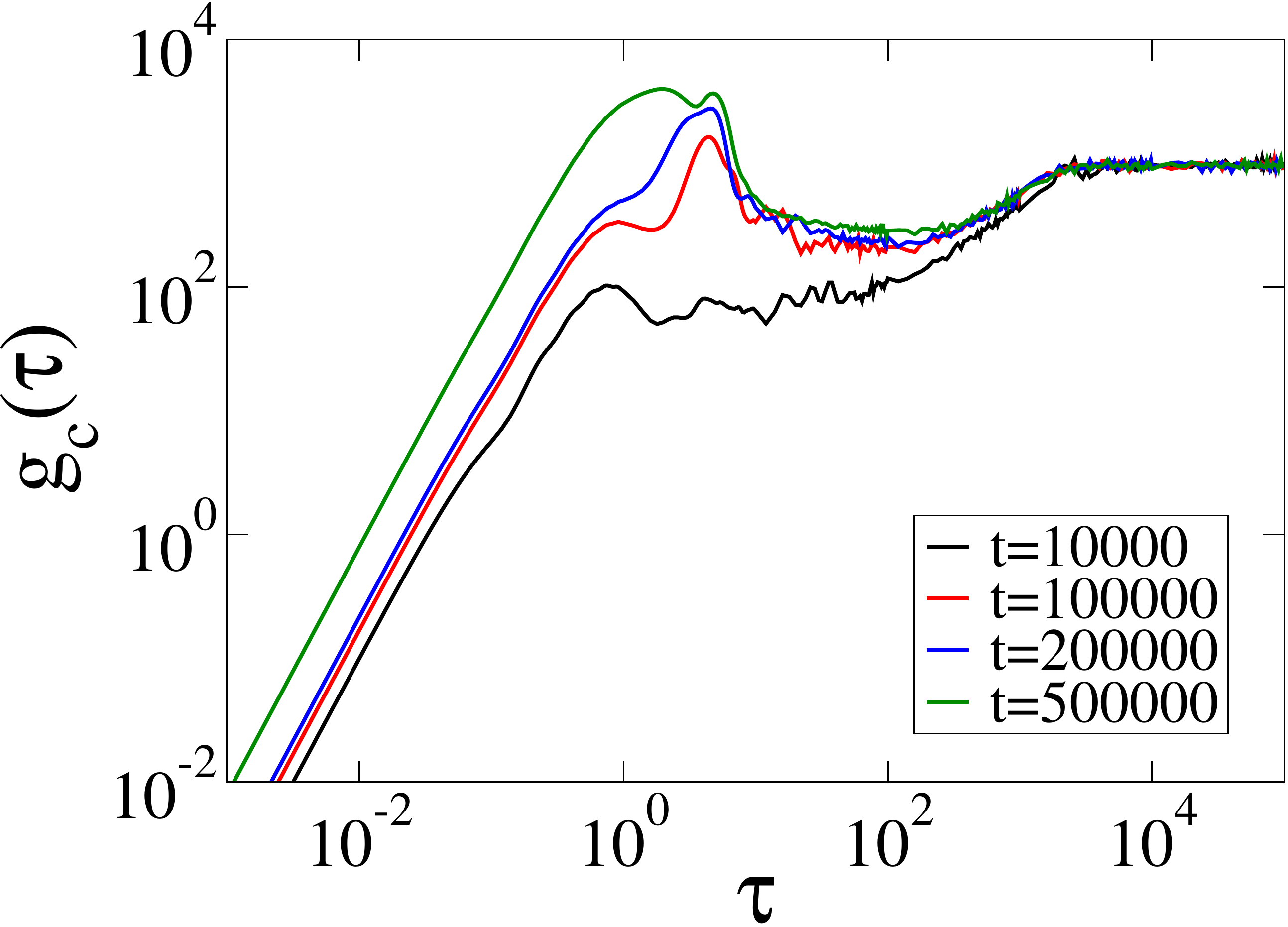}}
\caption{\label{fig14}Evolution of connected spectral form factor obtained from the spectrum of EH corresponding to the $C_E$ matrix. (a)$\lambda=0.5$ and (b)$\lambda=1.0$. Here system size $N=1920$ and $N_A=N/2$. The average has been taken over $500$ random values of $\theta_p$. }
\vspace{4mm}
\end{figure} 

 As one goes from the product state to the maximally entangled state,
 the magnitude of EE increases at saturation [see
   Figs.~\ref{fig10}(a)--\ref{fig10}(b)]. This is due to decreasing
 magnitude of the eigenvalues of the EH. Since the smaller the
 magnitude of the eigenvalues, the larger will be the value of $\tau$
 at which their difference dephases, the length (in units of $\tau$)
 of the ramp will become longer with the increasing value of EE. The
 same is true when we compare the connected spectral form factor in
 the long-time limit for the various correlation matrices, in the
 delocalized and critical phases which are shown in
 Figs.~\ref{fig13}(a) and Fig.~\ref{fig13}(b) respectively. It can be
 concluded that there exists a correspondence between the magnitude of
 EE and the length of the ramp (in units of $\tau$) which is not
 apparent through the NN gap ratio. A systematic comparison of the
 states attained by the system starting from various initial states is
 useful. During the first timescale, the length of the ramp is always
 longer corresponding to a higher EE, irrespective of the magnitude of
 the NN gap ratio. However if EEs are comparable (eg compare $C_E$ and
 $C_S$ with $a=0.2$ and $b=0.8$), the NN gap ratio seems to come into
 play. We observe that the length of the ramp corresponding to a
 higher NN gap ratio is longer.  Also in the maximally entangled case
 i.e. when $a=0.5$ and $b=0.5$, we find both numerically and
 analytically that $g(\tau)=N_A^2$, while $g_c(\tau)=0$, on account of
 all the EH eigenvalues being $\approx 0$. In the second timescale
 since the NN gap ratios have already saturated, we see a clear
 monotonic relationship between the length of the ramp and the EE.

 When the initial state corresponds to the $C_E$ matrix, we see a
 distinct feature that is observable from
 Figs.~\ref{fig14}(a)--\ref{fig14}(b), which present the evolution of
 $g_{c}(\tau)$ in the delocalized phase and at the critical point
 respectively. Here due to the presence of correlations in the
 spectrum initially, the ramp structure is observed even at early
 times. This however also evolves with time and saturates for time
 $t>t_2$ i.e. when von-Neumann entropy saturates. The change in the
 length of the ramp as time evolves is not very clear here and
 requires further investigation. In particular, it would be
 interesting to study if the presence of initial higher-order
 correlations in the system is the reason for this anomalous
 behavior. In the localized phase, the ramp does not evolve,
 independent of the initial spectral correlations.

\section{\label{sec:level4}Conclusion}
We present a study of the non-equilibrium dynamics of the entanglement
spectrum of a disordered integrable system, namely the
Aubry-Andr\'e-Harper model which hosts a delocalization-localization
transition. For a system of non-interacting fermions in 1-D, we
numerically explore the presence and the development of spectral
correlations in the entanglement Hamiltonian spectrum.  We explore the
time evolution of the spectral form factor which is a measure of long
range correlations, along with other quantities such as the gap ratio
and the Renyi entropies for various initial (subsystem) correlation
matrices. The spectral form factor exhibits a characteristic `ramp'
structure whose length is intimately connected to the spectral
correlations.

We have analyzed various aspects of the EH spectrum when an initial
state of the system is density-wave (DW) type and is evolved under the
Hamiltonian.  One can see the presence of three distinct timescales in
the delocalized phase and at the critical point. We identify the time
$t_1$ at which the NN gap ratio saturates along with the zeroth-order
Renyi entropy $S_0$ and the time $t_2$ at which both the von Neumann
entropy $S_1$ and entanglement bandwidth $\delta_E$ saturate. The
three timescales are defined by $t<t_1$, $t_1<t<t_2$ and $t>t_2$. In
addition, we also find signatures of the timescales from the evolution
of the spectral density of the EH. We observe from the SFF that the
length of the ramp (in units of $\tau$) is nearly the same in the
delocalized phase and at the critical point which signifies that the
number of eigenvalues that exhibit universal spectral correlations are
approximately equal for both. However, in the localized phase, the
time evolution is seen to be completely flat, which is characteristic
of localization.

Also, a comparison between zeroth and first-order Renyi
entropies shows that at saturation though the number of non-zero
eigenvalues are the same for both the delocalized phase and at the
critical point their magnitudes are significantly different. The
length of the ramp of the SFF is also observed to decrease with the
system size.  We also study the time evolution of the higher-order gap
ratio. It is observed that the time taken for $n^{th}$ order gap ratio
to saturate increases initially with the order `$n$' till a certain
order, after which all higher orders saturate at the same time `$t$'
at which the SFF also saturates. A spectrum-resolved study of the
spread of correlations reveals that in the delocalized phase, the
spectrum becomes more correlated as we move towards the center from
the edges while at the phase transition point they are more evenly
spread out through the spectrum.

We also study the dynamics starting from a variety of initial states.
For all possible initial states, one can see that the gap ratio
eventually saturates approximately to its GUE value in the delocalized
phase and at the phase-transition point. The three timescales exist
for all these states, however, when the entanglement in the initial
state is high, the second timescale becomes smaller. It is also
observed that generically the length of the ramp (in units of $\tau$)
is greater when the corresponding entanglement entropy has a higher
magnitude. However, there are some exceptions to this trend when the
gap ratio also seems to play a role.

In our study, we are able to link the long-range spectral correlations
quantified by the spectral form factor to various physically relevant
observables in a disordered integrable model by studying the quench
dynamics of the system in different phases. As a future possibility,
one can also study interacting versions of this and other allied
models~\cite{prakash2021universal}. Another interesting direction
would be to probe higher moments of the spectral form factor. The SFF
remains to be explored in several contexts where the spectral
correlations of the entanglement spectrum are important. An
inexhaustive list of systems where this quantity may potentially find
application includes periodically driven
systems~\cite{roy2020random,bertini2018exact}, open quantum
systems~\cite{carollo2020entanglement,kolovsky2020quantum},
non-Hermitian~\cite{chang2020entanglement,li2021spectral} and
topological~\cite{gong2018topological,su2020quench} systems.

\begin{acknowledgments}
A.A. is grateful to the Council of Scientific and Industrial Research
(CSIR), India for her Ph.D. fellowship.  N.R acknowledges financial
support from the Indian Institute of Science under the IOE-IISc fellowship
program.  A.S acknowledges financial support from SERB via the grant
(File Number: CRG/2019/003447), and from DST via the DST-INSPIRE
Faculty Award [DST/INSPIRE/04/2014/002461].

\end{acknowledgments}

\bibliography{SFF}

\providecommand{\noopsort}[1]{}\providecommand{\singleletter}[1]{#1}%
\begin{thebibliography}{68}%
\makeatletter
\providecommand \@ifxundefined [1]{%
 \@ifx{#1\undefined}
}%
\providecommand \@ifnum [1]{%
 \ifnum #1\expandafter \@firstoftwo
 \else \expandafter \@secondoftwo
 \fi
}%
\providecommand \@ifx [1]{%
 \ifx #1\expandafter \@firstoftwo
 \else \expandafter \@secondoftwo
 \fi
}%
\providecommand \natexlab [1]{#1}%
\providecommand \enquote  [1]{``#1''}%
\providecommand \bibnamefont  [1]{#1}%
\providecommand \bibfnamefont [1]{#1}%
\providecommand \citenamefont [1]{#1}%
\providecommand \href@noop [0]{\@secondoftwo}%
\providecommand \href [0]{\begingroup \@sanitize@url \@href}%
\providecommand \@href[1]{\@@startlink{#1}\@@href}%
\providecommand \@@href[1]{\endgroup#1\@@endlink}%
\providecommand \@sanitize@url [0]{\catcode `\\12\catcode `\$12\catcode
  `\&12\catcode `\#12\catcode `\^12\catcode `\_12\catcode `\%12\relax}%
\providecommand \@@startlink[1]{}%
\providecommand \@@endlink[0]{}%
\providecommand \url  [0]{\begingroup\@sanitize@url \@url }%
\providecommand \@url [1]{\endgroup\@href {#1}{\urlprefix }}%
\providecommand \urlprefix  [0]{URL }%
\providecommand \Eprint [0]{\href }%
\providecommand \doibase [0]{https://doi.org/}%
\providecommand \selectlanguage [0]{\@gobble}%
\providecommand \bibinfo  [0]{\@secondoftwo}%
\providecommand \bibfield  [0]{\@secondoftwo}%
\providecommand \translation [1]{[#1]}%
\providecommand \BibitemOpen [0]{}%
\providecommand \bibitemStop [0]{}%
\providecommand \bibitemNoStop [0]{.\EOS\space}%
\providecommand \EOS [0]{\spacefactor3000\relax}%
\providecommand \BibitemShut  [1]{\csname bibitem#1\endcsname}%
\let\auto@bib@innerbib\@empty
\bibitem [{\citenamefont {{R. Dubertrand and S. Muller}}(2016)}]{chaos1}%
  \BibitemOpen
  \bibfield  {author} {\bibinfo {author} {\bibnamefont {{R. Dubertrand and S.
  Muller}}},\ }\href {https://doi.org/10.1088/1367-2630/18/3/033009} {\bibfield
   {journal} {\bibinfo  {journal} {New J. Phys.}\ }\textbf {\bibinfo {volume}
  {18}},\ \bibinfo {pages} {033009} (\bibinfo {year} {2016})}\BibitemShut
  {NoStop}%
\bibitem [{\citenamefont {{J. Cotlera and N. H. Jones}}(2020)}]{chaos2}%
  \BibitemOpen
  \bibfield  {author} {\bibinfo {author} {\bibnamefont {{J. Cotlera and N. H.
  Jones}}},\ }\href {https://doi.org/10.1007/JHEP12(2020)205} {\bibfield
  {journal} {\bibinfo  {journal} {J. High Energy Phys.}\ }\textbf {\bibinfo
  {volume} {12}},\ \bibinfo {pages} {205(2020)}}\BibitemShut {NoStop}%
\bibitem [{\citenamefont {{F. Borgonovi, F. M. Izrailev, and L. F.
  Santos}}(2019{\natexlab{a}})}]{chaos3}%
  \BibitemOpen
  \bibfield  {author} {\bibinfo {author} {\bibnamefont {{F. Borgonovi, F. M.
  Izrailev, and L. F. Santos}}},\ }\href
  {https://link.aps.org/doi/10.1103/PhysRevE.99.010101} {\bibfield  {journal}
  {\bibinfo  {journal} {Phys. Rev. E}\ }\textbf {\bibinfo {volume} {99}},\
  \bibinfo {pages} {010101(R)} (\bibinfo {year}
  {2019}{\natexlab{a}})}\BibitemShut {NoStop}%
\bibitem [{\citenamefont {{F. Borgonovi, F. M. Izrailev, and L. F.
  Santos}}(2019{\natexlab{b}})}]{chaos4}%
  \BibitemOpen
  \bibfield  {author} {\bibinfo {author} {\bibnamefont {{F. Borgonovi, F. M.
  Izrailev, and L. F. Santos}}},\ }\href
  {https://link.aps.org/doi/10.1103/PhysRevE.99.052143} {\bibfield  {journal}
  {\bibinfo  {journal} {Phys. Rev. E}\ }\textbf {\bibinfo {volume} {99}},\
  \bibinfo {pages} {052143} (\bibinfo {year} {2019}{\natexlab{b}})}\BibitemShut
  {NoStop}%
\bibitem [{\citenamefont {{T. Xu, T. Scaffidi, and X. Cao}}(2020)}]{chaos5}%
  \BibitemOpen
  \bibfield  {author} {\bibinfo {author} {\bibnamefont {{T. Xu, T. Scaffidi,
  and X. Cao}}},\ }\href
  {https://link.aps.org/doi/10.1103/PhysRevLett.124.140602} {\bibfield
  {journal} {\bibinfo  {journal} {Phys.\ Rev.Lett.}\ }\textbf {\bibinfo
  {volume} {124}},\ \bibinfo {pages} {140602} (\bibinfo {year}
  {2020})}\BibitemShut {NoStop}%
\bibitem [{\citenamefont {{É. L. Hurtubise, S. Plugge, O. Can, and M.
  Franz}}(2020)}]{chaos6}%
  \BibitemOpen
  \bibfield  {author} {\bibinfo {author} {\bibnamefont {{É. L. Hurtubise, S.
  Plugge, O. Can, and M. Franz}}},\ }\href
  {https://link.aps.org/doi/10.1103/PhysRevResearch.2.013254} {\bibfield
  {journal} {\bibinfo  {journal} {Phys.\ Rev.Research.2}\ ,\ \bibinfo {pages}
  {013254}} (\bibinfo {year} {2020})}\BibitemShut {NoStop}%
\bibitem [{\citenamefont {{L. F. Santos , F. Perez-Bernal , and E. J.
  Torres-Herrera }}(2020)}]{chaos7}%
  \BibitemOpen
  \bibfield  {author} {\bibinfo {author} {\bibnamefont {{L. F. Santos , F.
  Perez-Bernal , and E. J. Torres-Herrera }}},\ }\href
  {https://link.aps.org/doi/10.1103/PhysRevResearch.2.043034} {\bibfield
  {journal} {\bibinfo  {journal} {Phys.\ Rev.Research.2}\ ,\ \bibinfo {pages}
  {043034}} (\bibinfo {year} {2020})}\BibitemShut {NoStop}%
\bibitem [{\citenamefont {{A. Chan, A. DeLuca, and J. T.
  Chalker}}(2018{\natexlab{a}})}]{chaos8}%
  \BibitemOpen
  \bibfield  {author} {\bibinfo {author} {\bibnamefont {{A. Chan, A. DeLuca,
  and J. T. Chalker}}},\ }\href
  {https://link.aps.org/doi/10.1103/PhysRevX.8.041019} {\bibfield  {journal}
  {\bibinfo  {journal} {Phys.\ Rev.X}\ }\textbf {\bibinfo {volume} {8}},\
  \bibinfo {pages} {041019} (\bibinfo {year} {2018}{\natexlab{a}})}\BibitemShut
  {NoStop}%
\bibitem [{\citenamefont {{T. Zhou, and A. Nahum}}(2020)}]{chaos9}%
  \BibitemOpen
  \bibfield  {author} {\bibinfo {author} {\bibnamefont {{T. Zhou, and A.
  Nahum}}},\ }\href {https://link.aps.org/doi/10.1103/PhysRevX.10.031066}
  {\bibfield  {journal} {\bibinfo  {journal} {Phys.\ Rev.X}\ }\textbf {\bibinfo
  {volume} {10}},\ \bibinfo {pages} {031066} (\bibinfo {year}
  {2020})}\BibitemShut {NoStop}%
\bibitem [{\citenamefont {{E. J. T. Herrera and L. F.
  Santos}}(2019)}]{chaos10}%
  \BibitemOpen
  \bibfield  {author} {\bibinfo {author} {\bibnamefont {{E. J. T. Herrera and
  L. F. Santos}}},\ }\href {https://doi.org/10.1140/epjst/e2019-800057-8}
  {\bibfield  {journal} {\bibinfo  {journal} {Eur. Phys. J. Spec. Top.}\
  }\textbf {\bibinfo {volume} {227}},\ \bibinfo {pages} {1897–1910} (\bibinfo
  {year} {2019})}\BibitemShut {NoStop}%
\bibitem [{\citenamefont {{E. J. T. Herrera and L. F.
  Santos}}(2020)}]{chaos11}%
  \BibitemOpen
  \bibfield  {author} {\bibinfo {author} {\bibnamefont {{E. J. T. Herrera and
  L. F. Santos}}},\ }\href {https://www.mdpi.com/2410-3896/5/1/7} {\bibfield
  {journal} {\bibinfo  {journal} {Condens. Matter}\ }\textbf {\bibinfo {volume}
  {5}},\ \bibinfo {pages} {7} (\bibinfo {year} {2020})}\BibitemShut {NoStop}%
\bibitem [{\citenamefont {Haake}()}]{Haake}%
  \BibitemOpen
  \bibfield  {author} {\bibinfo {author} {\bibfnamefont {F.}~\bibnamefont
  {Haake}},\ }\href {https://doi.org/10.1007/978-1-4899-3698-1_38} {\emph
  {\bibinfo {title} {Quantum Signatures of Chaos}}},\ 3rd ed.,Springer Series
  in Synergetics,Vol. 54\ (\bibinfo  {publisher} {Springer-Verlag,
  Berlin,2010})\BibitemShut {NoStop}%
\bibitem [{\citenamefont {Mehta}()}]{Mehta}%
  \BibitemOpen
  \bibfield  {author} {\bibinfo {author} {\bibfnamefont {M.~L.}\ \bibnamefont
  {Mehta}},\ }\href@noop {} {\emph {\bibinfo {title} {Random Matrices}}},\ 3rd
  ed.,Pure and Applied Mathematics,Vol. 142\ (\bibinfo  {publisher} {Elsevier,
  New York, 2004})\BibitemShut {NoStop}%
\bibitem [{\citenamefont {{E. J. T. Herrera, J. Karp, M. Távora and L. F.
  Santos}}(2016)}]{RMT1}%
  \BibitemOpen
  \bibfield  {author} {\bibinfo {author} {\bibnamefont {{E. J. T. Herrera, J.
  Karp, M. Távora and L. F. Santos}}},\ }\href
  {https://www.mdpi.com/1099-4300/18/10/359} {\bibfield  {journal} {\bibinfo
  {journal} {Entropy}\ }\textbf {\bibinfo {volume} {18}},\ \bibinfo {pages}
  {359} (\bibinfo {year} {2016})}\BibitemShut {NoStop}%
\bibitem [{\citenamefont {{E.B. Rozenbaum, Sriram Ganeshan, and Victor
  Galitski}}(2019)}]{OTOC1}%
  \BibitemOpen
  \bibfield  {author} {\bibinfo {author} {\bibnamefont {{E.B. Rozenbaum, Sriram
  Ganeshan, and Victor Galitski}}},\ }\href
  {https://link.aps.org/doi/10.1103/PhysRevB.100.035112} {\bibfield  {journal}
  {\bibinfo  {journal} {Phys.\ Rev.B}\ }\textbf {\bibinfo {volume} {100}},\
  \bibinfo {pages} {035112} (\bibinfo {year} {2019})}\BibitemShut {NoStop}%
\bibitem [{\citenamefont {{P. D. Bergamasco, G. G. Carlo and A. M. F.
  Rivas}}(2019)}]{OTOC2}%
  \BibitemOpen
  \bibfield  {author} {\bibinfo {author} {\bibnamefont {{P. D. Bergamasco, G.
  G. Carlo and A. M. F. Rivas}}},\ }\href
  {https://link.aps.org/doi/10.1103/PhysRevResearch.1.033044} {\bibfield
  {journal} {\bibinfo  {journal} {Phys.\ Rev.Research.1}\ ,\ \bibinfo {pages}
  {033044}} (\bibinfo {year} {2019})}\BibitemShut {NoStop}%
\bibitem [{\citenamefont {{J. Lee, D. Kim, and D.H. Kim}}(2019)}]{OTOC3}%
  \BibitemOpen
  \bibfield  {author} {\bibinfo {author} {\bibnamefont {{J. Lee, D. Kim, and
  D.H. Kim}}},\ }\href {https://link.aps.org/doi/10.1103/PhysRevB.99.184202}
  {\bibfield  {journal} {\bibinfo  {journal} {Phys.\ Rev.B}\ }\textbf {\bibinfo
  {volume} {99}},\ \bibinfo {pages} {184202} (\bibinfo {year}
  {2019})}\BibitemShut {NoStop}%
\bibitem [{\citenamefont {{E. Brezin and S. Hikami}}(1997)}]{SFF1}%
  \BibitemOpen
  \bibfield  {author} {\bibinfo {author} {\bibnamefont {{E. Brezin and S.
  Hikami}}},\ }\href {https://link.aps.org/doi/10.1103/PhysRevE.55.4067}
  {\bibfield  {journal} {\bibinfo  {journal} {Phys. Rev. E}\ }\textbf {\bibinfo
  {volume} {55}},\ \bibinfo {pages} {4067} (\bibinfo {year}
  {1997})}\BibitemShut {NoStop}%
\bibitem [{\citenamefont {{Y. Chen, H. Zhaia and P. Zhang}}(2017)}]{OTOC4}%
  \BibitemOpen
  \bibfield  {author} {\bibinfo {author} {\bibnamefont {{Y. Chen, H. Zhaia and
  P. Zhang}}},\ }\href {https://doi.org/10.1007/JHEP07(2017)150} {\bibfield
  {journal} {\bibinfo  {journal} {J. High Energy Phys.}\ }\textbf {\bibinfo
  {volume} {07}},\ \bibinfo {pages} {150(2017)}}\BibitemShut {NoStop}%
\bibitem [{\citenamefont {{T. Nosaka, D. Rosa and J. Yoon}}(2018)}]{OTOC5}%
  \BibitemOpen
  \bibfield  {author} {\bibinfo {author} {\bibnamefont {{T. Nosaka, D. Rosa and
  J. Yoon}}},\ }\href {https://doi.org/10.1007/JHEP09(2018)041} {\bibfield
  {journal} {\bibinfo  {journal} {J. High Energy Phys.}\ }\textbf {\bibinfo
  {volume} {09}},\ \bibinfo {pages} {041(2018)}}\BibitemShut {NoStop}%
\bibitem [{\citenamefont {{B. Lian, S.L. Sondhib and Z. Yang}}(2019)}]{OTOC6}%
  \BibitemOpen
  \bibfield  {author} {\bibinfo {author} {\bibnamefont {{B. Lian, S.L. Sondhib
  and Z. Yang}}},\ }\href {https://doi.org/10.1007/JHEP09(2019)067} {\bibfield
  {journal} {\bibinfo  {journal} {J. High Energy Phys.}\ }\textbf {\bibinfo
  {volume} {09}},\ \bibinfo {pages} {067(2019)}}\BibitemShut {NoStop}%
\bibitem [{\citenamefont {{J. S. Cotler, G. Gur-Ari, M. Hanada, J. Polchinski,
  P. Saad, S. H. Shenker, D. Stanford, A. Streichera and M.
  Tezuka}}(2017)}]{SFF2}%
  \BibitemOpen
  \bibfield  {author} {\bibinfo {author} {\bibnamefont {{J. S. Cotler, G.
  Gur-Ari, M. Hanada, J. Polchinski, P. Saad, S. H. Shenker, D. Stanford, A.
  Streichera and M. Tezuka}}},\ }\href
  {https://link.springer.com/content/pdf/10.1007/JHEP05(2017)118.pdf}
  {\bibfield  {journal} {\bibinfo  {journal} {J. High Energy Phys.}\ }\textbf
  {\bibinfo {volume} {05}},\ \bibinfo {pages} {118(2017)}}\BibitemShut
  {NoStop}%
\bibitem [{\citenamefont {{X. Chen and A. W. W. Ludwig}}(2018)}]{SFF3}%
  \BibitemOpen
  \bibfield  {author} {\bibinfo {author} {\bibnamefont {{X. Chen and A. W. W.
  Ludwig}}},\ }\href {https://link.aps.org/doi/10.1103/PhysRevB.98.064309}
  {\bibfield  {journal} {\bibinfo  {journal} {Phys.\ Rev.B}\ }\textbf {\bibinfo
  {volume} {98}},\ \bibinfo {pages} {064309} (\bibinfo {year}
  {2018})}\BibitemShut {NoStop}%
\bibitem [{\citenamefont {{P. Sierant and J. Zakrzewski}}(2020)}]{SFF5}%
  \BibitemOpen
  \bibfield  {author} {\bibinfo {author} {\bibnamefont {{P. Sierant and J.
  Zakrzewski}}},\ }\href {https://link.aps.org/doi/10.1103/PhysRevB.101.104201}
  {\bibfield  {journal} {\bibinfo  {journal} {Phys.\ Rev.B}\ }\textbf {\bibinfo
  {volume} {101}},\ \bibinfo {pages} {104201} (\bibinfo {year}
  {2020})}\BibitemShut {NoStop}%
\bibitem [{\citenamefont {{A. J. Friedman, A. Chan, A. DeLuca ,and J. T.
  Chalker}}(2019)}]{SFF8}%
  \BibitemOpen
  \bibfield  {author} {\bibinfo {author} {\bibnamefont {{A. J. Friedman, A.
  Chan, A. DeLuca ,and J. T. Chalker}}},\ }\href
  {https://link.aps.org/doi/10.1103/PhysRevLett.123.210603} {\bibfield
  {journal} {\bibinfo  {journal} {Phys.\ Rev.Lett.}\ }\textbf {\bibinfo
  {volume} {123}},\ \bibinfo {pages} {210603} (\bibinfo {year}
  {2019})}\BibitemShut {NoStop}%
\bibitem [{\citenamefont {{P. Kos, M. Ljubotina, and T. Prosen}}(2018)}]{SFF9}%
  \BibitemOpen
  \bibfield  {author} {\bibinfo {author} {\bibnamefont {{P. Kos, M. Ljubotina,
  and T. Prosen}}},\ }\href
  {https://link.aps.org/doi/10.1103/PhysRevX.8.021062} {\bibfield  {journal}
  {\bibinfo  {journal} {Phys.\ Rev.X}\ }\textbf {\bibinfo {volume} {8}},\
  \bibinfo {pages} {021062} (\bibinfo {year} {2018})}\BibitemShut {NoStop}%
\bibitem [{\citenamefont {{H. Gharibyan, M. Hanada, S. H. Shenkera and M.
  Tezukae}}(2018)}]{SFF4}%
  \BibitemOpen
  \bibfield  {author} {\bibinfo {author} {\bibnamefont {{H. Gharibyan, M.
  Hanada, S. H. Shenkera and M. Tezukae}}},\ }\href
  {https://doi.org/10.1007/JHEP07(2018)124} {\bibfield  {journal} {\bibinfo
  {journal} {J. High Energy Phys.}\ }\textbf {\bibinfo {volume} {07}},\
  \bibinfo {pages} {124(2018)}}\BibitemShut {NoStop}%
\bibitem [{\citenamefont {{A. Gaikwad,and R. Sinha}}(2019)}]{SFF6}%
  \BibitemOpen
  \bibfield  {author} {\bibinfo {author} {\bibnamefont {{A. Gaikwad,and R.
  Sinha}}},\ }\href {https://link.aps.org/doi/10.1103/PhysRevD.100.026017}
  {\bibfield  {journal} {\bibinfo  {journal} {Phys.\ Rev.D}\ }\textbf {\bibinfo
  {volume} {100}},\ \bibinfo {pages} {026017} (\bibinfo {year}
  {2019})}\BibitemShut {NoStop}%
\bibitem [{\citenamefont {{A. Chan, A. DeLuca, and J. T.
  Chalker}}(2018{\natexlab{b}})}]{SFF7}%
  \BibitemOpen
  \bibfield  {author} {\bibinfo {author} {\bibnamefont {{A. Chan, A. DeLuca,
  and J. T. Chalker}}},\ }\href
  {https://link.aps.org/doi/10.1103/PhysRevLett.121.060601} {\bibfield
  {journal} {\bibinfo  {journal} {Phys.\ Rev.Lett.}\ }\textbf {\bibinfo
  {volume} {121}},\ \bibinfo {pages} {060601} (\bibinfo {year}
  {2018}{\natexlab{b}})}\BibitemShut {NoStop}%
\bibitem [{\citenamefont {{Junyu Liu}}(2018)}]{SFF10}%
  \BibitemOpen
  \bibfield  {author} {\bibinfo {author} {\bibnamefont {{Junyu Liu}}},\ }\href
  {https://link.aps.org/doi/10.1103/PhysRevD.98.086026} {\bibfield  {journal}
  {\bibinfo  {journal} {Phys.\ Rev.D}\ }\textbf {\bibinfo {volume} {98}},\
  \bibinfo {pages} {086026} (\bibinfo {year} {2018})}\BibitemShut {NoStop}%
\bibitem [{\citenamefont {{S. Choudhury, and A. Mukherjee}}(2019)}]{SFF11}%
  \BibitemOpen
  \bibfield  {author} {\bibinfo {author} {\bibnamefont {{S. Choudhury, and A.
  Mukherjee}}},\ }\href {https://doi.org/10.1007/JHEP05(2019)149} {\bibfield
  {journal} {\bibinfo  {journal} {J. High Energy Phys.}\ }\textbf {\bibinfo
  {volume} {05}},\ \bibinfo {pages} {149(2019)}}\BibitemShut {NoStop}%
\bibitem [{\citenamefont {{M. Pouranvari and J. Abouie}}(2019)}]{model1}%
  \BibitemOpen
  \bibfield  {author} {\bibinfo {author} {\bibnamefont {{M. Pouranvari and J.
  Abouie}}},\ }\href {https://link.aps.org/doi/10.1103/PhysRevB.100.195109}
  {\bibfield  {journal} {\bibinfo  {journal} {Phys.\ Rev.B}\ }\textbf {\bibinfo
  {volume} {100}},\ \bibinfo {pages} {195109} (\bibinfo {year}
  {2019})}\BibitemShut {NoStop}%
\bibitem [{\citenamefont {Roy}\ and\ \citenamefont {Sharma}(2018)}]{EE6}%
  \BibitemOpen
  \bibfield  {author} {\bibinfo {author} {\bibfnamefont {N.}~\bibnamefont
  {Roy}}\ and\ \bibinfo {author} {\bibfnamefont {A.}~\bibnamefont {Sharma}},\
  }\href {https://link.aps.org/doi/10.1103/PhysRevB.97.125116} {\bibfield
  {journal} {\bibinfo  {journal} {Physical Review B}\ }\textbf {\bibinfo
  {volume} {97}},\ \bibinfo {pages} {125116} (\bibinfo {year}
  {2018})}\BibitemShut {NoStop}%
\bibitem [{\citenamefont {{S. Aubry and G. Andre}}(1980)}]{model2}%
  \BibitemOpen
  \bibfield  {author} {\bibinfo {author} {\bibnamefont {{S. Aubry and G.
  Andre}}},\ }\href@noop {} {\bibfield  {journal} {\bibinfo  {journal} {Israel
  Phys. Soc.}\ }\textbf {\bibinfo {volume} {3}},\ \bibinfo {pages} {18}
  (\bibinfo {year} {1980})}\BibitemShut {NoStop}%
\bibitem [{\citenamefont {{P. G. Harper}}(1955)}]{model3}%
  \BibitemOpen
  \bibfield  {author} {\bibinfo {author} {\bibnamefont {{P. G. Harper}}},\
  }\href {https://doi.org/10.1088/0370-1298/68/10/304} {\bibfield  {journal}
  {\bibinfo  {journal} {Proc. Phys. Soc. A}\ }\textbf {\bibinfo {volume}
  {68}},\ \bibinfo {pages} {874} (\bibinfo {year} {1955})}\BibitemShut
  {NoStop}%
\bibitem [{\citenamefont {{N. Roy and A. Sharma}}(2019)}]{model4}%
  \BibitemOpen
  \bibfield  {author} {\bibinfo {author} {\bibnamefont {{N. Roy and A.
  Sharma}}},\ }\href {https://link.aps.org/doi/10.1103/PhysRevB.100.195143}
  {\bibfield  {journal} {\bibinfo  {journal} {Phys.\ Rev.B}\ }\textbf {\bibinfo
  {volume} {100}},\ \bibinfo {pages} {195143} (\bibinfo {year}
  {2019})}\BibitemShut {NoStop}%
\bibitem [{\citenamefont {{P. Y. Chang, X. Chen, S. Gopalakrishnan,and J. H.
  Pixley}}(2019)}]{model5}%
  \BibitemOpen
  \bibfield  {author} {\bibinfo {author} {\bibnamefont {{P. Y. Chang, X. Chen,
  S. Gopalakrishnan,and J. H. Pixley}}},\ }\href
  {https://link.aps.org/doi/10.1103/PhysRevLett.123.190602} {\bibfield
  {journal} {\bibinfo  {journal} {Phys.\ Rev.Lett.}\ }\textbf {\bibinfo
  {volume} {123}},\ \bibinfo {pages} {190602} (\bibinfo {year}
  {2019})}\BibitemShut {NoStop}%
\bibitem [{\citenamefont {{R. Vosk and E. Altman}}(2013)}]{EED1}%
  \BibitemOpen
  \bibfield  {author} {\bibinfo {author} {\bibnamefont {{R. Vosk and E.
  Altman}}},\ }\href {https://link.aps.org/doi/10.1103/PhysRevLett.110.067204}
  {\bibfield  {journal} {\bibinfo  {journal} {Phys.\ Rev.Lett.}\ }\textbf
  {\bibinfo {volume} {110}},\ \bibinfo {pages} {067204} (\bibinfo {year}
  {2013})}\BibitemShut {NoStop}%
\bibitem [{\citenamefont {{P. Calabrese and J. Cardy}}(2016)}]{EED2}%
  \BibitemOpen
  \bibfield  {author} {\bibinfo {author} {\bibnamefont {{P. Calabrese and J.
  Cardy}}},\ }\href {https://doi.org/10.1088/1742-5468/2016/06/064003}
  {\bibfield  {journal} {\bibinfo  {journal} {J. Stat. Mech.}\ ,\ \bibinfo
  {pages} {064003}} (\bibinfo {year} {2016})}\BibitemShut {NoStop}%
\bibitem [{\citenamefont {{P. Calabrese and J. Cardy}}(2005)}]{EED3}%
  \BibitemOpen
  \bibfield  {author} {\bibinfo {author} {\bibnamefont {{P. Calabrese and J.
  Cardy}}},\ }\href {https://doi.org/10.1088/1742-5468/2005/04/p04010}
  {\bibfield  {journal} {\bibinfo  {journal} {J. Stat. Mech.}\ ,\ \bibinfo
  {pages} {P04010}} (\bibinfo {year} {2005})}\BibitemShut {NoStop}%
\bibitem [{\citenamefont {{M. Modugno}}(2009)}]{model7}%
  \BibitemOpen
  \bibfield  {author} {\bibinfo {author} {\bibnamefont {{M. Modugno}}},\ }\href
  {https://doi.org/10.1088/1367-2630/11/3/033023} {\bibfield  {journal}
  {\bibinfo  {journal} {New J. Phys.}\ }\textbf {\bibinfo {volume} {11}},\
  \bibinfo {pages} {033023} (\bibinfo {year} {2009})}\BibitemShut {NoStop}%
\bibitem [{\citenamefont {Evers}\ and\ \citenamefont {Mirlin}(2008)}]{model8}%
  \BibitemOpen
  \bibfield  {author} {\bibinfo {author} {\bibfnamefont {F.}~\bibnamefont
  {Evers}}\ and\ \bibinfo {author} {\bibfnamefont {A.~D.}\ \bibnamefont
  {Mirlin}},\ }\bibfield  {title} {\bibinfo {title} {Anderson transitions},\
  }\href {https://doi.org/10.1103/RevModPhys.80.1355} {\bibfield  {journal}
  {\bibinfo  {journal} {Rev. Mod. Phys.}\ }\textbf {\bibinfo {volume} {80}},\
  \bibinfo {pages} {1355} (\bibinfo {year} {2008})}\BibitemShut {NoStop}%
\bibitem [{\citenamefont {{D. J. Thouless}}(1983)}]{model6}%
  \BibitemOpen
  \bibfield  {author} {\bibinfo {author} {\bibnamefont {{D. J. Thouless}}},\
  }\href {https://link.aps.org/doi/10.1103/PhysRevB.28.4272} {\bibfield
  {journal} {\bibinfo  {journal} {Phys.\ Rev.B}\ }\textbf {\bibinfo {volume}
  {28}},\ \bibinfo {pages} {4272} (\bibinfo {year} {1983})}\BibitemShut
  {NoStop}%
\bibitem [{\citenamefont {{R. E. Prange}}(1997)}]{avg1}%
  \BibitemOpen
  \bibfield  {author} {\bibinfo {author} {\bibnamefont {{R. E. Prange}}},\
  }\href {https://link.aps.org/doi/10.1103/PhysRevLett.78.2280} {\bibfield
  {journal} {\bibinfo  {journal} {Phys.\ Rev. Lett.}\ }\textbf {\bibinfo
  {volume} {78}},\ \bibinfo {pages} {2280} (\bibinfo {year}
  {1997})}\BibitemShut {NoStop}%
\bibitem [{\citenamefont {{V. Balasubramanian, B. Craps, B. L. Czechc and G.
  Sarosia}}(2017)}]{avg2}%
  \BibitemOpen
  \bibfield  {author} {\bibinfo {author} {\bibnamefont {{V. Balasubramanian, B.
  Craps, B. L. Czechc and G. Sarosia}}},\ }\href
  {https://doi.org/10.1007/JHEP03(2017)154} {\bibfield  {journal} {\bibinfo
  {journal} {J. High Energy Phys.}\ }\textbf {\bibinfo {volume} {03}},\
  \bibinfo {pages} {154(2017)}}\BibitemShut {NoStop}%
\bibitem [{\citenamefont {{T. Guhr, A. MllerGroeling, and H. A.
  Weidenmller}}(1998)}]{avg3}%
  \BibitemOpen
  \bibfield  {author} {\bibinfo {author} {\bibnamefont {{T. Guhr, A.
  MllerGroeling, and H. A. Weidenmller}}},\ }\href
  {https://www.sciencedirect.com/science/article/pii/S0370157397000884}
  {\bibfield  {journal} {\bibinfo  {journal} {Phys.Rep.}\ }\textbf {\bibinfo
  {volume} {299}},\ \bibinfo {pages} {189} (\bibinfo {year}
  {1998})}\BibitemShut {NoStop}%
\bibitem [{\citenamefont {{I. Peschel}}(2003)}]{EE1}%
  \BibitemOpen
  \bibfield  {author} {\bibinfo {author} {\bibnamefont {{I. Peschel}}},\ }\href
  {https://doi.org/10.1088/0305-4470/36/14/101} {\bibfield  {journal} {\bibinfo
   {journal} {J. Phys. A: Math. Gen.}\ }\textbf {\bibinfo {volume} {36}},\
  \bibinfo {pages} {L205} (\bibinfo {year} {2003})}\BibitemShut {NoStop}%
\bibitem [{\citenamefont {{I. Peschel and V. Eisler}}(2009)}]{EE2}%
  \BibitemOpen
  \bibfield  {author} {\bibinfo {author} {\bibnamefont {{I. Peschel and V.
  Eisler}}},\ }\href {https://doi.org/10.1088/1751-8113/42/50/504003}
  {\bibfield  {journal} {\bibinfo  {journal} {J. Phys. A: Math. Theor.}\
  }\textbf {\bibinfo {volume} {42}},\ \bibinfo {pages} {504003} (\bibinfo
  {year} {2009})}\BibitemShut {NoStop}%
\bibitem [{\citenamefont {{I. Peschel}}(2012)}]{EE3}%
  \BibitemOpen
  \bibfield  {author} {\bibinfo {author} {\bibnamefont {{I. Peschel}}},\ }\href
  {https://doi.org/10.1007/s13538-012-0074-1} {\bibfield  {journal} {\bibinfo
  {journal} {Braz. J. Phys.}\ }\textbf {\bibinfo {volume} {42}},\ \bibinfo
  {pages} {267} (\bibinfo {year} {2012})}\BibitemShut {NoStop}%
\bibitem [{\citenamefont {Eisler}\ and\ \citenamefont
  {Peschel}(2012)}]{Eisler_2012}%
  \BibitemOpen
  \bibfield  {author} {\bibinfo {author} {\bibfnamefont {V.}~\bibnamefont
  {Eisler}}\ and\ \bibinfo {author} {\bibfnamefont {I.}~\bibnamefont
  {Peschel}},\ }\href {https://doi.org/10.1209/0295-5075/99/20001} {\bibfield
  {journal} {\bibinfo  {journal} {{EPL} (Europhysics Letters)}\ }\textbf
  {\bibinfo {volume} {99}},\ \bibinfo {pages} {20001} (\bibinfo {year}
  {2012})}\BibitemShut {NoStop}%
\bibitem [{\citenamefont {{Y. Y. Atas, E. Bogomolny, O. Giraud, and G.
  Roux}}(2013)}]{gap-ratio}%
  \BibitemOpen
  \bibfield  {author} {\bibinfo {author} {\bibnamefont {{Y. Y. Atas, E.
  Bogomolny, O. Giraud, and G. Roux}}},\ }\href
  {https://link.aps.org/doi/10.1103/PhysRevLett.110.084101} {\bibfield
  {journal} {\bibinfo  {journal} {Phys.Rev. Lett.}\ }\textbf {\bibinfo {volume}
  {110}},\ \bibinfo {pages} {084101} (\bibinfo {year} {2013})}\BibitemShut
  {NoStop}%
\bibitem [{\citenamefont {{H. Casini and M. Huerta }}(2009)}]{Renyi}%
  \BibitemOpen
  \bibfield  {author} {\bibinfo {author} {\bibnamefont {{H. Casini and M.
  Huerta }}},\ }\href {https://doi.org/10.1088/1751-8113/42/50/504007}
  {\bibfield  {journal} {\bibinfo  {journal} {Phys. A: Math. Theor.}\ }\textbf
  {\bibinfo {volume} {42}},\ \bibinfo {pages} {504007} (\bibinfo {year}
  {2009})}\BibitemShut {NoStop}%
\bibitem [{\citenamefont {Rakovszky}\ \emph {et~al.}(2019)\citenamefont
  {Rakovszky}, \citenamefont {Gopalakrishnan}, \citenamefont {Parameswaran},\
  and\ \citenamefont {Pollmann}}]{model9}%
  \BibitemOpen
  \bibfield  {author} {\bibinfo {author} {\bibfnamefont {T.}~\bibnamefont
  {Rakovszky}}, \bibinfo {author} {\bibfnamefont {S.}~\bibnamefont
  {Gopalakrishnan}}, \bibinfo {author} {\bibfnamefont {S.~A.}\ \bibnamefont
  {Parameswaran}},\ and\ \bibinfo {author} {\bibfnamefont {F.}~\bibnamefont
  {Pollmann}},\ }\bibfield  {title} {\bibinfo {title} {Signatures of
  information scrambling in the dynamics of the entanglement spectrum},\ }\href
  {https://doi.org/10.1103/PhysRevB.100.125115} {\bibfield  {journal} {\bibinfo
   {journal} {Phys. Rev. B}\ }\textbf {\bibinfo {volume} {100}},\ \bibinfo
  {pages} {125115} (\bibinfo {year} {2019})}\BibitemShut {NoStop}%
\bibitem [{\citenamefont {{M. Pouranvari}}(2020)}]{EE4}%
  \BibitemOpen
  \bibfield  {author} {\bibinfo {author} {\bibnamefont {{M. Pouranvari}}},\
  }\href {https://doi.org/10.1140/epjb/e2020-10052-3} {\bibfield  {journal}
  {\bibinfo  {journal} {The Euro. Phys. Jour. B}\ }\textbf {\bibinfo {volume}
  {93}},\ \bibinfo {pages} {6} (\bibinfo {year} {2020})}\BibitemShut {NoStop}%
\bibitem [{\citenamefont {{R. Modak and T. Nag}}(2020)}]{EE5}%
  \BibitemOpen
  \bibfield  {author} {\bibinfo {author} {\bibnamefont {{R. Modak and T.
  Nag}}},\ }\href {https://link.aps.org/doi/10.1103/PhysRevResearch.2.012074}
  {\bibfield  {journal} {\bibinfo  {journal} {Phys.\ Rev.Research.2}\ ,\
  \bibinfo {pages} {012074(R)}} (\bibinfo {year} {2020})}\BibitemShut {NoStop}%
\bibitem [{\citenamefont {Beck}\ and\ \citenamefont {Schögl}(1993)}]{Beck}%
  \BibitemOpen
  \bibfield  {author} {\bibinfo {author} {\bibfnamefont {C.}~\bibnamefont
  {Beck}}\ and\ \bibinfo {author} {\bibfnamefont {F.}~\bibnamefont {Schögl}},\
  }\href {https://doi.org/10.1017/CBO9780511524585} {\emph {\bibinfo {title}
  {Thermodynamics of Chaotic Systems: An Introduction}}},\ Cambridge Nonlinear
  Science Series\ (\bibinfo  {publisher} {Cambridge University Press},\
  \bibinfo {year} {1993})\BibitemShut {NoStop}%
\bibitem [{\citenamefont {{A. Pandey and R. Ramaswamy }}(1991)}]{Harmonic}%
  \BibitemOpen
  \bibfield  {author} {\bibinfo {author} {\bibnamefont {{A. Pandey and R.
  Ramaswamy }}},\ }\href {https://link.aps.org/doi/10.1103/PhysRevA.43.4237}
  {\bibfield  {journal} {\bibinfo  {journal} {Phys.\ Rev. A}\ }\textbf
  {\bibinfo {volume} {43}},\ \bibinfo {pages} {4237} (\bibinfo {year}
  {1991})}\BibitemShut {NoStop}%
\bibitem [{\citenamefont {{S. H. Tekur,U. T. Bhosale,and M. S.
  Santhanam}}(2018)}]{Higher-order}%
  \BibitemOpen
  \bibfield  {author} {\bibinfo {author} {\bibnamefont {{S. H. Tekur,U. T.
  Bhosale,and M. S. Santhanam}}},\ }\href
  {https://link.aps.org/doi/10.1103/PhysRevB.98.104305} {\bibfield  {journal}
  {\bibinfo  {journal} {Phys.\ Rev.B}\ }\textbf {\bibinfo {volume} {98}},\
  \bibinfo {pages} {104305} (\bibinfo {year} {2018})}\BibitemShut {NoStop}%
\bibitem [{\citenamefont {{S. H. Tekur and M. S.
  Santhanam}}(2020)}]{Higher-order-P}%
  \BibitemOpen
  \bibfield  {author} {\bibinfo {author} {\bibnamefont {{S. H. Tekur and M. S.
  Santhanam}}},\ }\href
  {https://link.aps.org/doi/10.1103/PhysRevResearch.2.032063} {\bibfield
  {journal} {\bibinfo  {journal} {Phys.\ Rev.Research.2}\ ,\ \bibinfo {pages}
  {032063(R)}} (\bibinfo {year} {2020})}\BibitemShut {NoStop}%
\bibitem [{\citenamefont {Prakash}\ \emph {et~al.}(2021)\citenamefont
  {Prakash}, \citenamefont {Pixley},\ and\ \citenamefont
  {Kulkarni}}]{prakash2021universal}%
  \BibitemOpen
  \bibfield  {author} {\bibinfo {author} {\bibfnamefont {A.}~\bibnamefont
  {Prakash}}, \bibinfo {author} {\bibfnamefont {J.~H.}\ \bibnamefont
  {Pixley}},\ and\ \bibinfo {author} {\bibfnamefont {M.}~\bibnamefont
  {Kulkarni}},\ }\href
  {https://link.aps.org/doi/10.1103/PhysRevResearch.3.L012019} {\bibfield
  {journal} {\bibinfo  {journal} {Phys.\ Rev.Research}\ }\textbf {\bibinfo
  {volume} {3}},\ \bibinfo {pages} {L012019} (\bibinfo {year}
  {2021})}\BibitemShut {NoStop}%
\bibitem [{\citenamefont {{D. Roy and T. Prosen }}(2020)}]{roy2020random}%
  \BibitemOpen
  \bibfield  {author} {\bibinfo {author} {\bibnamefont {{D. Roy and T. Prosen
  }}},\ }\href {https://link.aps.org/doi/10.1103/PhysRevE.102.060202}
  {\bibfield  {journal} {\bibinfo  {journal} {Phys. Rev. E}\ }\textbf {\bibinfo
  {volume} {102}},\ \bibinfo {pages} {060202(R)} (\bibinfo {year}
  {2020})}\BibitemShut {NoStop}%
\bibitem [{\citenamefont {Bertini}\ \emph {et~al.}(2018)\citenamefont
  {Bertini}, \citenamefont {Kos},\ and\ \citenamefont
  {Prosen}}]{bertini2018exact}%
  \BibitemOpen
  \bibfield  {author} {\bibinfo {author} {\bibfnamefont {B.}~\bibnamefont
  {Bertini}}, \bibinfo {author} {\bibfnamefont {P.}~\bibnamefont {Kos}},\ and\
  \bibinfo {author} {\bibfnamefont {T.}~\bibnamefont {Prosen}},\ }\href
  {https://link.aps.org/doi/10.1103/PhysRevLett.121.264101} {\bibfield
  {journal} {\bibinfo  {journal} {Phys.\ Rev. Lett.}\ }\textbf {\bibinfo
  {volume} {121}},\ \bibinfo {pages} {264101} (\bibinfo {year}
  {2018})}\BibitemShut {NoStop}%
\bibitem [{\citenamefont {{F. Carollo and C.
  Perez-Espigares}}(2020)}]{carollo2020entanglement}%
  \BibitemOpen
  \bibfield  {author} {\bibinfo {author} {\bibnamefont {{F. Carollo and C.
  Perez-Espigares}}},\ }\href
  {https://link.aps.org/doi/10.1103/PhysRevE.102.030104} {\bibfield  {journal}
  {\bibinfo  {journal} {Phys. Rev. E}\ }\textbf {\bibinfo {volume} {102}},\
  \bibinfo {pages} {030104(R)} (\bibinfo {year} {2020})}\BibitemShut {NoStop}%
\bibitem [{\citenamefont {{A. R. Kolovsky}}(2020)}]{kolovsky2020quantum}%
  \BibitemOpen
  \bibfield  {author} {\bibinfo {author} {\bibnamefont {{A. R. Kolovsky}}},\
  }\href {https://link.aps.org/doi/10.1103/PhysRevE.101.062116} {\bibfield
  {journal} {\bibinfo  {journal} {Phys. Rev. E}\ }\textbf {\bibinfo {volume}
  {101}},\ \bibinfo {pages} {062116} (\bibinfo {year} {2020})}\BibitemShut
  {NoStop}%
\bibitem [{\citenamefont {{Chang, Po-Yao and You, Jhih-Shih and Wen, Xueda and
  Ryu, Shinsei}}(2020)}]{chang2020entanglement}%
  \BibitemOpen
  \bibfield  {author} {\bibinfo {author} {\bibnamefont {{Chang, Po-Yao and You,
  Jhih-Shih and Wen, Xueda and Ryu, Shinsei}}},\ }\href
  {https://link.aps.org/doi/10.1103/PhysRevResearch.2.033069} {\bibfield
  {journal} {\bibinfo  {journal} {Phys.\ Rev.Research}\ }\textbf {\bibinfo
  {volume} {2}},\ \bibinfo {pages} {033069} (\bibinfo {year}
  {2020})}\BibitemShut {NoStop}%
\bibitem [{\citenamefont {Li}\ \emph {et~al.}(2021)\citenamefont {Li},
  \citenamefont {Prosen},\ and\ \citenamefont {Chan}}]{li2021spectral}%
  \BibitemOpen
  \bibfield  {author} {\bibinfo {author} {\bibfnamefont {J.}~\bibnamefont
  {Li}}, \bibinfo {author} {\bibfnamefont {T.}~\bibnamefont {Prosen}},\ and\
  \bibinfo {author} {\bibfnamefont {A.}~\bibnamefont {Chan}},\ }\href
  {https://arxiv.org/abs/2103.05001} {\bibfield  {journal} {\bibinfo  {journal}
  {arXiv preprint arXiv:2103.05001}\ } (\bibinfo {year} {2021})}\BibitemShut
  {NoStop}%
\bibitem [{\citenamefont {Gong}\ and\ \citenamefont
  {Ueda}(2018)}]{gong2018topological}%
  \BibitemOpen
  \bibfield  {author} {\bibinfo {author} {\bibfnamefont {Z.}~\bibnamefont
  {Gong}}\ and\ \bibinfo {author} {\bibfnamefont {M.}~\bibnamefont {Ueda}},\
  }\href {https://link.aps.org/doi/10.1103/PhysRevLett.121.250601} {\bibfield
  {journal} {\bibinfo  {journal} {Phys.\ Rev. Lett.}\ }\textbf {\bibinfo
  {volume} {121}},\ \bibinfo {pages} {250601} (\bibinfo {year}
  {2018})}\BibitemShut {NoStop}%
\bibitem [{\citenamefont {Su}\ \emph {et~al.}(2020)\citenamefont {Su},
  \citenamefont {Sun},\ and\ \citenamefont {Fan}}]{su2020quench}%
  \BibitemOpen
  \bibfield  {author} {\bibinfo {author} {\bibfnamefont {K.}~\bibnamefont
  {Su}}, \bibinfo {author} {\bibfnamefont {Z.-H.}\ \bibnamefont {Sun}},\ and\
  \bibinfo {author} {\bibfnamefont {H.}~\bibnamefont {Fan}},\ }\href
  {https://link.aps.org/doi/10.1103/PhysRevA.101.063613} {\bibfield  {journal}
  {\bibinfo  {journal} {Phys.\ Rev.A}\ }\textbf {\bibinfo {volume} {101}},\
  \bibinfo {pages} {063613} (\bibinfo {year} {2020})}\BibitemShut {NoStop}%
\end{thebibliography}%

\end{document}